\documentclass[english]{article}
\usepackage[LGR,T1]{fontenc}
\usepackage[latin9]{inputenc}
\usepackage{color}
\usepackage{amsbsy}
\usepackage{amssymb}
\usepackage{esint}
\usepackage{graphicx}
\usepackage{subfigure}
\usepackage{float}
\newcommand{\kpar}{k_{\parallel}}
\makeatletter

\DeclareRobustCommand{\greektext}{%
  \fontencoding{LGR}\selectfont\def\encodingdefault{LGR}}
\DeclareRobustCommand{\textgreek}[1]{\leavevmode{\greektext #1}}
\ProvideTextCommand{\~}{LGR}[1]{\char126#1}

\newcommand{\lyxaddress}[1]{
	\par {\raggedright #1
	\vspace{1.4em}
	\noindent\par}
}

\makeatother

\usepackage{babel}
\begin{document}
\title{\textbf{Amplitude representation of Landau-Lifshitz equation and its
application to ferromagnetic films.}}
\author{Gang Li$^{1*}$ and Valery Pokrovsky$^{1,2}$}
\maketitle

\lyxaddress{$^{1}$Department of Physics and Astronomy, Texas A\&M University,
College Station, TX 77843-4242, USA.\\ $^{2}$Landau Institute for Theoretical
Physics of Russian Academy of Sciences, Chernogolovka, 142432, Russian
Federation.\\
$^{*}$dgzy03@gmail.com}

\section{Introduction}

In 1935 Lev Landau and Evgenii Lifshitz set the foundation of static
and dynamics of weakly anisotropic ferromagnets \cite{Landau 1935, Landau 1984}.
They formulated the famous Landau-Lifshitz equation (LLE) that regulates
the motion of the ferromagnet magnetization in the long-wave low-frequency
limit. The purpose of this article is to develop a systematic approach
to the solution of the LLE in terms of the magnon wave function $\psi\left(\mathbf{r}\right)$
and apply it to physical phenomena in a thin ferromagnetic film.

This problem has a long history. First such approach was proposed
by Schlöman in 1959 \cite{Schlöman 1959} for a bulk ferromagnet. It was developed
and improved by Carl Patton and his coworkers (see references in the
review article by Krivosik and Patton \cite{Patton 2010}). The applications
focused on the ferromagnetic resonance (FMR) and the spin momentum
transfer, i.e., spin currents.

The theoretical study of ferromagnetic films started also in the the
middle of 20-th century by the seminal work of Damon and Eshbach \cite{Damon 1961}.
They have found exact solution of the LLE equation for an infinite
ferromagnetic film in which spins interact only through the dipolar
forces. In sufficiently thick films the evanescent waves propagating
in opposite direction at the two surfaces appear. They create a mechanical
torque acting on the film.

Gann \cite{Gann 1967}, De Wames and Wolfram \cite{Wolfram 1970}, Kalinikos
and Slavin \cite{Kalinikos 1980,Kalinikos 1986} extended the Damon-Eshbach theory to
a more general situation in which the spins interact also through
the exchange forces. An extension of these exact solutions for the
tilted external magnetic field was found by Arias \cite{Arias 2016}. In
the work by the authors, Chen Sun and Thomas Nattermann \cite{Li 2018}
the solution was extended to the wide range of the film thickness.
It enabled us to follow the transition from the magnon spectrum with
two symmetric minima in thick films to one-minimum spectra in thin
films.

The latter result was inspired by the discovery of the Bose-Einstein
condensation of magnons (BECM) at room temperature under permanent
pumping of electromagnetic waves made in 2006 by Demokritov \textit{et
al}\cite{Demokritov 2006}. The BECM was found in the Yttrium Iron Garnet (YIG),
a strongly insulating ferrite. For long-wave excitations all spins
in the primitive cells move as a whole. It means that in this regime
the ferrite is indistinguishable from a ferromagnet.

The amplitude representation (AR) is ideally adjusted to describe
the condensation. The condensate amplitudes $\psi_{\pm}$ are the
Fourier components of the coordinate wave functions $\psi\left(\mathbf{r}\right)$
at the values of wave vector $\mathbf{k=\pm Q}$ corresponding to
the two symmetric minima of magnon energy. Since we are mostly interested
in the properties of the condensate and its interaction with excited
magnons, our focus in the study of the (AR) will be different that
in already cited works by Schlöman and Patton. Certainly, some overlapping
is unavoidable, but we try to minimize it.

This article has also a purpose to represent the modern state of art for the properties of ferromagnetic films and the pumping-induced BECM in them at room temperature. Thus, it can be considered as a review on basic principles and the recent advances in the field. 

\section{Hamiltonian formulation of the Landau-Lifshitz equation and Amplitude
representation.}

\subsection{Poisson brackets for spins, magnetic moments and magnetization
in discrete and continuous models.}

Let us start with a discrete 3d-model of the ferromagnet, in which
all spins $\mathbf{S_{r}}$ are located in the centers of cubic cells
of volume $v_{0}$ labeled by vectors $\mathbf{r}$. The Poisson brackets
for the components of spins are:
\begin{equation}
\left\{ S_{k}\left(\mathbf{r}\right),S_{l}\left(\mathbf{r^{\prime}}\right)\right\} =\delta_{\mathbf{r},\mathbf{r}^{\prime}}\varepsilon_{klm}S_{m}\left(\mathbf{r}\right),\label{eq:Poisson-spins}
\end{equation}
where Kronecker symbol $\delta_{\mathbf{r},\mathbf{r}^{\prime}}$
is equal to 1 when $\mathbf{r=r^{\prime}}$ and 0 otherwise; $\varepsilon_{klm}$
is absolutely antisymmetric 3d tensor with $k,l,m$ independently
taking values 1,2,3 or $x,y,z$ that is equal to +1 if the permutation
$k,l,m$ is even and -1 if it is odd. We use the Einstein convention
that the summation must be performed over repeated indices.

The magnetic moment of a primitive cell is 
\begin{equation}
\mathfrak{\mathbf{\mathcal{M}_{\mathit{k}}=\gamma\mathit{S_{k},}}}\label{eq:magn-mom-spin}
\end{equation}
where $\gamma=\frac{e}{2mc}$ is the classical gyromagnetic ratio.
The relation (\ref{eq:magn-mom-spin}) becomes evident if one remembers
that a spin projection, for example $S_{z}$, is quantized in units
$\hbar$. As a consequence, the magnetic moment projection is quantized
in units of the Bohr's magneton $\mu_{B}=\frac{e\hbar}{2mc}$. Eq. (\ref{eq:Poisson-spins})
implies that the Poisson brackets for the components of the magnetic moments
are:
\begin{equation}
\left\{ \mathcal{M}_{k}\left(\mathbf{r}\right),\mathcal{M}_{l}\left(\mathbf{r^{\prime}}\right)\right\} =\gamma\delta_{\mathbf{r},\mathbf{r}^{\prime}}\varepsilon_{klm}\mathcal{M}_{m}\left(\mathbf{r}\right).\label{eq:Poisson-magmoments}
\end{equation}
The magnetization is defined as magnetic moment of unit volume. It
is expressed in terms of magnetic moments as $\boldsymbol{M\left(\mathbf{r}\right)=\frac{\mathcal{M\left(\mathbf{r}\right)}}{\mathit{v_{0}}}}$.
Therefore the Poisson brackets for magnetization in the discrete model
are:
\begin{equation}
\left\{ M_{k}\left(\mathbf{r}\right),M_{l}\left(\mathbf{r^{\prime}}\right)\right\} =\frac{\gamma}{v_{0}}\delta_{\mathbf{r},\mathbf{r}^{\prime}}\varepsilon_{klm}M_{m}\left(\mathbf{r}\right).\label{eq:Poisson-Magn-d}
\end{equation}
In continuous approximation the ratio $\frac{\delta_{\mathbf{r},\mathbf{r}^{\prime}}}{v_{0}}$
transits into the Dirac $\delta$-function:
\begin{equation}
\lim_{v_{0}\rightarrow0}\frac{\delta_{\mathbf{r},\mathbf{r}^{\prime}}}{v_{0}}=\delta\left(\mathbf{r}-\mathbf{r}^{\prime}\right).\label{eq:cont-lim-delta}
\end{equation}
To prove this statement let us introduce an arbitrary continuous function
$f\left(\mathbf{r}\right)$. Let us consider a sum over the cites
of the discrete model:
\[
v_{0}\sum_{\mathbf{\mathbf{r}^{\prime}}}\frac{\delta_{\mathbf{r},\mathbf{r}^{\prime}}}{v_{0}}f\left(\mathbf{r}^{\prime}\right)=f(\mathbf{r)}.
\]
In continuous limit $v_{0}\sum_{\mathbf{\mathbf{r}^{\prime}}}\rightarrow\intop d^{3}r^{\prime}$,
which, together with previous equation, proves eq. (\ref{eq:cont-lim-delta}).

Thus, the Poisson brackets for components of magnetization in continuous
limit are:
\begin{equation}
\left\{ M_{k}\left(\mathbf{r}\right),M_{l}\left(\mathbf{r}^{\prime}\right)\right\} =\gamma\delta\left(\mathbf{r}-\mathbf{r}^{\prime}\right)\varepsilon_{klm}M_{m}\label{eq:Poisson-magntz-cont}
\end{equation}
It is convenient to rewrite these relations explicitly as;
\begin{equation}
\left\{ M_{x}\left(\mathbf{r}\right),M_{y}\left(\mathbf{r}^{\prime}\right)\right\} =\gamma\delta\left(\mathbf{r}-\mathbf{r}^{\prime}\right)M_{z}\left(\mathbf{r}\right)\label{eq:Poisson-magntz-xyz}
\end{equation}
Two other Poisson brackets can be obtained from (\ref{eq:Poisson-magntz-xyz})
by the cyclical permutation of the indices $x,y$ and $z$. For further
applications it is useful to introduce complex transverse magnetizations:
\begin{equation}
M_{\pm}\left(\mathbf{r}\right)=M_{x}\left(\mathbf{r}\right)\pm iM_{y}\left(\mathbf{r}\right)\label{eq:complex-M-tr}
\end{equation}
For them eq. (\ref{eq:Poisson-magntz-xyz}) implies the following
Poisson brackets:
\begin{equation}
\begin{array}{c}
\left\{ M_{+}\left(\mathbf{r}\right),M_{-}\left(\mathbf{r^{\prime}}\right)\right\} =-2i\gamma\delta\left(\mathbf{r}-\mathbf{r}^{\prime}\right)M_{z}\left(\mathbf{r}\right)\\
\left\{ M_{\pm}\left(\mathbf{r}\right),M_{z}\left(\mathbf{r^{\prime}}\right)\right\} =\pm i\gamma\delta\left(\mathbf{r}-\mathbf{r}^{\prime}\right)M_{\pm}\left(\mathbf{r}\right)
\end{array}\label{eq:Poisson+-}
\end{equation}

\subsection{Amplitude representation and Poisson brackets for the magnon
wave function.}
\begin{figure}[!htb]
\centering
{\includegraphics[width=10cm]{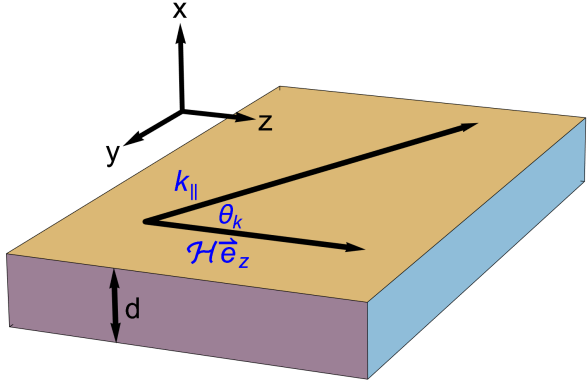}}
\caption{
The coordinate system for a ferromagnetic film of thickness $d$: $z-$axis is chosen along the common direction of the magnetic field and
static magnetization, $x-$axis is perpendicular to the film, $\theta_k$ is the angle between the magnon wave vector and magnetic field.}
	\label{fig:geometry}
\end{figure}

Let the spontaneous magnetization and external magnetic field be
directed along $z-$axis, perpendicular to its direction in the
plane of film be $y$ and direction perpendicular
to the film $x$ as shown in Fig. {\ref{fig:geometry}}. The wave function of magnons $\psi\left(\mathbf{r}\right)$ is determined
by the magnon classical Holstein-Primakoff transformation:
\begin{equation}
\begin{array}{c}
M_{+}\left(\mathbf{r}\right)=\sqrt{\mu_{B}}\psi\left(\mathbf{r}\right)\sqrt{2M-\mu_{B}\psi^{*}\left(\mathbf{r}\right)\psi\left(\mathbf{r}\right)}\\
M_{-}\left(\mathbf{r}\right)=\sqrt{\mu_{B}}\psi^{*}\left(\mathbf{r}\right)\sqrt{2M-\mu_{B}\psi^{*}\left(\mathbf{r}\right)\psi\left(\mathbf{r}\right)}\\
M_{z}=M-\mu_{B}\psi^{*}\left(\mathbf{r}\right)\psi\left(\mathbf{r}\right)
\end{array},\label{eq:AR}
\end{equation}
where $M$ is the magnitude of magnetization vector that is assumed
to be constant. The third equation (\ref{eq:AR}) shows that the physical
meaning of the square of modulus $\psi^{*}\left(\mathbf{r}\right)\psi\left(\mathbf{r}\right)$
is the density of magnons $n\left(\mathbf{r}\right)$. Note that the
order of factors in eqs. (\ref{eq:AR}) is not important. The second
useful remark is that $\sqrt{2M-\mu_{B}\psi^{*}\left(\mathbf{r}\right)\psi\left(\mathbf{r}\right)}=\sqrt{M+M_{z}}$.

The equations (\ref{eq:Poisson+-}) are compatible with the amplitude
representation (\ref{eq:AR}) if and only if the wave functions satisfy
the following permutation relations:
\begin{equation}
\begin{array}{c}
\left\{ \psi\left(\mathbf{r}\right),\psi^{*}\left(\mathbf{r^{\prime}}\right)\right\} =-\frac{i}{\hbar}\delta\left(\mathbf{r}-\mathbf{r}^{\prime}\right)\\
\left\{ \psi\left(\mathbf{r}\right),\psi\left(\mathbf{r^{\prime}}\right)\right\} =\left\{ \psi^{*}\left(\mathbf{r}\right),\psi^{*}\left(\mathbf{r^{\prime}}\right)\right\} =0
\end{array}\label{eq:Poisson-wf}
\end{equation}
Let us prove this theorem for the second equation (\ref{eq:Poisson+-}).
We will use the algebraic identity valid for any algebra of operators
with defined operations of addition and non-commutative multiplication:
\begin{equation}
\left\{ AB,C\right\} =A\left\{ B,C\right\} C+\{A,C\}B\label{eq:commutator A-BC}
\end{equation}
Employing this rule and the third equation (\ref{eq:AR}), we find:
\[
\begin{array}{c}
\left\{ M_{+},M_{z}\right\} =\sqrt{\mu_{B}}\left\{ \psi\left(\mathbf{r}\right)\sqrt{M+M_{z}}\left(\mathbf{r}\right),M_{z}\left(\mathbf{r^{\prime}}\right)\right\} =\\
\sqrt{\mu_{B}\left(M+M_{z}\left(\mathbf{r}\right)\right)}\left\{ \psi\left(\mathbf{r}\right),M_{z}\left(\mathbf{r^{\prime}}\right)\right\} =-\sqrt{\mu_{B}^{3}\left(M+M_{z}\left(\mathbf{r}\right)\right)}\left\{ \psi\left(\mathbf{r}\right),\psi^{*}\left(\mathbf{r^{\prime}}\right)\psi\left(\mathbf{r}^{\prime}\right)\right\} 
\end{array}
\]
Applying again the identity (\ref{eq:commutator A-BC}) and assuming
that $\left\{ \psi\left(\mathbf{r}\right),\psi\left(\mathbf{r^{\prime}}\right)\right\} =0$,
we arrive at relation
\[
\left\{ M_{+},M_{z}\right\} =-\sqrt{\mu_{B}^{3}\left(M+M_{z}\left(\mathbf{r}\right)\right)}\psi\left(\mathbf{r}^{\prime}\right)\left\{ \psi\left(\mathbf{r}\right),\psi^{*}\left(\mathbf{r^{\prime}}\right)\right\} 
\]
The right-hand
side of this equation must be equal to $i\gamma\delta\left(\mathbf{r}-\mathbf{r}^{\prime}\right)M_{+}\left(\mathbf{r}\right)$ according to the second equation (\ref{eq:Poisson+-}) .
The necessary and sufficient requirement to satisfy this condition
is given by eqs. (\ref{eq:Poisson-wf}). The validity of the first
equation (\ref{eq:Poisson+-}) can be checked by a similar calculation.

\subsection{Landau-Lifshitz Hamiltonian.}

The Landau-Lifshitz Hamiltonian $H_{LL}$ for our problem contains
several parts: the exchange interaction $H_{ex}$, the dipolar interaction
$H_{dip}$ and the Zeeman interaction $H_{Z}$. It can also may contain
the anisotropy (spin-orbit) energy $H_{an}$ . First we write them in terms of
magnetization:
\begin{equation}
H_{LL}=H_{ex}+H_{dip}+H_{Z}+H_{an},\label{eq:H-LL-sum}
\end{equation}
where:
\begin{equation}
H_{ex}=\frac{D}{2}\int\left(\nabla\mathbf{M}\right)^{2}dV\equiv\frac{D}{2}\int\partial_{i}M_{j}\partial_{i}M_{j}dV;\label{eq:H-ex}
\end{equation}
\begin{equation}
H_{z}=-\mathcal{H}\intop M_{z}dV\label{eq:H-z}
\end{equation}
\begin{equation}
H_{dip}=\frac{1}{2}\iint\left(\mathbf{M\nabla}\right)\left(\mathbf{M^{\prime}\nabla^{\prime}}\right)\frac{1}{\left|\mathbf{r-r}^{\prime}\right|}dVdV^{\prime},\label{eq:H-dip}
\end{equation}
In eq. (\ref{eq:H-dip}) we omitted for brevity the arguments in functions
denoting $\mathbf{M=M\left(r\right)};\\
\mathbf{M}^{\prime}=\mathbf{M}\left(\mathbf{r}^{\prime}\right);\boldsymbol{\nabla=\nabla}_{\mathbf{r}};\boldsymbol{\nabla}^{\prime}=\boldsymbol{\nabla}_{\mathbf{r}^{\prime}}$.
When employing the amplitude representation for the components of
magnetization (\ref{eq:AR}), we similarly use abbreviations $\psi\equiv\psi\left(\mathbf{r}\right)$,
$\psi^{\prime}\equiv\psi\left(\mathbf{r}^{\prime}\right)$ and $\partial_{\pm}\equiv\partial_{x}\pm i\partial_{y}$,
$\partial_{\pm}^{\prime}\equiv\partial_{x^{\prime}}\pm i\partial_{y^{\prime}}$.
The exchange constant $\mathit{D}$ determines the exchange length $\ell =\sqrt{D}$ that separates 
the length range, in which the dipolar interaction dominates $l\gg\ell$,  from the range 
$l\ll\ell$ where exchange interaction dominates.

The LL equation assumes that the magnitude of the magnetization vector
rapidly relaxes to its equilibrium value. Thus, the LL equation describes
the relatively slow motion of the vector $\mathbf{M}\left(\mathbf{r},t\right)$
on the sphere. The slowness of this motion in space and time is controlled
by two small parameters $a/\lambda$ and $\omega\tau_{M}$, where
$a$ is the lattice constant, $\lambda$ is the wave-length or another
characteristic length of the magnetization motion, $\omega$ is its
characteristic frequency and $\tau_{M}$ is the relaxation time of
the magnetization magnitude. All magnetic phenomena in this limit
are dominantly classical since the number of magnons in the volume
with the linear size of the order of $\lambda$ is large and the change
of this number by $1$ produces negligibly small change of magnetization.

In terms of amplitudes the three parts of the Hamiltonian given by
equations (\ref{eq:H-ex},\ref{eq:H-z},\ref{eq:H-dip}) are
\begin{equation}
\begin{array}{c}
H_{ex}=\frac{\mu_{B}^{2}\ell^{2}}{2}\intop\left(\nabla\left|\psi\right|^{2}\right)^{2}dV+\\
\frac{\mu_{B}\ell^{2}}{2}\intop\left|\nabla\left(\psi\sqrt{2M-\mu_{B}\left|\psi\right|^{2}}\right)\right|^{2}dV
\end{array}\label{eq:H-ex-psi}
\end{equation}
\begin{equation}
H_{z}=\mu_{B}\mathcal{H}\intop\left|\psi\right|^{2}dV\label{eq:H-z-psi}
\end{equation}
\begin{equation}
H_{dip}=\frac{1}{2}\iint\hat{\Omega}\left(\mathbf{r}\right)\hat{\Omega}\left(\mathbf{r}^{\prime}\right)\frac{dVdV^{\prime}}{\left|\mathbf{r-r}^{\prime}\right|},\label{eq:H-dip-psi}
\end{equation}
where
\begin{equation}
\hat{\Omega}\left(\mathbf{r}\right)=\left(M-\mu_{B}\left|\psi\right|^{2}\right)\partial_{z}+\frac{\sqrt{\mu_{B}\left(2M-\mu_{B}\left|\psi\right|^{2}\right)}}{2}\left(\psi\partial_{-}+\psi^{*}\partial_{+}\right)\label{eq:Omega-hat}
\end{equation}

\section{Spectrum and wave functions of magnons.}

In this section we consider the approximation of free magnons and
find their spectrum and wave function. For that purpose it is necessary
to separate the part of the total Hamiltonian quadratic in amplitudes
$\psi,\psi^{*}$ and diagonalize it.

\subsection{Quadratic part of the Hamiltonian.}

The Zeeman part of the Hamiltonian $H_{Z}$ given by eq. (\ref{eq:H-z-psi})
is naturally quadratic. The quadratic parts of the exchange and dipolar
Hamiltonians are:
\begin{equation}
H_{ex}^{(2)}=\mu_{B}M\ell^{2}\int\left|\nabla\psi\right|^{2}dV\label{eq:H-ex-quad}
\end{equation}
\begin{equation}
H_{dip}^{\left(2\right)}=\frac{\mu_{B}M}{4}\iint\left(\psi\partial_{-}+\psi^{*}\partial_{+}\right)\left(\psi^{\prime}\partial_{-}^{\prime}+\psi^{\prime*}\partial_{+}^{\prime}\right)\frac{1}{\left|\mathbf{r-r}\prime\right|}dVdV^{\prime}\label{eq:H-dip-quad}
\end{equation}
Note that quadratic parts of the exchange Hamiltonian is local in
space and it conserves the total number of magnons $N=\int\left|\psi\right|^{2}dV$,
whereas the quadratic part of dipolar Hamiltonian is non-local and
it violates the conservation of the magnon number. All three parts
of the quadratic Hamiltonian are invariant with respect to any translation
in the film plane. Therefore, it is natural to describe the motion
in plane as a superposition of running plane waves. In other words,
the problem must be partly diagonalized by the Fourier-transformation:
\begin{equation}
\psi\left(\mathbf{r}\right)=\frac{1}{\sqrt{A}}\sum_{\mathbf{q}}\chi_{\mathbf{q}}\left(x\right)e^{i\mathbf{qr}},\label{eq:psi-Fourier}
\end{equation}
where $\mathbf{q}=iq_{y}\hat{y}+iq_{z}\hat{z}$ is the in-plane wave
vector; the Fourier-coefficients $\chi_{\mathbf{q}}\left(x\right)$
depend on the transverse-to-plane coordinate $x$; $A$ is the area
of any film cross-section parallel to its surfaces. The
inverse Fourier transformation gives the amplitude of a magnon with the
wave vector $\mathbf{q}$ in a general state with the wave function
$\psi\left(\mathbf{r}\right)$:
\begin{equation}
\chi_{\mathbf{q}}\left(x\right)=\frac{1}{\sqrt{A}}\iintop\psi\left(\mathbf{r}\right)e^{i\mathbf{qr}}dydz\label{eq:chi-psi}
\end{equation}
Employing the Poisson brackets for $\psi\left(\mathbf{r}\right)$
eq. (\ref{eq:Poisson-wf}), the Poisson
brackets for the amplitudes $\chi_{\mathbf{q}}\left(x\right)$ are:
\begin{equation}
\left\{ \chi_{\mathbf{q}}\left(x\right),\chi_{\mathbf{q}^{\prime}}^{*}\left(x^{\prime}\right)\right\} =-\frac{i}{\hbar}\delta_{\mathbf{q},\mathbf{q}^{\prime}}\delta\left(x-x^{\prime}\right).\label{eq:Poisson-chi}
\end{equation}
In terms of the variables $\chi_{\mathbf{q}}\left(x\right)$ the three
parts of the Hamiltonian are:Q
\begin{equation}
H_{ex}^{\left(2\right)}=\mu_{B}M\ell^{2}\sum_{\mathbf{q}}\intop_{-d/2}^{d/2}\left(\left|\frac{d\chi_{\mathbf{q}}\left(x\right)}{dx}\right|^{2}+\mathbf{q}^{2}\left|\chi_{\mathbf{q}}\left(x\right)\right|^{2}\right)dx\label{eq:H-ex-quad-chi}
\end{equation}
\begin{equation}
H_{Z}^{(2)}=\mu_{B}\mathcal{H}\sum_{\mathbf{q}}\intop_{-d/2}^{d/2}\left|\chi_{\mathbf{q}}\left(x\right)\right|^{2}dx\label{eq:H-Z-quad-chi}
\end{equation}
\begin{equation}
\begin{array}{c}
H_{dip}^{\left(2\right)}=\pi\mu_{B}M\sum_{\mathbf{\mathbf{q}}}\iintop_{-d/2}^{d/2}\left[\chi_{\mathbf{q}}\left(d_{x}-q_{y}\right)+\chi_{\mathbf{-q}}^{*}\left(d_{x}+q_{y}\right)\right]\\
\times\left[\chi_{\mathbf{-q}}^{\prime}\left(d_{x^{\prime}}+q_{y}\right)+\chi_{\mathbf{q}}^{\prime*}\left(d_{x^{\prime}}-q_{y}\right)\right]G_{q}\left(x-x^{\prime}\right),
\end{array}\label{eq:H-dip-quad-chi}
\end{equation}
where we omitted for brevity the arguments $x$ and $x^{\prime}$
writing $\chi_{\mathbf{q}}$ instead of $\chi_{\mathbf{q}}\left(x\right)$
and $\chi_{\mathbf{q}}^{\prime}$ instead of $\chi_{\mathbf{q}}\left(x^{\prime}\right)$
and employed the abbreviation $d_x\equiv\frac{d}{dx}$.
The symbol $G_{q}\left(x\right)$ stays for for the Green function
of the 1d Helmholtz equation:
\begin{equation}
G_{q}\left(x\right)=\frac{e^{-q\left|x\right|}}{2q}\label{eq:Green-1d}
\end{equation}
It obeys the 1d Helmholtz equation with a point source at origin:
\begin{equation}
\left(d_{x}^{2}-q^{2}\right)G_{q}\left(x\right)=-\delta\left(x\right).\label{eq:Helmholtz-eq}
\end{equation}

\subsection{Bogoliubov transformation.}

The exchange and Zeeman parts of the quadratic Hamiltonian are diagonal
in the variables $\chi_{\mathbf{q}}\left(x\right)$, but the dipolar
part mixes $\chi_{\mathbf{q}}\left(x\right)$ with $\chi_{\mathbf{-q}}^{*}\left(x\right)$.
To diagonalize the total quadratic Hamiltonian we apply the extended
Bogoliubov transformation introducing for each $\mathbf{q}$ an infinite
series of variables $\eta_{\mathbf{q}n}$ associated with $\chi_{\mathbf{q}}\left(x\right)$
and $\chi_{\mathbf{-q}}^{*}\left(x\right)$ by a linear transformation:
\begin{equation}
\eta_{\mathbf{q}n}=\intop_{-d/2}^{d/2}\left[u_{\mathbf{q}n}\left(x\right)\chi_{\mathbf{q}}\left(x\right)+v_{\mathbf{q}n}\left(x\right)\chi_{\mathbf{-q}}^{*}\left(x\right)\right]dx.\label{eq:Bogoliubov-tr}
\end{equation}
To be canonical this transformation must produce correct Poisson brackets
for variables $\eta_{\mathbf{q}n}$:
\begin{equation}
\left\{ \eta_{\mathbf{q}n},\eta_{\mathbf{q^{\prime}}n^{\prime}}^{*}\right\} =-\frac{i}{\hbar}\delta_{\mathbf{q},\mathbf{q}^{\prime}}\delta_{n,n^{\prime}}\label{Poisson-eta}
\end{equation}
This requirement is equivalent to the condition of canonical transformation
in classical mechanics \cite{Goldstein 2011} or unitary transformation in
quantum mechanics. Therefore we will also use the word "unitarity"
or "unitary" as equivalent to "canonical". The requirement (\ref{Poisson-eta})
together with the Bogoliubov transformation (\ref{eq:Bogoliubov-tr})
and Poisson brackets for $\chi_{\mathbf{q}}\left(x\right)$ (\ref{eq:Poisson-chi})
implies a series of constraints:
\begin{equation}
\intop_{-d/2}^{d/2}\left[u_{\mathbf{q}n}\left(x\right)u_{\mathbf{q}n^{\prime}}^{*}\left(x\right)-v_{\mathbf{q}n}\left(x\right)v_{\mathbf{q}n^{\prime}}^{*}\left(x\right)\right]dx=\delta_{n,n^{\prime}}\label{eq:unitarity-u-v}
\end{equation}
The inverse Bogoliubov transformation determines $\chi_{\mathbf{q}}\left(x\right)$
as a linear combination of $\eta_{\mathbf{q}n}$:
\begin{equation}
\chi_{\mathbf{q}}\left(x\right)=\sum_{n}\left[U_{\mathbf{q}n}\left(x\right)\eta_{\mathbf{q}n}+V_{\mathbf{q}n}\left(x\right)\eta_{\mathbf{-q}n}^{*}\right]\label{eq:inverse-Bog}
\end{equation}
Replacing the amplitudes $\eta_{\mathbf{q}n}$, $\eta_{\mathbf{-q}n}^{*}$in
eq. (\ref{eq:inverse-Bog}) by their Bogoliubov representation (\ref{eq:Bogoliubov-tr}),
we arrive at equations relating direct and inverse Bogolyubov transformations:
\begin{equation}
\begin{array}{c}
\sum_{n}\left[U_{\mathbf{q}n}\left(x\right)u_{\mathbf{q}n}\left(x^{\prime}\right)+V_{\mathbf{q}n}\left(x\right)v_{\mathbf{-q}n}^{*}\left(x^{\prime}\right)\right]=\delta\left(x-x^{\prime}\right)\\
\sum_{n}\left[U_{\mathbf{q}n}\left(x\right)v_{\mathbf{q}n}\left(x^{\prime}\right)+V_{\mathbf{q}n}\left(x\right)u_{\mathbf{-q}n}^{*}\left(x^{\prime}\right)\right]=0
\end{array}\label{eq:unitarity-UV-uv}
\end{equation}
On the other hand, the unitarity of the inverse
Bogoliubov transformation requires
\begin{equation}
\sum_{n}\left[U_{\mathbf{q}n}\left(x\right)U_{\mathbf{q}n}^{*}\left(x^{\prime}\right)-V_{\mathbf{q}n}\left(x\right)V_{\mathbf{q}n}^{*}\left(x^{\prime}\right)\right]=\delta\left(x-x^{\prime}\right)\label{eq:untarity-UU-VV}
\end{equation}
Comparing this equation with the first eq. (29), we arrive at conclusion
that\enskip $U_{\mathbf{q}n}\left(x\right)=u_{\mathbf{q}n}^{*}\left(x\right)$
and $V_{\mathbf{q}n}\left(x\right)=-v_{\mathbf{-q}n}\left(x\right)$.
Thus, the inverse Bogolyubov transformation can be rewritten as
\begin{equation}
\chi_{\mathbf{q}}\left(x\right)=\sum_{n}\left[u_{\mathbf{q}n}^{*}\left(x\right)\eta_{\mathbf{q}n}-v_{\mathbf{-q}n}\left(x\right)\eta_{\mathbf{-q}n}^{*}\right]\label{eq:inv-bog-uv}
\end{equation}
In addition from the $U-V$ unitarity condition (\ref{eq:untarity-UU-VV})
we find the dual unitarity condition in terms of the initial Bogolyubov
coefficients:
\begin{equation}
\sum_{n}\left[u_{\mathbf{q}n}^{*}\left(x\right)u_{\mathbf{q}n}\left(x^{\prime}\right)-v_{\mathbf{-q}n}\left(x\right)v_{\mathbf{-q}n}^{*}\left(x^{\prime}\right)\right]=\delta\left(x-x^{\prime}\right)\label{eq:unitarity-uv-dual}
\end{equation}

\subsection{The wave functions and spectrum of magnons.}

\subsubsection{Spectrum of magnons.}

The magnon amplitudes must satisfy the stationary Schrödinger equation
whose classical analogue is
\begin{equation}
\left\{ H^{(2)},\eta_{\mathbf{q},n}\right\} =-i\omega_{\mathbf{q},n}\eta_{\mathbf{q},n}.\label{eq:SE-class}
\end{equation}
The Poisson brackets of the quadratic Hamiltonian and the vector of
amplitudes is a linear anti-Hermitian operator acting on this vector.
Thus, the vector of amplitudes $\eta_{\mathbf{q},n}$ is the eigenvector
and the frequency of a magnon is the corresponding eigenvalue of the
Hermitian operator $i\left\{ H^{(2)},\right\} $. In this subsection
we express these equations in terms of the Bogoliubov
coefficients. Their solutions in some limiting cases will be found
in the next subsection.

In order to write the left part of eq. (\ref{eq:SE-class}) explicitly,
we employ eqs. (\ref{eq:H-ex-quad-chi},\ref{eq:H-Z-quad-chi},\ref{eq:H-dip-quad-chi})
for the three parts of the quadratic Hamiltonian, equation (\ref{Poisson-eta})
for the Poisson brackets of the two amplitude vectors and the Bogoliubov
transformation (\ref{eq:inv-bog-uv}) from the amplitudes $\chi_{\mathbf{q},n}$
to magnon amplitudes $\eta_{\mathbf{q},n}$. In resulting equations
we omit for brevity the subscripts $\mathbf{q}$ and $n$ since they
are invariant under the Bogoliubov transformation. Thus, equations (\ref{eq:SE-class})
can be rewritten as:
\begin{equation}
\begin{array}{c}
\left[\omega+\gamma\left(\mathcal{H}+M\ell^{2}\left(\mathbf{q}^{2}-d_{x}^{2}\right)\right)\right]u=-2\pi\gamma M\left[\left(q_{y}^{2}-d_{x}^{2}\right)\zeta_{u}+\left(q_{y}-d_{x}\right)^{2}\zeta_{v}\right];\\
\left[\omega-\gamma\left(\mathcal{H}+M\ell^{2}\left(\mathbf{q}^{2}-d_{x}^{2}\right)\right)\right]v=2\pi\gamma M\left[\left(q_{y}^{2}-d_{x}^{2}\right)\zeta_{v}+\left(q_{y}+d_{x}\right)^{2}\zeta_{u}\right],
\end{array}\label{eq:SE-class-detail}
\end{equation}
where we denoted $\gamma=|e|/(2mc)$ is the classical gyromagnetic constant and 
\begin{equation}
\zeta_{u,v}\left(x\right)=\intop_{-d/2}^{d/2}G\left(x-x^{\prime}\right)\begin{array}{c}
u\left(x^{\prime}\right)\\
v\left(x^{\prime}\right)
\end{array}dx^{\prime}.\label{eq:zeta-u-v}
\end{equation}
The physical meaning of the integral terms in the r.-h. side of eqs.
(\ref{eq:SE-class-detail}) is the magnetic field $\mathbf{h}$ generated
by magnon magnetization $\mathbf{m}$. The magnetic field can be expressed
in terms of magnetostatic potential $\phi$ as $\mathbf{h=-\nabla\phi}$.
If it is generated by the magnetization $\mathbf{m\left(r\right)},$
then 
\begin{equation}
\phi\left(\mathbf{r}\right)=-\boldsymbol{\nabla}\cdot\intop\mathbf{m}\left(\mathbf{r}^{\prime}\right)\left|\mathbf{r}-\mathbf{r}^{\prime}\right|^{-1}d^{3}x^{\prime}\label{eq:pot-vs-magn}
\end{equation}
The coefficients $u$ and $v$ should be identified with the $x$- and
$y$-components of magnetization, the operators $\pm iq_{y}-d_{x}$ with the
complex presentation of gradient and divergence. Then equation (\ref{eq:zeta-u-v})
is equivalent to (\ref{eq:pot-vs-magn}) integrated over $y$ and
$z$.

The reference (\ref{eq:SE-class-detail}) is a system of two integral-differential
equations. However, they can be transformed in the purely differential
linear equations by employing operator $q^{2}-d_{x}^{2}$ (Laplacian)
to both sides of equations (\ref{eq:SE-class-detail}) and employing
eq. (\ref{eq:Helmholtz-eq}) to eliminate the Green function $G\left(x-x^{\prime}\right)$.
The application of this operator to $\zeta_{u,v}\left(x\right)$ transforms
these integrals into $u\left(x\right)$ and $v\left(x\right),$ respectively.

Thus, we obtain a system ordinary linear differential equations of
the fourth order:
\begin{equation}
\begin{array}{c}
\begin{array}{c}
\left[\omega+\gamma\left(\mathcal{H}+M\ell^{2}\left(q^{2}-d_{x}^{2}\right)\right)\right]\left(q^{2}-d_{x}^{2}\right)u\\
=-2\pi\gamma M\left[\left(q_{y}^{2}-d_{x}^{2}\right)u+\left(q_{y}-d_{x}\right)^{2}v\right];
\end{array}\\
\begin{array}{c}
\left[\omega-\gamma\left(\mathcal{H}+M\ell^{2}\left(q^{2}-d_{x}^{2}\right)\right)\right]\left(q^{2}-d_{x}^{2}\right)v\\
=2\pi\gamma M\left[\left(q_{y}^{2}-d_{x}^{2}\right)v+\left(q_{y}+d_{x}\right)^{2}u\right].
\end{array}
\end{array}\label{eq:4-th-diff}
\end{equation}
Their solutions must be a superposition of exponents $e^{i\kappa x}$
with $\kappa$\enskip being a root of the secular polynomial. To find
this polynomial, it is convenient to introduce the vector $\mathbf{k}$ with the components $k_{x}=\left|\kappa\right|$,
$k_{y,z}=q_{y,z}$ whose square if magnitude is $k^2=q^2+\kappa^2$. Let us define a simplest solution of the system
(\ref{eq:4-th-diff}) is:
\begin{equation}
u\left(x\right)=u_{0}e^{i\kappa x};v\left(x\right)=v_{0}e^{i\kappa x}\label{eq:exp-x-solution}
\end{equation}
Substituting this solution into eq.(\ref{eq:4-th-diff}), we obtain a
system of two linear homogeneous equations for $u_{0},v_{0}$.
The condition of its solvability is the nullification of their determinant
(secular equation):
\begin{equation}
\begin{array}{c}
\omega^{2}k^{2}=\gamma^{2}\left(\mathcal{H}+M\ell^{2}k^{2}\right)\times\\
\left[\left(\mathcal{H}+M\ell^{2}k^{2}\right)k^{2}+4\pi M\left(k^{2}-k_{z}^{2}\right)\right]
\end{array}.\label{eq:secular}
\end{equation}
This equation can be interpreted as dispersion relation for magnons:
\begin{equation}
\omega=\gamma\sqrt{\left(\mathcal{H}+M\ell^{2}k^{2}\right)\left[\mathcal{H}+M\ell^{2}k^{2}+\frac{4\pi M\left(k_{x}^{2}+k_{y}^{2}\right)}{k^{2}}\right]}\label{eq:spectrum}
\end{equation}
It is valid if $a/\lambda=ka/(2\pi)\ll1$. At room temperature the
thermal wavelength $\lambda=\hbar/\sqrt{2mk_{B}T}$. For effective
mass of magnon for YIG of the order of magnitude $m\approx3m_{e}$,
$\lambda$ is about $0.7nm$, whereas the lattice constant $a=1.2nm.$
Therefore, eq. (\ref{eq:spectrum}) is invalid for thermal magnons.
The calculation of the magnon spectrum at high energies for YIG were
given in the seminal article by Kolokolov, L'vov and Cherepanov \cite{Kolokolov 1983}.

\subsubsection{Bulk and evanescent waves.}

At fixed parameters $\ell,M,\mathcal{H},k_{z}=q_{z}$ and frequency
$\omega$, eq. (\ref{eq:secular}) is a cubic equation for the variable
$k^{2}$. Note that its coefficients do not depend not
only on the film thickness $d$ but also on the value $k_{y}$.
Inspection of the coefficients of the cubic equation shows that the
product of three roots is positive, whereas their sum is negative.
Therefore, there are two opportunities: i) one roots $k^{2}$ is positive
and two others are negative or ii) one root is positive and two others
are complex conjugated with negative real part. Sonin proved \cite{Sonin 2017}
that in thick films $d\gg\ell$ and for $kl\ll1$, the opportunity
i) is realized. Gang. Li \textit{et al. }\cite{Li 2018} proved that the
opportunity ii) leads to negative $\omega^{2}$ and therefore is forbidden.

For thick films $d\gg\ell$ and $k_{z}\ll1/\ell$ and $\omega^{2}<\gamma^{2}\mathcal{H}\left(\mathcal{H}+4\pi M\right)$,
the positive root $k_{1}^{2}$ can be found approximately. In this
case it is possible to retain in eq. (\ref{eq:secular}) only terms
linear in $k^{2}$ and independent on $k^{2}$ and neglect the terms
quadratic and cubic in $k^{2}$. The result is:
\begin{equation}
k_{1}^{2}=k_{z}^{2}\frac{4\pi\gamma^{2}\mathcal{H}M}{\gamma^{2}\mathcal{H}\left(\mathcal{H}+4\pi M\right)-\omega^{2}}\label{eq:positive-root}
\end{equation}
Two others negative solutions $k^{2}=-\mathfrak{k}_{1,2}^{2}$ are
determined by equation:
\begin{equation}
\mathfrak{k}_{1,2}^{2}=\left(2\pi+\frac{\mathcal{H}}{M}\pm\sqrt{4\pi^{2}+\frac{\omega^{2}}{\gamma^{2}M^{2}}}\right)\ell^{-2}\label{eq:k-square-neg}
\end{equation}
When frequency approaches the ferromagnetic resonance value $\omega_{FR}=\gamma\sqrt{\mathcal{H}\left(\mathcal{H}+4\pi M\right)}$
to the distance $\omega_{FR}-\omega\lesssim\frac{k_{z}^{2}\ell^{2}}{\sqrt{1+\frac{4\pi \mathcal{H}}{M}}}2\pi\gamma M$,
the inequality $k_{1}\ell\ll1$ becomes invalid and instead of quadratic
the cubic equation must be solved. At large frequency $\omega\gg\omega_{FR}$,\enskip the
exchange energy dominates and $\omega\approx\gamma M\ell^{2}k^{2}$.
It corresponds to the region of large wave vectors $k\ell\gg1$. For
thick films $d\gg\ell$, the four wave functions of the type $\chi_{\mathbf{q}}\left(x\right)\propto\exp\left[-\mathfrak{k}_{1,2}\left(\frac{d}{2}\pm x\right)\right]$
correspond to the four evanescent waves localized in a layer of the
depth $\sim\ell$ near the surfaces of the film $x=\pm d/2$.

\subsection{Self-consistency.}

We proved that any propagating in-plane excitation is a superposition
of several transverse modes. The transverse modes may be either superposition
of $\cos k_{x}x$ and $\sin k_{x}x$ or the evanescent waves. However,
the inverse statement that any such superposition is a solution of
the initial equations of motion is wrong. This happens because the
initial equations of motion were integral-differential. The system
of ordinary differential equations was obtained from them by application
of additional differential operators. This operation introduces additional
solutions of resulting system of equations that are not solutions
of the initial problem. Below we derive the selection rules that separate
only solutions of the initial integral-differential equations (\ref{eq:SE-class-detail},\ref{eq:zeta-u-v}).

Equations for the Bogoliubov transformation functions (\ref{eq:SE-class-detail},\ref{eq:zeta-u-v})
permit real solution. Therefore the Bogoliubov functions can be searched
in the form:
\begin{equation}
\begin{array}{c}
u_{\mathbf{q},n}\left(x\right)=a_{n}\cos k_{x}x+b_{n}\sin k_{x}x+\\
\sum_{m=1,2}\left(A_{nm}\frac{\cosh\mathfrak{k}_{m}x}{\cosh\mathfrak{k}_{m}d/2}+B_{nm}\frac{\sinh\mathfrak{k}_{m}x}{\sinh\mathfrak{k}_{m}d/2}\right);
\end{array}\label{eq:u-ansatz}
\end{equation}
\begin{equation}
\begin{array}{c}
v_{\mathbf{q},n}\left(x\right)=c_{n}\cos k_{x}x+d_{n}\sin k_{x}x+\\
\sum_{m=1,2}\left(C_{nm}\frac{\cosh\mathfrak{k}_{m}x}{\cosh\mathfrak{k}_{m}d/2}+D_{nm}\frac{\sinh\mathfrak{k}_{m}x}{\sinh\mathfrak{k}_{m}d/2}\right),
\end{array}\label{eq:v-ansatz}
\end{equation}
where all coefficients $a_{n},b_{n},A_{nm},B_{nm,}c_{n},d_{n},C_{nm},D_{nm}$
are real numbers. In further calculations we omit the subscripts $n$
and $\mathbf{q}$ since they are fixed. All evanescent waves exponentially decrease far from boundaries
on the scale $\sim\ell$ as $\exp\left[-\mathfrak{k}_{m}\left|\frac{d}{2}\pm x\right|\right]$.

Substitution of expressions (\ref{eq:u-ansatz},\ref{eq:v-ansatz})
to the integral-differential equations (\ref{eq:SE-class-detail},\ref{eq:zeta-u-v})
leads to appearance of exponential functions that do not belong to
the 6 exponents permitted by the secular equation (\ref{eq:secular}).
They are produced by the integrals (\ref{eq:zeta-u-v}). Their explicit
calculation can be reduced to the four basic integrals:
\begin{equation}
\begin{array}{c}
I_{c}\left(x\right)\equiv\intop_{-d/2}^{d/2}\frac{e^{-q\left|x-x^{\prime}\right|}}{2q}\cos k_{x}x^{\prime}dx^{\prime}\\
=\frac{\cos k_{x}x}{k^{2}}-\frac{e^{-qd/2}}{qk^{2}}\cosh qx\,f_{1};
\end{array}\label{eq:I-c}
\end{equation}
\begin{equation}
\begin{array}{c}
I_{s}\left(x\right)\equiv\intop_{-d/2}^{d/2}\frac{e^{-q\left|x-x^{\prime}\right|}}{2q}\sin k_{x}x^{\prime}dx^{\prime}\\
=\frac{\sin k_{x}x}{k^{2}}-\frac{e^{-qd/2}}{qk^{2}}\sinh qx\,f_{2};
\end{array}\label{eq:I-s}
\end{equation}
\begin{equation}
\begin{array}{c}
J_{cm}\left(x\right)\equiv\intop_{-d/2}^{d/2}\frac{e^{-q\left|x-x^{\prime}\right|}}{2q}\cosh\mathfrak{k}_{m}x^{\prime}dx^{\prime}\\
=\frac{\cosh\mathfrak{k}_{m}x}{q^{2}-\mathfrak{k}_{m}^{2}}-\frac{e^{-qd/2}}{q\left(q^{2}-\mathfrak{k}_{m}^{2}\right)}\cosh qx\,g_{1m};
\end{array}\label{eq:J-c}
\end{equation}
\begin{equation}
\begin{array}{c}
J_{sm}\left(x\right)\equiv\intop_{-d/2}^{d/2}\frac{e^{-q\left|x-x^{\prime}\right|}}{2q}\sinh\mathfrak{k}_{m}x^{\prime}dx^{\prime}\\
=\frac{\sinh\mathfrak{k}_{m}x}{q^{2}-\mathfrak{k}_{m}^{2}}-\frac{e^{-qd/2}}{q\left(q^{2}-\mathfrak{k}_{m}^{2}\right)}\sinh qx\,g_{2m},
\end{array}\label{eq:J-s}
\end{equation}
where the notations $f_{1,2}$, $g_{1,2}$ are used for the following
functions:
\begin{equation}
f_{1}=q\cos\frac{k_{x}d}{2}-k_{x}\sin\frac{k_{x}d}{2};\label{eq:f-c}
\end{equation}
\begin{equation}
f_{2}=q\sin\frac{k_{x}d}{2}+k_{x}\cos\frac{k_{x}d}{2};\label{eq:f-s}
\end{equation}
\begin{equation}
g_{1m}=q\cosh\frac{\mathfrak{k}_{m}d}{2}+\mathfrak{k}_{m}\sinh\frac{\mathfrak{k}_{m}d}{2};\label{eq:g-c}
\end{equation}
\begin{equation}
g_{2m}=q\sinh\frac{\mathfrak{k}_{m}d}{2}+\mathfrak{k}_{m}\cosh\frac{\mathfrak{k}_{m}d}{2}.\label{eq:g-s}
\end{equation}
Employing these results, it is possible to calculate $\zeta_{u}\left(x\right)$
and $\zeta_{v}\left(x\right)$ defined by eq. (\ref{eq:zeta-u-v}):
\begin{equation}
\zeta_{u}\left(x\right)=aI_{c}+bI_{s}+\sum_{m=1}^{2}\left(\frac{A_{m}J_{cm}}{\cosh\frac{\mathfrak{k}_{m}d}{2}}+\frac{B_{m}J_{sm}}{\sinh\frac{\mathfrak{k}_{m}d}{2}}\right);\label{eq:zeta-u-expl}
\end{equation}
\begin{equation}
\zeta_{v}\left(x\right)=cI_{c}+dI_{s}+\sum_{m=1}^{2}\left(\frac{C_{m}J_{cm}}{\cosh\frac{\mathfrak{k}_{m}d}{2}}+\frac{D_{m}J_{sm}}{\sinh\frac{\mathfrak{k}_{m}d}{2}}\right).\label{eq:zeta-v-expl}
\end{equation}
The terms with $I_{c}$ and $I_{s}$ in these equations contain the
functions $\cosh qx$ and $\sinh qx$ or equivalently $\exp\left(\pm qx\right)$.
The wave vector $k=q$ does not satisfy the secular equation (\ref{eq:secular}).
Therefore, they should vanish in the r.-h. side of eqs. (\ref{eq:SE-class-detail}).
These requirements represent four constraints onto 12 coefficients
$a,b,c,d,A_{1},B_{1},C_{1},D_{1},A_{2},B_{2},C_{2},D_{2}$ \cite{Li 2018}.
Neglecting evanescent waves in the integrals, we obtain 4 equations
for 4 coefficients $a,b,c,d$ at ``bulk'' waves: 
\begin{equation}
\begin{array}{ccccc}
\left(q_{y}^{2}-q^{2}\right)af_{1} & + & \left(q_{y}^{2}+q^{2}\right)cf_{1} & +2q_{y}qdf_{2} & =0\\
 & \left(q_{y}^{2}-q^{2}\right)bf_{2} & +2q_{y}qcf_{1} & +\left(q_{y}^{2}+q^{2}\right)df_{2} & =0\\
\left(q_{y}^{2}+q^{2}\right)af_{1} & -2q_{y}qbf_{2} & +\left(q_{y}^{2}-q^{2}\right)cf_{1} &  & =0\\
-2q_{y}qaf_{1} & +\left(q_{y}^{2}+q^{2}\right)bf_{2} & + & \left(q_{y}^{2}-q^{2}\right)df_{2} & =0
\end{array},\label{eq:self-consist}
\end{equation}
The determinant of this system is identically zero . Thus, this system
does not determine quantization of $k_{x}$. A simple  reason why any $4 \times 4$ minor of the $4 \times 24$ matrix formed by  coefficients at $e^{\pm qx}$ in each of the mentioned above twelve coefficients 
has zero determinant is that all of them obey an inhomogeneous Helmholtz equation, for example, 
\begin{equation}\label{inh-Helm}
\frac{d^2 I_c}{dx^2} - q^2 I_c =\cos (k_x x);
\frac{d^2 J_{cm}}{dx^2} - q^2 J_{cm} = \cosh (k_x x).
\end{equation}
Since the solutions of such equations can include any linear combination of 
$e^{\pm qx}$, the condition of zero coefficients at these function cannot put any restriction of the $4\times24$ matrix.
It means that any its $4\times 4$ minor has zero determinant.

The self-consistency equations are equivalent to the MBC, but they
simplify calculations.

\subsection{Boundary conditions and the quantization of transverse modes.}

\subsubsection{Spin boundary conditions.}

There are two kinds of boundary conditions: magnetostatic (MBC) associated
with the variation of the magnetic field and induction near the boundary
and the spin boundary conditions (SBC) associated with variation of
spin (magnetization) at the boundary. The MBC requires continuity
of tangential component of magnetic field $\mathbf{h}$ and the normal
component of the induction $\mathbf{b=h}+4\pi\mathbf{m}$ at two surfaces
$x=\pm d/2$ of the film. The MBC are satisfied automatically if the
magnetic potential is related to the magnetization by the equation
(\ref{eq:pot-vs-magn}). Therefore, only the SBC must be taken into
account.

Let us consider the simplest possibility that spins on the surfaces
are free. The variation of the exchange energy (\ref{eq:H-ex}) gives
the surface term:
\begin{equation}
\begin{array}{c}
\delta H_{ex}=\ell^{2}\intop_{-d/2}^{d/2}dx\iintop_{-\infty}^{\infty}dydz\partial_{i}\delta m_{\alpha}\cdot\partial_{i}m_{\alpha}=\\
\ell^{2}\left.\iintop_{-\infty}^{\infty}dydz\delta m_{\alpha}\partial_{x}m_{\alpha}\right|_{-d/2}^{d/2}+\mathrm{volume\,}\mathrm{terms}
\end{array}.\label{eq:H-ex-var}
\end{equation}
The volume terms contribute exchange terms in equations of motion,
whereas the surface term in this equation implies that on both surfaces
magnetization obeys the spin boundary condition:
\begin{equation}
\partial_{x}\mathbf{m}|_{x=\pm d/2}=0.\label{eq:SBC}
\end{equation}
The variation of the Zeeman and dipolar Hamiltonians does not give
the surface term since they do not contain derivatives of magnetization.

Returning to the amplitude representation, we identify as before the
two components of magnetization with the Bogolyubov coefficients $u$
and $v$ at fixed $\mathbf{q}$. Thus, eq. (\ref{eq:SBC}) in amplitude
representation is:
\begin{equation}
\partial_{x}u|_{x=\pm d/2}=\partial_{x}v|_{x=\pm d/2}=0\label{eq:SBC-uv}
\end{equation}
For the thick film and $k_{x}\ell\ll1$, these equations imply that
the magnitudes of coefficients at the evanescent waves $A_{m},B_{m},C_{m},D_{m}$
are less than the magnitudes of amplitudes of the bulk waves $a,b,c,d$
by the factor $\sim k_{x}\ell$.\cite{Sonin 2017} To see that, let us
put all coefficients except of $a,c$ and $A_{1},C_{1}$ equal to
zero. Then equation (\ref{eq:SBC-uv}) takes form:
\begin{equation}
\left(c-a\right)k_{x}\sin\frac{k_{x}d}{2}=\left(A_{1}+C_{1}\right)\mathfrak{k}_{1}\label{eq:SBC-art}
\end{equation}
This equation proves the Sonin's statement since $\mathfrak{k}_{1}\sim1/\ell$.
Nevertheless the evanescent waves allow to satisfy the MBC at fixed
amplitudes of the bulk waves.

Neglecting in equations of motion (\ref{eq:SE-class-detail})
evanescent waves, we can rewrite them as:
\begin{equation}
\hat{\mathcal{M}}\left(\begin{array}{c}
a\\
b\\
c\\
d
\end{array}\right)=0,\label{eq:motion-abcd}
\end{equation}
where the $4\times4$ matrix $\hat{\mathcal{M}}$ is:
\begin{equation}
\hat{\mathcal{M}}=\left(\begin{array}{cccc}
\omega-\mathcal{A} & 0 & \mathcal{B} & \mathcal{C}\\
0 & \omega-\mathcal{A} & -\mathcal{C} & \mathcal{B}\\
-\mathcal{B} & \mathcal{C} & \omega+\mathcal{A} & 0\\
-\mathcal{C} & -\mathcal{B} & 0 & \omega+\mathcal{A}
\end{array}\right),\label{eq:M-matrix}
\end{equation}
and
\begin{equation}
\begin{array}{c}
\mathcal{A}=\gamma\left(\mathcal{H}+M\ell^{2}k^{2}+\frac{2\pi M\left(k_{x}^{2}+k_{y}^{2}\right)}{k^{2}}\right)\\
B=\frac{2\pi\gamma M\left(k_{x}^{2}-k_{y}^{2}\right)}{k^{2}}\\
\mathcal{C}=\frac{4\pi\gamma Mk_{x}k_{y}}{k^{2}}
\end{array}.\label{eq:ABC}
\end{equation}
The determinant of the matrix $\hat{\mathcal{M}}$ is
\begin{equation}
\det\hat{\mathcal{M}}=\left(\omega^{2}-\mathcal{A}^{2}+\mathcal{B}^{2}+\mathcal{C}^{2}\right)^{2}.\label{eq:detM}
\end{equation}
It turns into zero at $\omega=\sqrt{\mathcal{A}^{2}-\mathcal{B}^{2}-\mathcal{C}^{2}}$
that gives the obtained earlier dispersion relation (\ref{eq:spectrum}).
The eigenvalues $\pm\omega$ of the matrix $\hat{\mathcal{M}}$ 
are double degenerate. Therefore, their eigenvectors contain two independent
coordinates, for example the amplitudes $a$ and $b$, whereas two
others are expressed as their linear combination as it follows from
the equations (\ref{eq:motion-abcd}):
\begin{equation}
\begin{array}{ccc}
c & = & \frac{\mathcal{B}}{\omega\pm\mathcal{A}}a-\frac{\mathcal{C}}{\omega\pm\mathcal{A}}b\\
d & = & \frac{\mathcal{C}}{\omega\pm\mathcal{A}}a+\frac{\mathcal{B}}{\omega\pm\mathcal{A}}b
\end{array}\label{eq:cd-ab}
\end{equation}
Note that the two eigenvectors corresponding to different signs in denominators
are orthogonal at mass shell, i.e., at $\omega=\sqrt{\mathcal{A}^{2}-\mathcal{B}^{2}-\mathcal{C}^{2}}$
and any choice of coordinates $a$ and $b$.

Let us substitute the amplitudes $c$ and $d$ from eqs. (\ref{eq:cd-ab})
for the sign + into the first two of self-consistency equations (\ref{eq:self-consist}).
Then we find a system of two homogeneous equations of the form:
\begin{equation}
\begin{array}{c}
Pa+Qb=0\\
Ra+Sb=0
\end{array},\label{eq:ab-system}
\end{equation}
where
\begin{equation}
\begin{array}{c}
P=\left[q_{y}^{2}-q^{2}+\frac{\left(q_{y}^{2}+q^{2}\right)\mathcal{B}}{\omega+\mathcal{A}}\right]f_{1}-\frac{2q_{y}q\mathcal{C}}{\omega+\mathcal{A}}f_{2}\\
Q=\frac{\left(q_{y}^{2}+q^{2}\right)\mathcal{C}}{\omega+\mathcal{A}}f_{1}+\frac{2q_{y}q\mathcal{B}}{\omega+\mathcal{A}}f_{2}\\
R=-\frac{\left(q_{y}^{2}+q^{2}\right)\mathcal{C}}{\omega+\mathcal{A}}f_{1}+\frac{2q_{y}q\mathcal{C}}{\omega+\mathcal{A}}f_{1}\\
S=\left[q_{y}^{2}-q^{2}+\frac{\left(q_{y}^{2}+q^{2}\right)\mathcal{B}}{\omega+\mathcal{A}}\right]f_{2}+\frac{2q_{y}q\mathcal{C}}{\omega+\mathcal{A}}f_{1}
\end{array}\label{eq:PQRS}
\end{equation}
The determinant of the system (\ref{eq:ab-system}) $PS-QR$ must
be zero. It determines the quantization of $k_{x}.$ Equation $PS-QR=0$
gives:
\begin{equation}
f_{1}^{2}-f_{2}^{2}=2\Gamma f_{1}f_{2};\,\Gamma=\frac{\left(q_{y}^{2}-q^{2}\right)\omega+\left(q_{y}^{2}+q^{2}\right)\mathcal{B}}{2q_{y}q\mathcal{B}}.\label{eq:f-12-Gamma}
\end{equation}
From this equation we find:
\begin{equation}
\frac{f_{1}}{f_{2}}=\Lambda\equiv\Gamma\pm\sqrt{\Gamma^{2}+1}.\label{eq:ratio-f-12}
\end{equation}
Note that the change of sign in front of square root turns $\Lambda$
into $-1/\Lambda$. Employing equations (\ref{eq:f-c},\ref{eq:f-s}),
we represent the quantization condition in a more explicit form:
\begin{equation}
\tan\frac{k_{x}d}{2}=\frac{q-\Lambda k_{x}}{\Lambda q+k_{x}}.\label{eq:quantization}
\end{equation}
The change $\Lambda\rightarrow-1/\Lambda$ transforms the fraction
$\frac{q-\Lambda k_{x}}{\Lambda q+k_{x}}$ into inverse value with
opposite sign, i.e., $-\frac{\Lambda q+k_{x}}{q-\Lambda k_{x}}$.
For the waves propagating along spontaneous magnetization ($k_{y}=0$),
the quantization condition becomes 
\begin{equation}
\tan\frac{k_{x}d}{2}=\frac{q}{k_{x}}\mathrm{or}\tan\frac{k_{x}d}{2}=-\frac{k_{x}}{q}\label{eq:quantization-parallel}
\end{equation}
The first of them was first found by Damon and Eshbach \cite{Damon 1961}
for purely dipolar interaction and reproduced by Sonin.\cite{Sonin 2017}
It corresponds to the pure cosine solution ($b=0$). The second sign
at $k_{y}=0$ corresponds to the pure sine solution ($a=0$).\cite{Li 2018}
For general direction of propagation in-plane the two different signs
in front of square root in eq. (\ref{eq:ratio-f-12}) correspond to
two different branches of discrete solutions. We denote them by discrete
index $\nu$ accepting two values $\pm$.

\subsubsection{Quantization of transverse wave vectors. Parallel propagation.}

\begin{figure}
\centering
{\includegraphics[width=7cm]{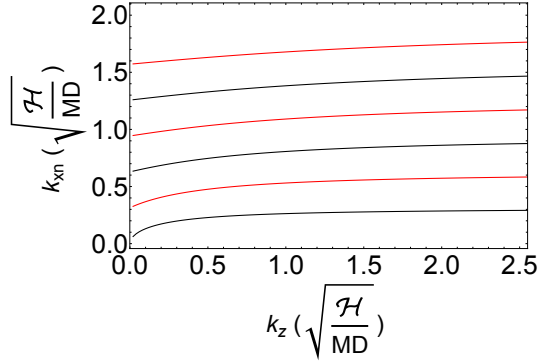}}
\caption{Plots of the dependence of quantized transverse wave vectors $k_{xn}$ on $k_z$ in units $\sqrt{\frac{\mathcal{H}}{MD}}$ for $d =10$ in units $\sqrt{\frac{MD}{\mathcal{H}}}$ . Black and red curves correspond to even and odd transverse modes, respectively}
	\label{quantization}
\end{figure}

Equations (\ref{eq:quantization-parallel}) have a discrete set of
solutions for $k_{xn}$ in the intervals $\left(\frac{\pi n}{d},\frac{\pi\left(n+1/2\right)}{d}\right)$
for the cosine  and in the intervals $\left(\frac{\pi\left(n+1/2\right)}{d},\frac{\pi\left(n+1\right)}{d}\right)$ for the sine transverse magnetization,
where $n$ is any non-negative integer. It is clearly seen from Fig. {\ref{quantization}}.
In the limit $qd\gg1$ the approximate analytical solution is possible
for $n\ll qd.$ In this case  $k_{x}\ll q$ so the
ratio$\frac{q}{k_{x}}\gg1$ for the first series of quantized $k_{x}.$
Therefore, $\frac{k_{x}d}{2}$ in the first equation (\ref{eq:quantization-parallel}) must be
close to $\left(n+\frac{1}{2}\right)\pi$ and
\begin{equation}
k_{xn}^{\left(+\right)}\approx\frac{\left(2n+1\right)\pi}{d}\left(1-\frac{2}{qd}\right)\label{eq:quant-kx-parallel+}
\end{equation}
Here we used the index $+$ as notation of the first series (even
transverse distribution of magnetization). For large $n$ and $qd\gg1$
the approximate equation for the quantized values of the first series
is:
\begin{equation}
k_{xn}^{\left(+\right)}\approx\frac{2n\pi}{d}+\frac{2}{d}\arctan\frac{qd}{2n\pi}\label{eq:quant-kx-par+large-n}
\end{equation}
It accurately matches the result (\ref{eq:quant-kx-parallel+}) for
$1\ll n\ll qd$.

For the second series the quantized transverse wave vectors for $qd\gg1$
and $n\ll qd$ are
\begin{equation}
k_{xn}^{\left(-\right)}\approx\frac{2n\pi}{d}\left(1-\frac{2}{qd}\right)\label{eq:quant-qx-par-}
\end{equation}
and for $n\gg1$ 
\begin{equation}
k_{xn}^{\left(-\right)}\approx\frac{\left(2n+1\right)\pi}{d}+\frac{2}{d}\arctan\frac{qd}{2n\pi}\label{eq:quant-kx-par-large-n-}
\end{equation}

\subsubsection{Wave vectors and effective masses at minimum energy. \label{subsec:Wave-vectors-and}}

Two energy minima $\pm Q$ are located on $z-$axis and correspond
to minimal value $n=0$ and symmetric branch of the transverse momentum
quantization, i.e. $k_{x}\approx\frac{\pi}{d}$. Let us minimize explicitly
the energy or frequency eq. (\ref{eq:spectrum}). For a thick film
$d\gg\ell,$ the energy is $\varepsilon=\hbar\omega\left(\mathbf{q},k_{x}\right)$.
It is more convenient to minimize the square of energy
\begin{equation}
\varepsilon^{2}\left(\mathbf{q},k_{x}\right)=\mu_{B}^{2}\left(\mathcal{H}^{2}+2\mathcal{H}M\ell^{2}k^{2}+\frac{4\pi\mathcal{H}M\left(k_{x}^{2}+k_{y}^{2}\right)}{k_{z}^{2}}\right).\label{eq:energy-square-qz-qx}
\end{equation}
We first minimize square of energy over $q_{y}$ putting $q_{y}=0$
and in the square of total momentum $k^{2}=k_{x}^{2}+q_{y}^{2}+q_{z}^{2}$
neglect $k_{x}^{2}$. Taking derivative over $q_{z}$ from $\varepsilon^{2}\left(q_{z},0,k_{x}\right)$
at $k_{x}=\frac{\pi}{d}$, we get:
\begin{equation}
2\varepsilon\frac{\partial\varepsilon}{\partial q_{z}}=4\mu_{B}^{2}\mathcal{H}M\left(\ell^{2}q_{z}-\frac{2\pi^{3}}{q_{z}^{3}d^{2}}\right).\label{eq:energy-square-der}
\end{equation}
At minimum energy the derivative $\frac{\partial\varepsilon}{\partial q_{z}}=0$.
From this requirement we find, that two minima are located at $q_{z}=\pm Q,$ where
\begin{equation}
Q=\frac{\left(2\pi^{3}\right)^{1/4}}{\sqrt{\ell d}}.\label{eq:Q}
\end{equation}
This result was obtained by E. Sonin.\cite{Sonin 2017}

The main value of the mass tensor $m_{z}$ in $z$ direction relates
to the second derivative $\frac{\partial^{2}\varepsilon}{\partial q_{z}^{2}}$
for $q_{z}=\pm Q$ as $m_{z}=\hbar^{2}/\left.\frac{\partial^{2}\varepsilon}{\partial q_{z}^{2}}\right|_{q_{z}=Q}$.
By differentiation of eq. (\ref{eq:energy-square-der}) and putting
$q_{z}=Q,$ $\varepsilon_{\min}=\mu_{B}\mathcal{H},$ we find:
\begin{equation}
m_{z}=\frac{\hbar^{2}}{8\mu_{B}M\ell^{2}}\label{eq:m-z.}
\end{equation}
To find $m_{y}$, we need to take the second derivative of $\varepsilon^{2}\left(\mathbf{q},k_{x}\right)$
given by eq. (\ref{eq:energy-square-qz-qx}) over $q_{y}$ at $q_{y}=0,q_{z}=Q$
neglecting $k_{x}$. The searched effective mass is $m_{y}=\hbar^{2}/\left.\frac{\partial^{2}\varepsilon}{\partial q_{y}^{2}}\right|_{q_{y}=0}$.
An elementary calcualtion gives:
\begin{equation}
m_{y}=\frac{\hbar^{2}Q^{2}}{8\pi\mu_{B}M}\label{eq:m-y}
\end{equation}
The mass $m_{y}$ is much less than $m_{z}$: their ratio is $m_{y}/m_{z}=\ell/\left(\pi d\right)\ll1$.
For the film of YIG 5\textgreek{m}m thick $Q\approx6.44\times10^{5}cm^{-1}$
, $m_{z}=7.37\times10^{-27}g;$ $m_{y}=1.78\times10^{-29}g$.

\subsubsection{Quantization of transverse wave vector: arbitrary direction
of propagation.}

Despite of rather involved structure of quantization condition (\ref{eq:quantization})
its solution can be written explicitly in the limit $d\gg\ell$, and
$qd\gg1$. The roots of this equation are $k_{x\nu n}$,
where $n=0,1,2...$ is the number of quantized value $k_{x}$, $\nu=\pm$ stays for even or odd transverse distribution of magnetization.
The explicit analytical expression for these roots in the asymptotic
region and large $n\gg1$ is
\begin{equation}
k_{x\nu n}=\frac{2n\pi}{d}+\frac{2}{d}\arctan\frac{qd-2\pi n\Lambda_{\nu n}}{qd\Lambda_{\nu n}+2\pi n}.\label{eq:quant-large-qd}
\end{equation}
To find parameters $\Lambda_{\nu n}=\Gamma+\nu\sqrt{\Gamma^{2}+1}$
it is necessary to replace $k_{x}$ by $2\pi n/d$ in the equations
(\ref{eq:ratio-f-12}) for $\Lambda$ and (\ref{eq:f-12-Gamma}) in
all functions containing $k_{x}$ in its arguments. Equation (\ref{eq:quant-large-qd})
has precision $1/qd$ and is valid for $1\ll n\ll qd$. In the entire
this region the difference between the quantized values of $k_{x}$
with the same number in the two branches is
\begin{equation}
k_{x+n}-k_{x-n}=\frac{\pi}{d}\label{eq:k-difference}
\end{equation}
The ratio of amplitudes in this range of variables is
\begin{equation}
\frac{b_{\nu n}}{a_{\nu n}}=-\frac{q\mathcal{A}}{\omega}\Lambda_{\nu}-\frac{\mathcal{C}}{\omega}\label{eq:ampl-ratio}
\end{equation}
At fixed direction of in-plane propagation given by the angle $\theta$
between the wave vector and direction of the spontaneous magnetization
$\mathbf{M}$, the frequency as function of the wave vector magnitude
has minimum at 
\begin{equation}
q_{0}=\frac{2\sqrt{\pi}\chi^{3/4}\sqrt{\cos\theta}}{\left(2+\chi\sin^{2}\theta\right)^{1/4}}\sqrt{\frac{k_{x\nu n}}{\ell}},\label{eq:min-freq-loc}
\end{equation}
where $\chi=\frac{4\pi M}{\mathcal{H}}$. From this equation and strong
inequality $q\ell\ll1$ it follows that $q_{0}\gg k_{x\nu n}\approx\frac{2\pi}{d}n$.

\subsubsection{Motion of energy minimum vs. $\mathbf{\mathit{k}_{\mathit{x}}}$.}

At very large $n\gg\frac{d}{\ell}$ the value $k^{2}$ becomes so
large that the exchange interactions dominates and the frequency of
a magnon becomes equal to $\omega=\gamma M\ell^{2}k^{2}.$ Then the
minimum energy occurs at $q=0$. It means that the position of minimum
of frequency $q_{0}$ first grows with $k_{x}$ and reaches its maximum
at some specific $k_{x1}\sim1/\ell$. At further growth of $k_{x}$
the position of frequency minimum $q_{0}\left(k_{x}\right)$ decreases
and reaches zero at another specific value of $k_{x}=k_{x2}$. At
further growth of $n$ it remains zero. Theory gives exact analytical
answers for all these values, namely:
\begin{equation}
k_{x1}^{2}=\frac{1}{3}k_{1}^{2}+\frac{2+\chi}{12\pi}\tan^{2}\theta k_{1}^{4}\ell^{2},\label{eq:kx1}
\end{equation}
where
\begin{equation}
k_{1}^{2}=\frac{\mathcal{H}}{6M\ell^{2}}\left[\sqrt{\left(2+\chi\sin^{2}\theta\right)^{2}+6\chi\cos\theta}-2-\chi\sin^{2}\theta\right].\label{eq:k1}
\end{equation}
The maximal value of $q_{0}$ is given by
\begin{equation}
q_{0\max}^{2}=k_{1}^{2}+k_{x1}^{2}.\label{eq:q0-max}
\end{equation}
Finally the value of $k_{x}^{2}$ at which the minimum of frequency
merges with maximum located at $q=0$ is
\begin{equation}
k_{x2}^{2}=\frac{\mathcal{H}}{4M\ell^2}\left[\sqrt{\left(2+\chi\sin^{2}\theta\right)^{2}+8\chi\cos\theta}-2-\chi\sin^{2}\theta\right].\label{eq:kx2}
\end{equation}
The position of maximum $k_{0}\left(k_{x}\right)$ for $k_{x}\ell\gtrsim1$
is given by
\begin{equation}
k_{0}^{2}\left(k_{x}\right)=\frac{\mathcal{H}}{M\ell^{2}}\frac{2+\chi\sin^{2}\theta}{2}w\left(\xi\right),\label{eq:k0-large-kx}
\end{equation}
where $w\left(\xi\right)$ is the solution of a cubic equation:
\begin{equation}
w^{3}+w^{2}=\xi\label{eq:w-cubic}
\end{equation}
and 
\begin{equation}
\xi=\frac{\chi^{2}\cos^{2}\theta k_{x}^{2}\ell^{2}}{\pi\left(2+\chi\sin^{2}\theta\right)^{3}}\label{eq:xi-kx-theta}
\end{equation}
 Details of these calculations can be found in the Appendix{[}motion
of minima{]}. In the analysis of this subsection we followed the work
\cite{Li 2018}.

\subsection{Comparison with other calculations and experiment.}

\begin{figure*}
\centering
\subfigure[]
	{\label{fig:spectra:a}
	\includegraphics[width=3.8cm]{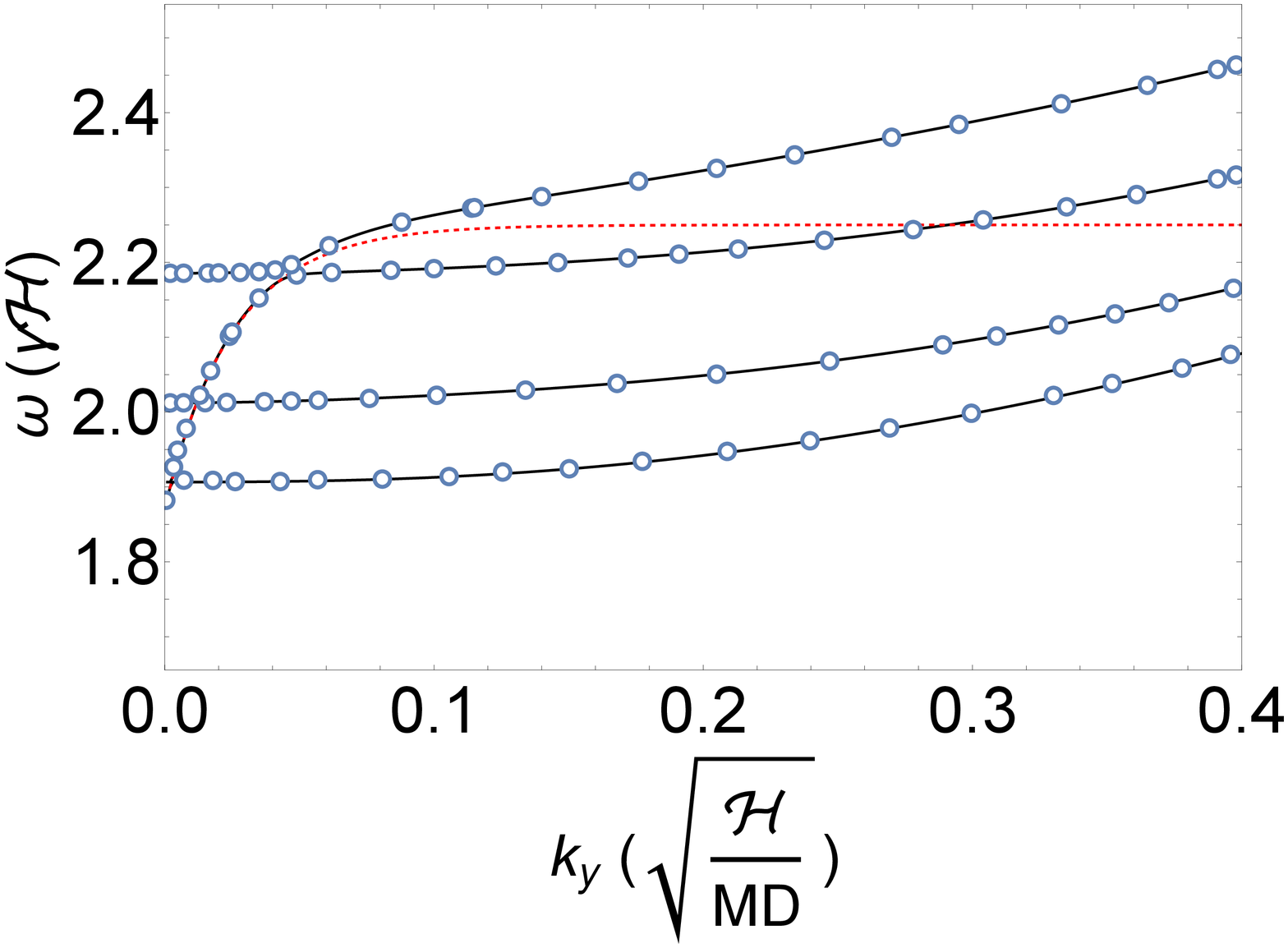}}	
\subfigure[]
	{\label{fig:spectra:b}
	\includegraphics[width=3.8cm]{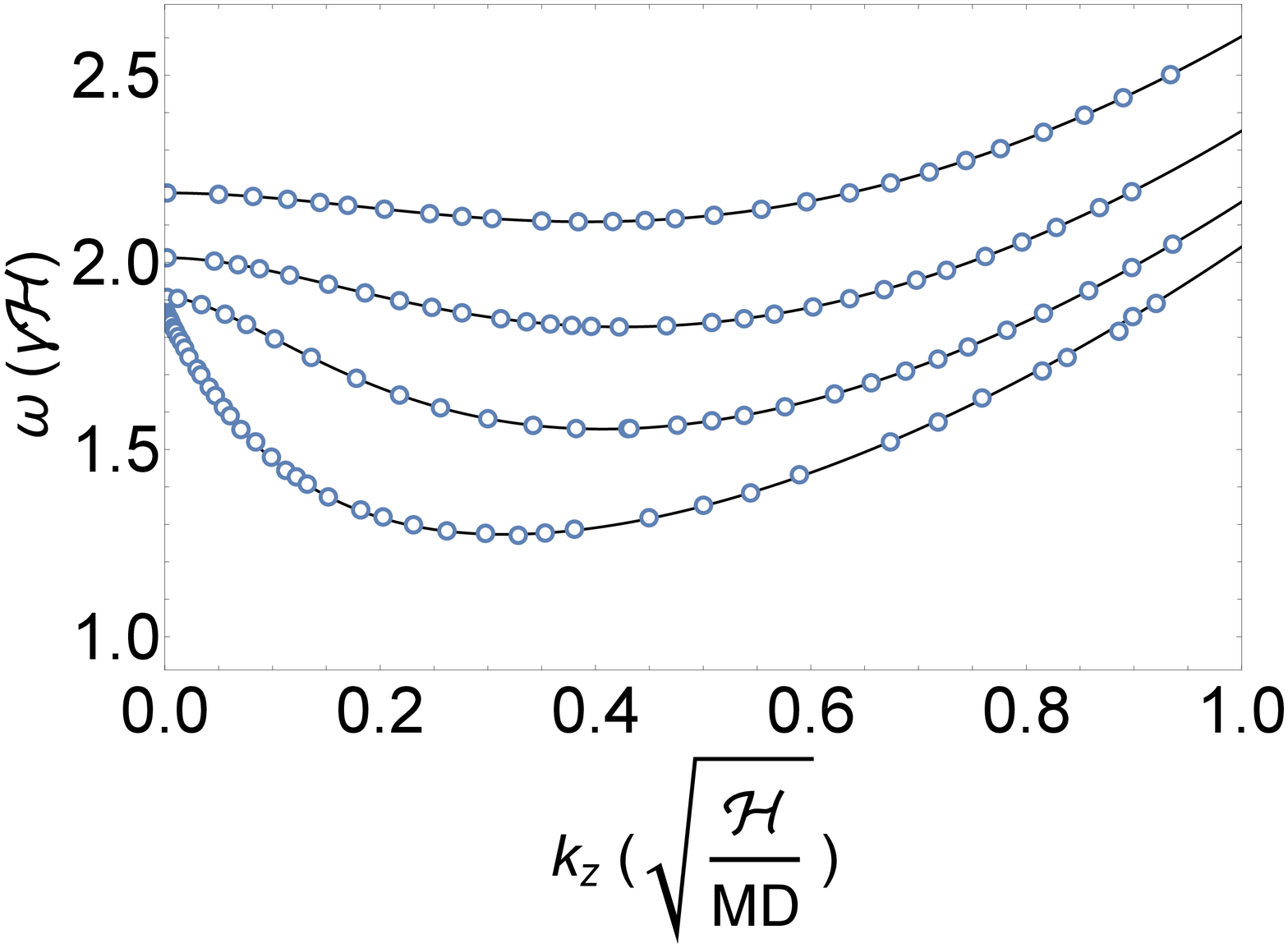}}
\subfigure[]
	{\label{fig:spectra:c}
	\includegraphics[width=3.75cm]{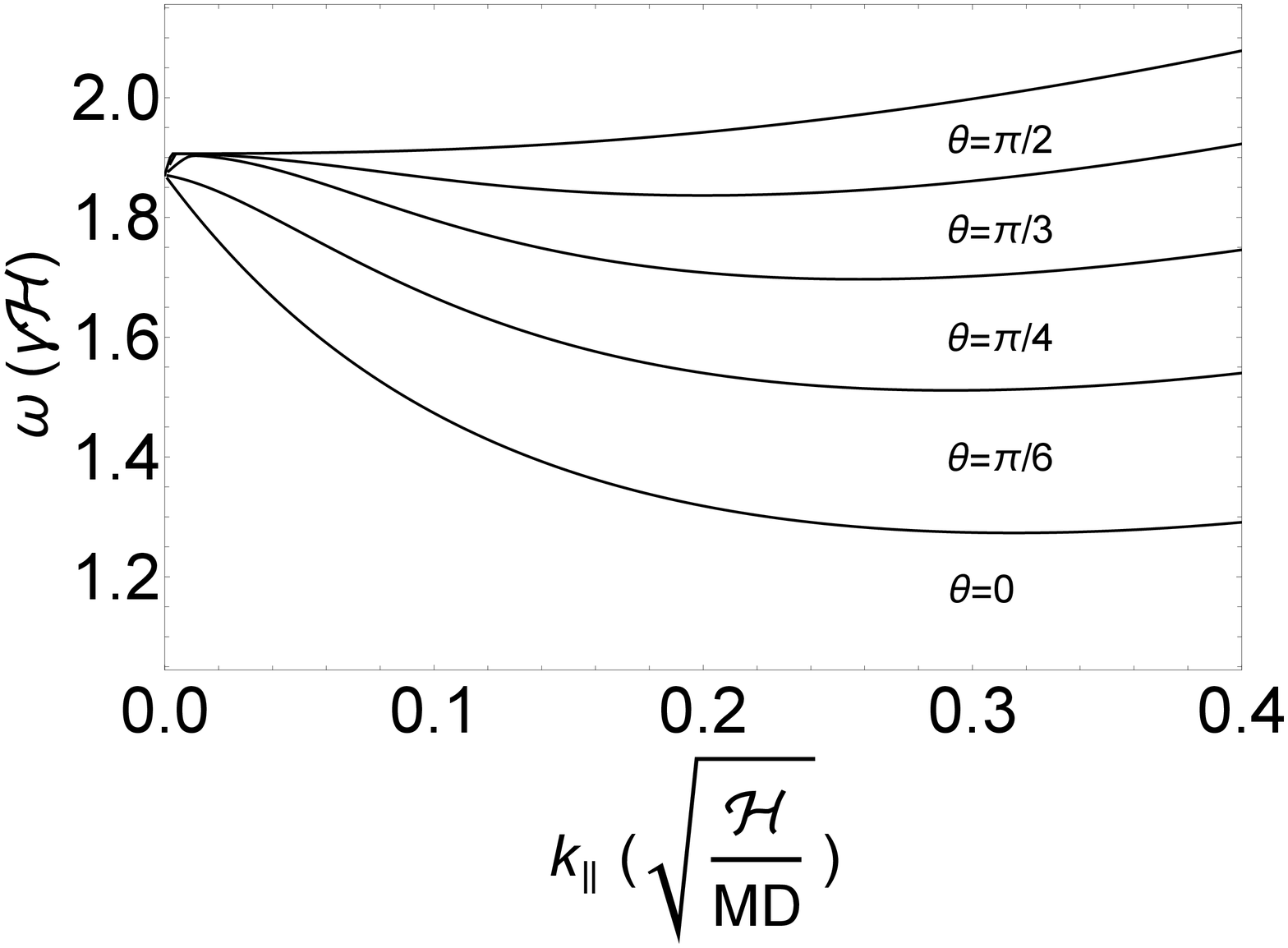}}
	\caption{Results of numerical calculations for the case $d=18.2$ in units $\sqrt{\frac{MD}{\mathcal{H}}}$ and $\chi=2.5$. (a) The spectra of first four quantized modes  for direction of propagation  perpendicular to magnetization.
	(b) Spectra of the first four modes  for direction of propagation  parallel to magnetization.
	(c) Spectra of the first transverse modes for $\theta=0,\frac{\pi}{6},\frac{\pi}{4},\frac{\pi}{3} ,\frac{\pi}{2}$. Black solid curves correspond to our numerical calculations, red dashed line is the Damon-Eshbach surface mode, circles are numerical calculations by Kreisel \textit{et al..} \cite{Kreisel 2009}. These figures agree with the figures from \cite{Li 2018}.}
	\label{fig:spectra}
\end{figure*}

The results of numerical calculations of quantized spectra eq. (\ref{eq:energy-square-qz-qx}) with quantized $k_{xn}$ for propagation perpendicular and parallel to magnetization and  $d=18.2$ in units $\sqrt{\frac{MD}{\mathcal{H}}}$, $\chi=2.5$ are shown in Fig \ref{fig:spectra:a} and \ref{fig:spectra:b}, spectra of the first transverse modes for a number of different directions of propagation specified by the angle $\theta = \arctan \frac{k_y}{k_z}$ are shown in Fig. \ref{fig:spectra:c}.

The spectra for parallel and perpendicular propagation (Fig. \ref{fig:spectra:a} and \ref{fig:spectra:b}) agree very well with the numerical calculations of the work \cite{Kreisel 2009} based on diagonalization of a large matrix. We also discovered an excellent agreement with similar calculations of the same work made for the YIG film with a thickness of 5 $\mu m$.

\begin{figure*}
\centering
\subfigure[]
	{\includegraphics[width=3.8cm]{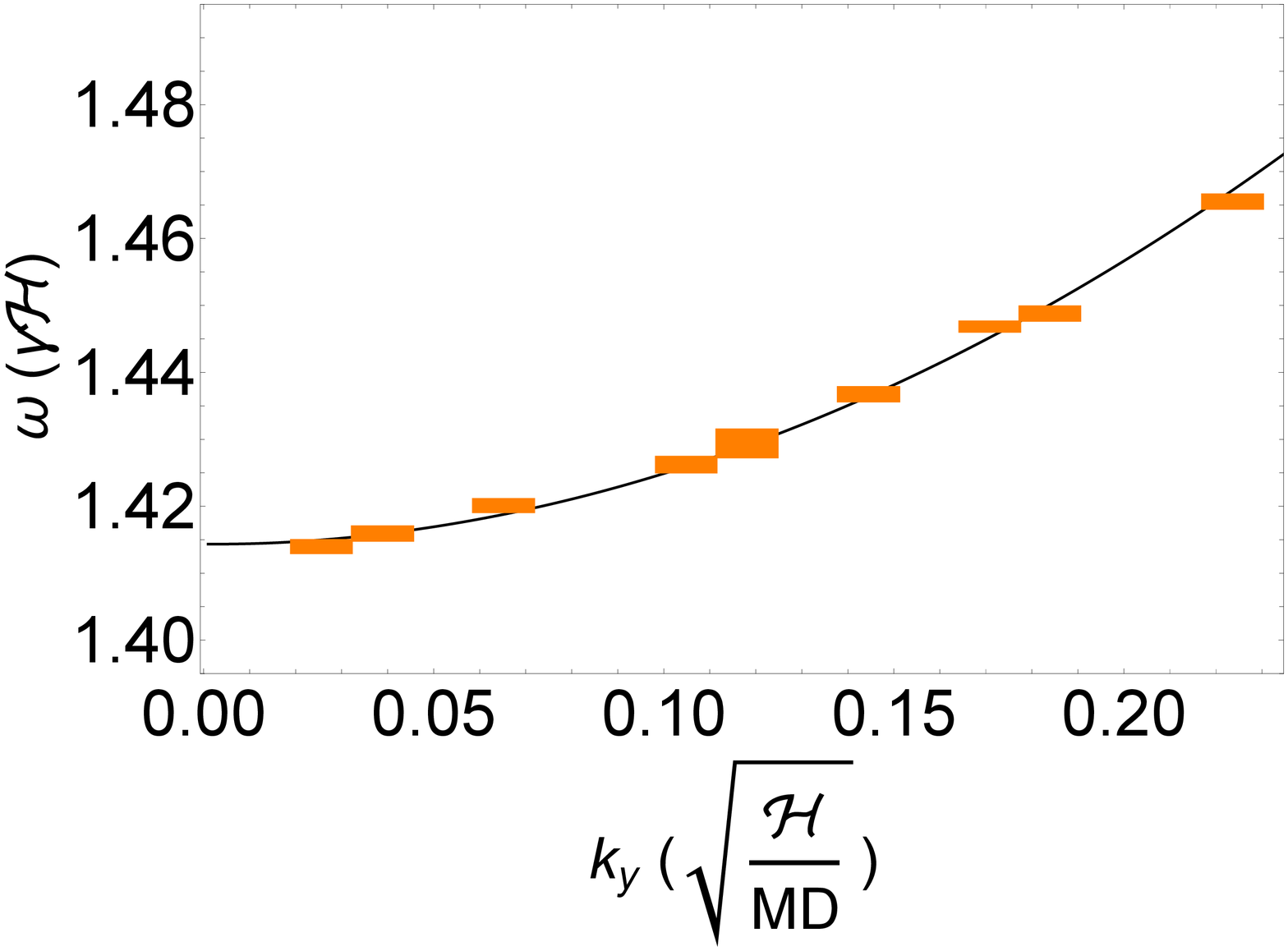}}	
\subfigure[]
	{\includegraphics[width=3.7cm]{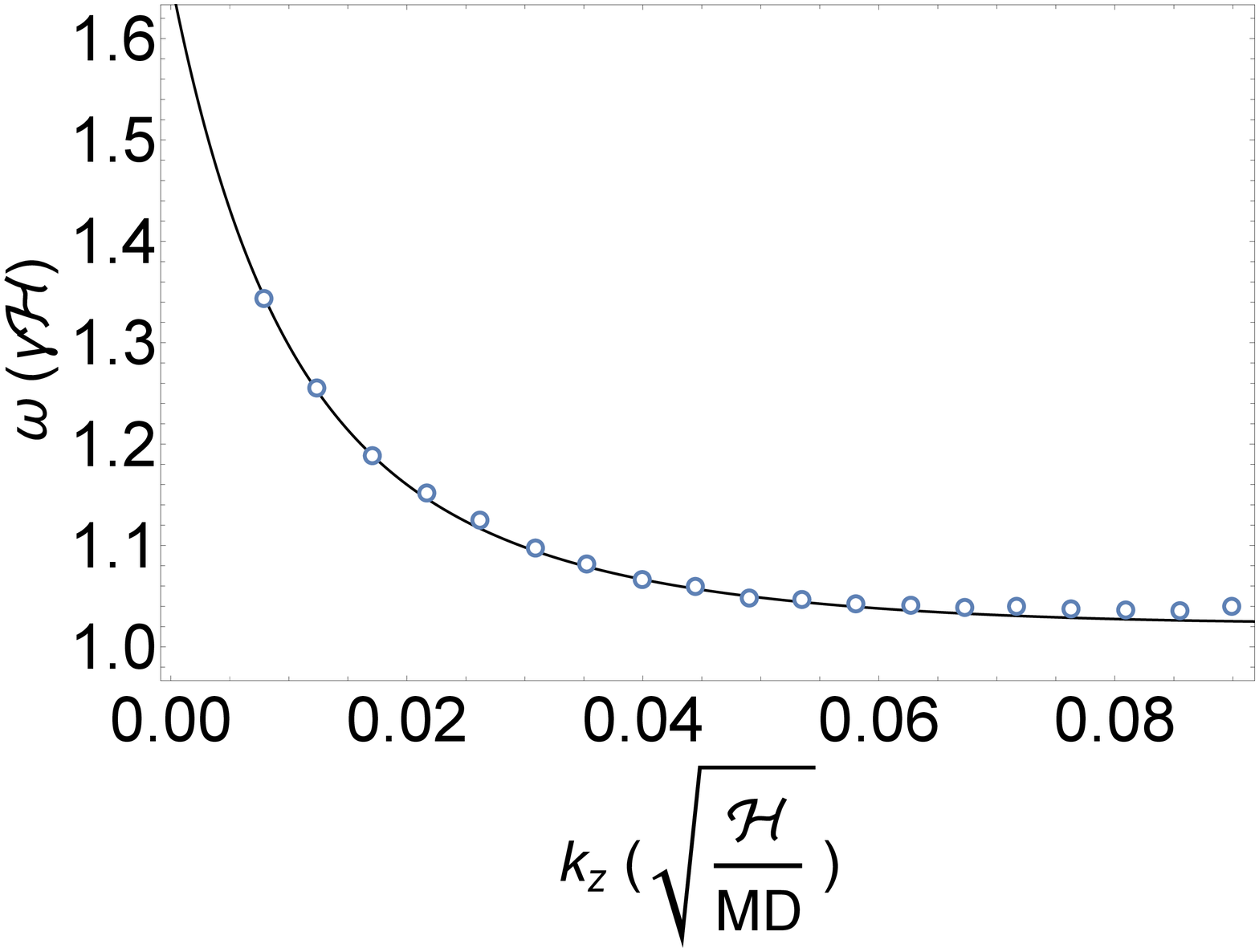}}
\subfigure[]
	{\includegraphics[width=3.8cm]{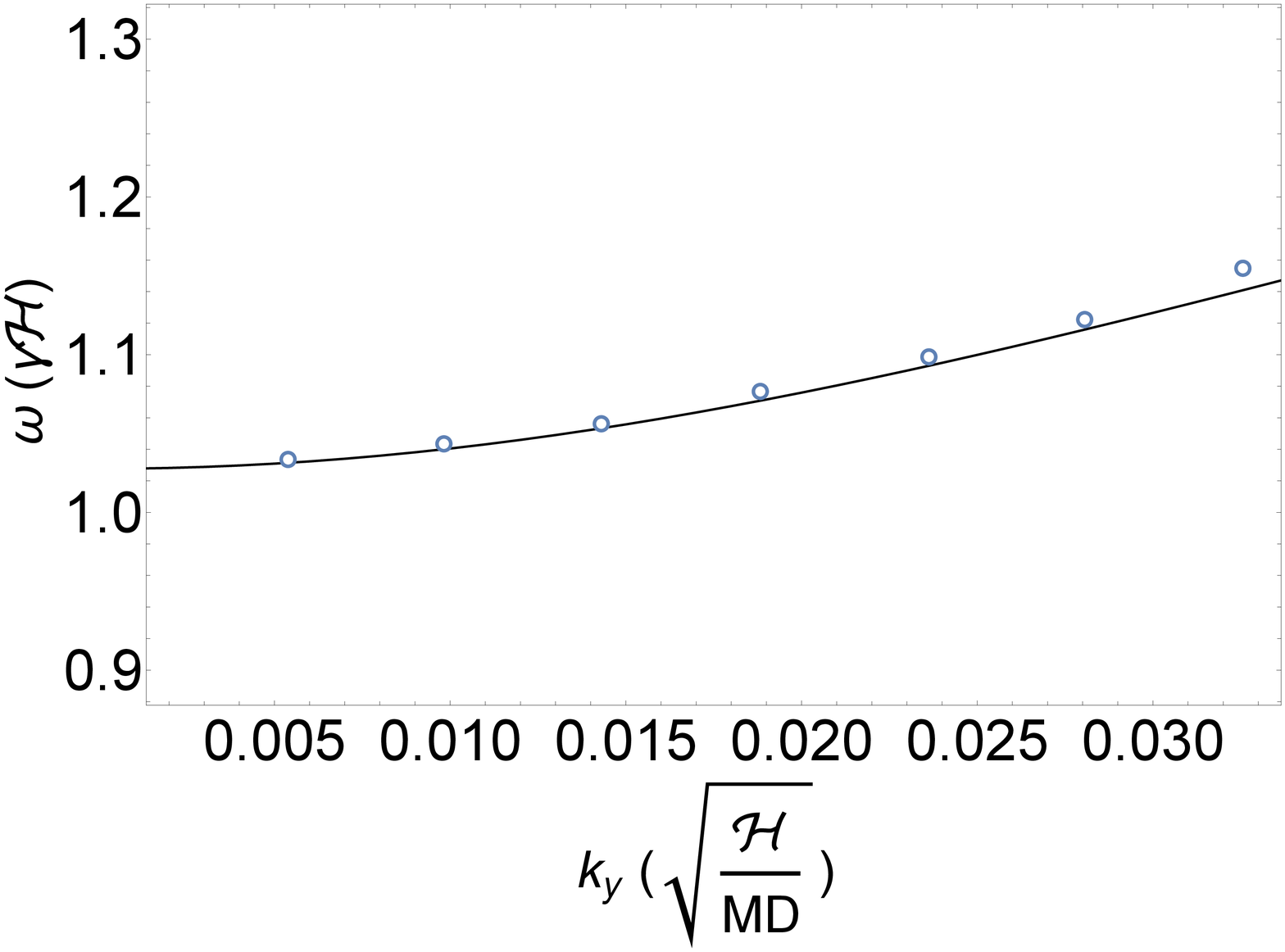}}
	\caption{Comparison  of theoretical spectrum with experiments. In experiments the Brillouin light scattering spectroscopy was used.(a)  Comparison with A. A. Serga \textit{et al.}\cite{Serga} $d=5$ $\mu m$, H=1750 Oe .
	(b) Comparison with V. E. Demidov \textit{et al.}\cite{Demidov:2008cg} $d=5.1$ $\mu m$, H=1000 Oe for direction of propagation parallel to magnetization.
	(c) Comparison with V. E. Demidov \textit{et al.}\cite{Demidov:2008cg} $d=5.1$ $\mu m$, H=1000 Oe. for fixed $k_z=3.4\times 10^4 cm^{-1}$. These figures agree with the figures from \cite{Li 2018}.}
	\label{experiment}
\end{figure*}

\begin{figure}
\centering
{\includegraphics[width=5cm]{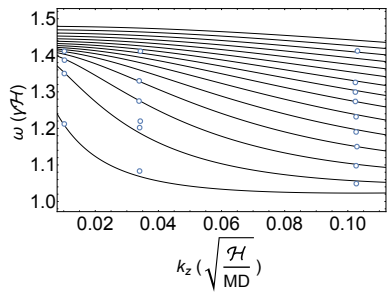}}
\caption{Comparison of theoretical spectrum with experiment. Solid curves are our calculations of the first 15 transverse modes for the YIG film of thickness 5$\mu$m, 4$\pi$M= 1940 Oe and  H= 1960 Oe . Circles on them are frequencies measured by J. Lim \textit{et al.} \cite{Lim 2018} at three fixed wavelengths for different transverse mode. This figure agrees with the figure from \cite{Li 2018}.}
	\label{modes}
\end{figure}

Figure \ref{experiment} shows a comparison of the theoretical spectrum with the experiment \cite{Serga, Demidov:2008cg}. Brillouin scattering spectroscopy was used in the experiment. Its precision is not sufficient for resolution of excited states. A dramatic increase in precision was achieved by an experimental group led by J. Ketterson \cite{Lim 2018}. His method makes use of direct microwave excitation of magnons via a specially designed antenna. It is made up of periodically repeated emitters that are powered by an adjustable frequency generator. The excited magnon wave-length coincides with the distance between emitters $\lambda$.
The magnon frequency at this wave vector $k_z=(2\pi)/\lambda$ is a frequency at which the resonance adsorption of microwave radiation reaches maximum. The increased resolution allowed for the observation of multiple magnon modes (up to nine). This is the first time that different transverse magnon modes have been experimentally observed. Figure \ref{modes}  shows a comparison of theoretical spectrum with experimental results \cite{Lim 2018}. 
The agreement between theory and experiment is excellent.

\begin{figure*}
\centering
\subfigure[]
	{\label{fig:spectra2:a}
	\includegraphics[width=3.6cm]{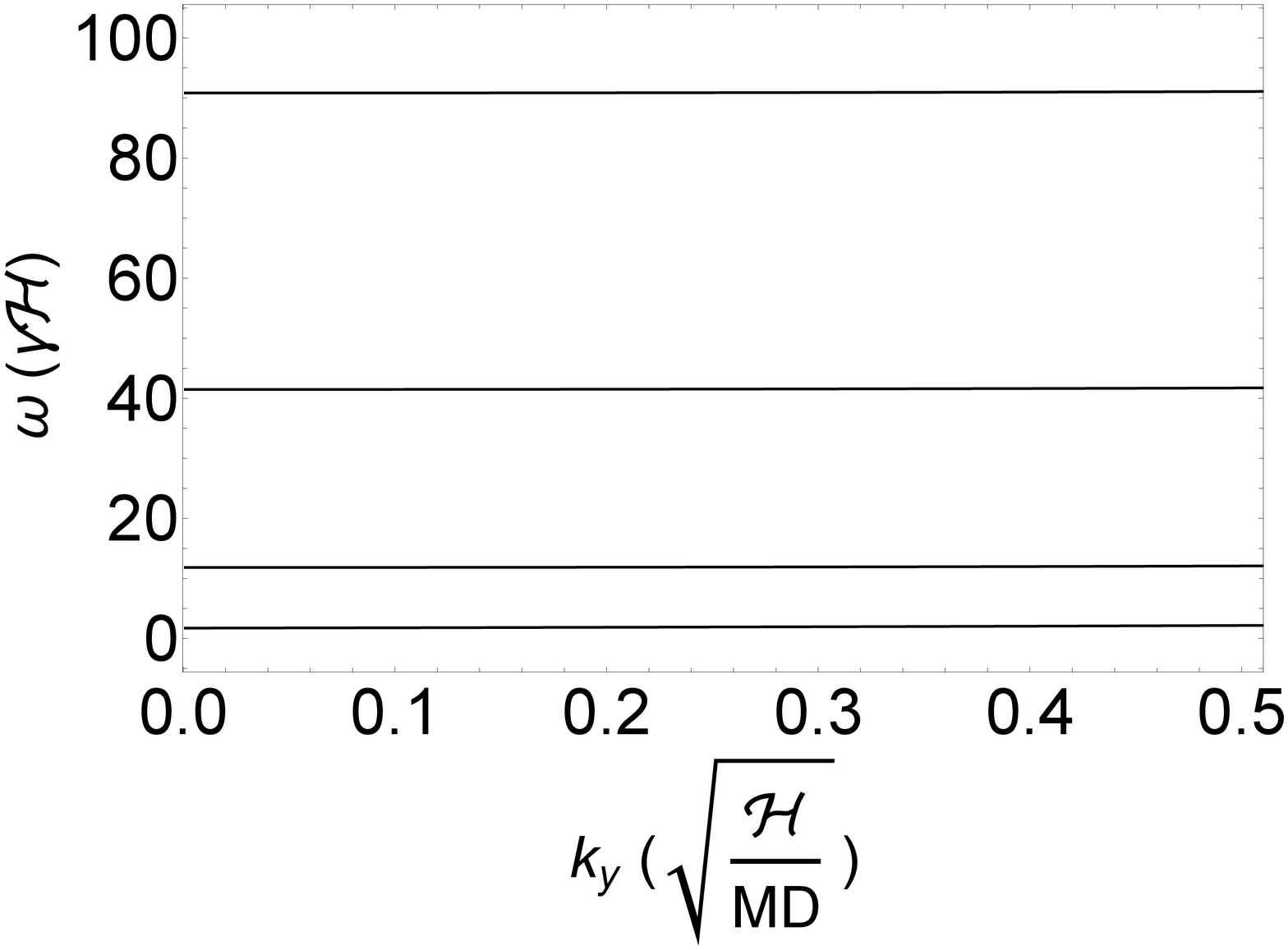}}	
\subfigure[]
	{\label{fig:spectra2:b}
	\includegraphics[width=3.6cm]{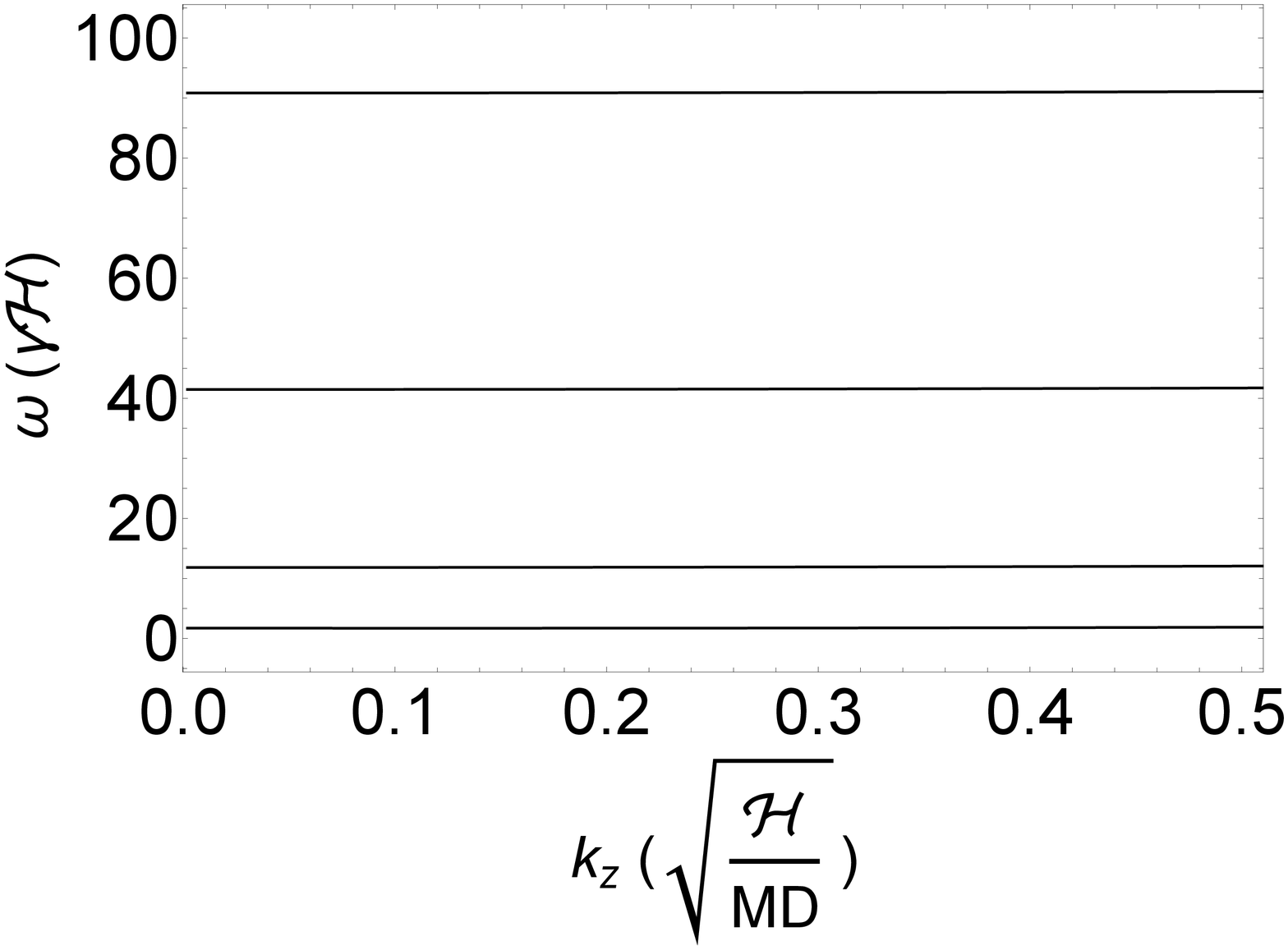}}
\subfigure[]
	{\label{fig:spectra2:c}
	\includegraphics[width=3.56cm]{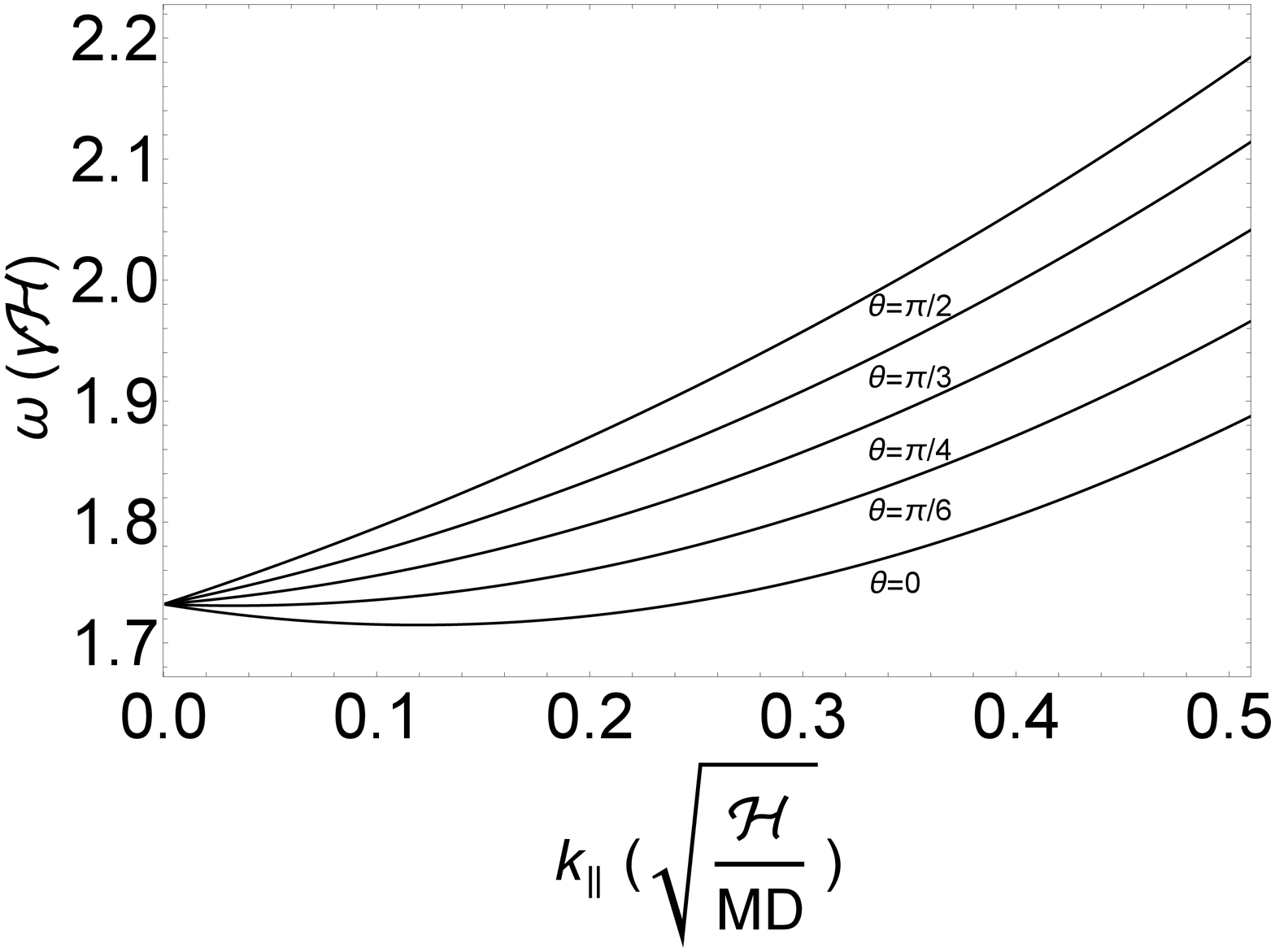}}
	\caption{Results of numerical calculations for a thin film $d=1$ in units $\sqrt{\frac{MD}{\mathcal{H}}}$ and $\chi=2$. (a) The spectra of first four quantized modes  for direction of propagation  perpendicular to magnetization.
	(b) Spectra of the first four modes  for direction of propagation  parallel to magnetization.
	(c) Spectra of the first transverse modes for $\theta=0,\frac{\pi}{6},\frac{\pi}{4},\frac{\pi}{3} ,\frac{\pi}{2}$. These figures agree with the figures from \cite{Li 2018}.}
	\label{fig:spectra2}
\end{figure*}

\begin{figure*}
\centering
\subfigure[]
	{\label{fig:frequency:a}
	\includegraphics[width=3.8cm]{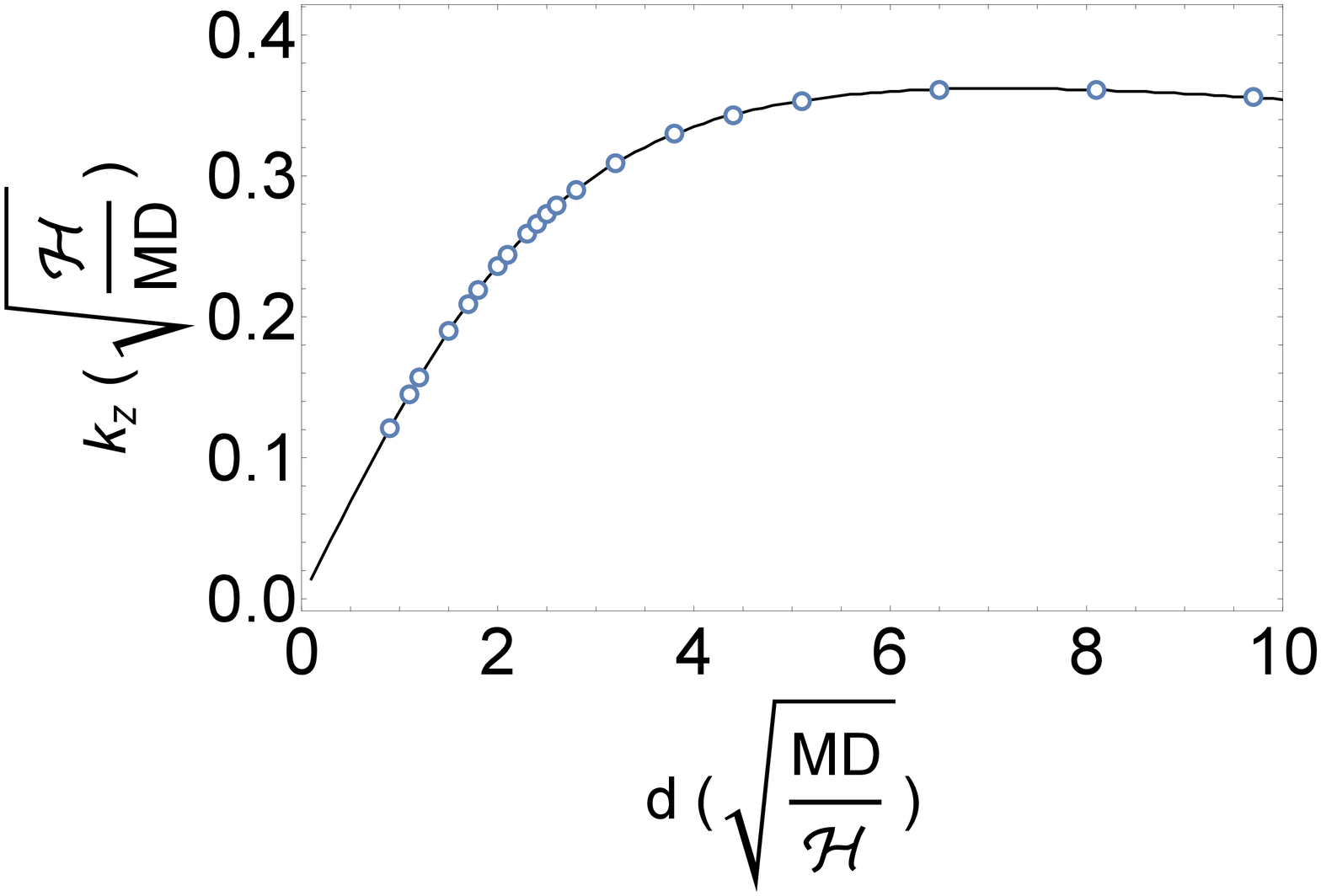}}	
\subfigure[]
	{\label{fig:frequency:b}
	\includegraphics[width=3.45cm]{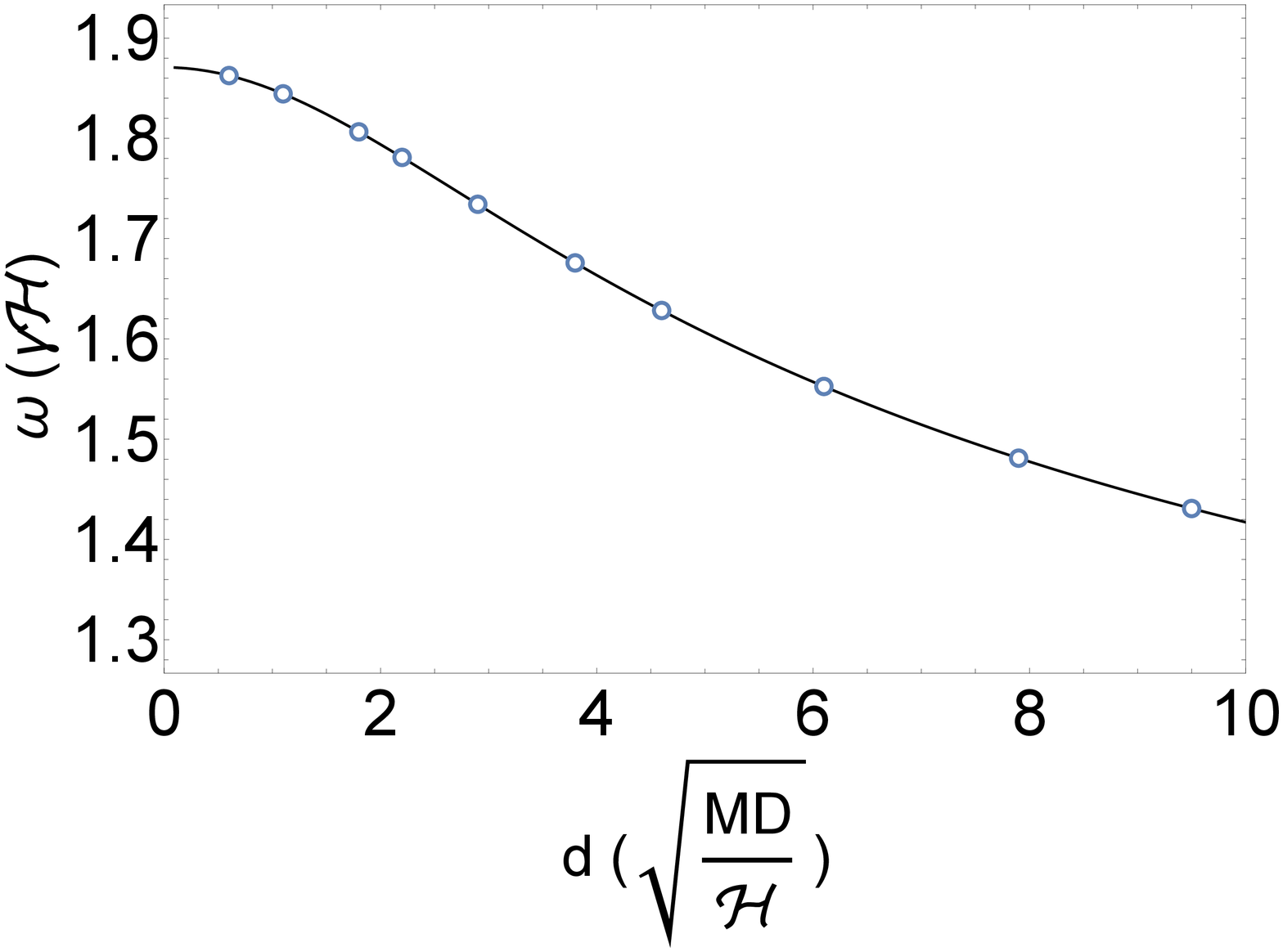}}
\subfigure[]
	{\label{fig:frequency:c}
	\includegraphics[width=3.75cm]{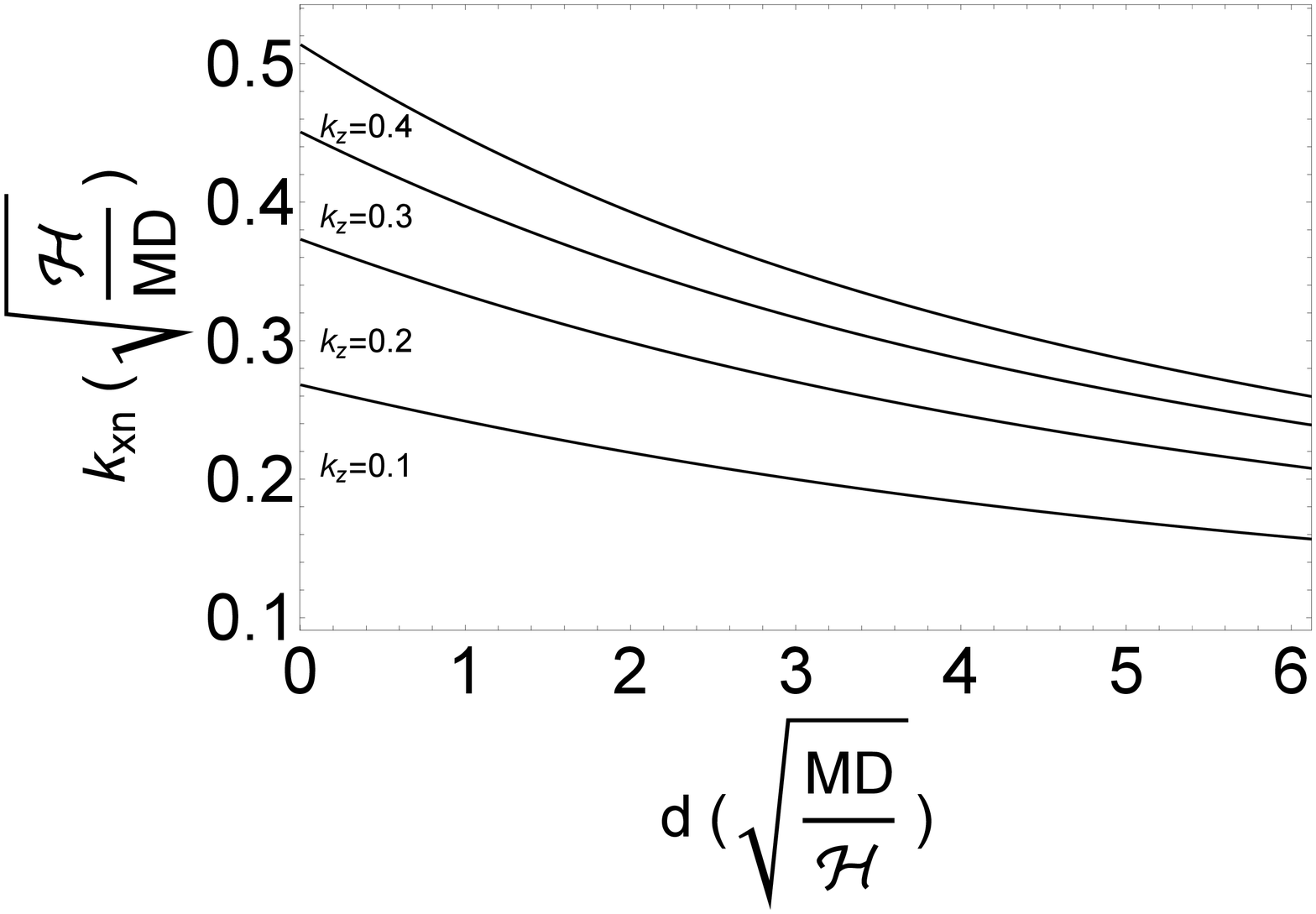}}
	\caption{Results of numerical calculations for the case $\chi=2.5$ and $\theta=0$. (a) Position of minima for the lowest mode vs $d$ for thin films.
	(b) The value of frequency in minimum for the lowest mode vs $d$ for thin films.
	(c) $k_{xn}$ for the lowest mode vs $d$ at fixed $k_z=0.1, 0.2, 0.3, 0.4$. Black solid curves correspond to our numerical calculations, circles are numerical calculations by Kreisel \textit{et al..} \cite{Kreisel 2009}.These figures agree with the figures from \cite{Li 2018}.} 
	\label{fig:frequency}
\end{figure*}
\subsection{Thin films.}
In what follows till the end of this section we use $\sqrt{M/\mathcal{H}}\ell$ as unit of length and $\left(\gamma\mathcal{H}\right)^{-1}$ as unit of time. 
In this part we discuss the case of thin films.
If the film's thickness is of the order of one or less ($\ell$ in dimensional units), it is regarded as thin. The experimental realization of ultrathin films of YIG with $d \ll 1$ looks very improbable since the typical value of $\ell$  (in YIG) is a few tens of nanometers. It may be accomplished in thin, monolayer-thick ferromagnetic materials. 
Transverse modes with high $n$ in thin films with $d\sim 1$ have $k_{xn}\approx \pi n/d \gg 1$ in the exchange dominance area. Thus, only a few modes with the lowest frequencies are of theoretical and experimental relevance. In these modes, evanescent waves penetrate to the film at a depth of the same order of magnitude as its thickness. They therefore play an equally essential role in spectral characteristics and TDM as the oscillating wave.

A compact analytic expression has been found only for frequency as function of the wave vector (see eq. (\ref{eq:spectrum})). 

Fig. \ref{fig:spectra2} shows examples of spectra in thin films that are qualitatively similar to spectra in thick films. Each mode determined by numbers $\nu,n$ at not very big $n$ has a frequency minimum at some $\kpar\neq 0$, but it does not follow equation $\frac{\partial\omega^2}{\partial\kpar^2}=0$ since $k_{xn}$ also depends on $\kpar$. Fig. \ref{fig:spectra2:a} and Fig. \ref{fig:spectra2:b} show that at $d=1$, the energy of transverse excitation weakly depends on $k_z$, a feature that could be expected for ultrathin films.

\begin{figure*}
\centering
\subfigure[]
	{\label{fig:TDM:a}
	\includegraphics[width=5cm]{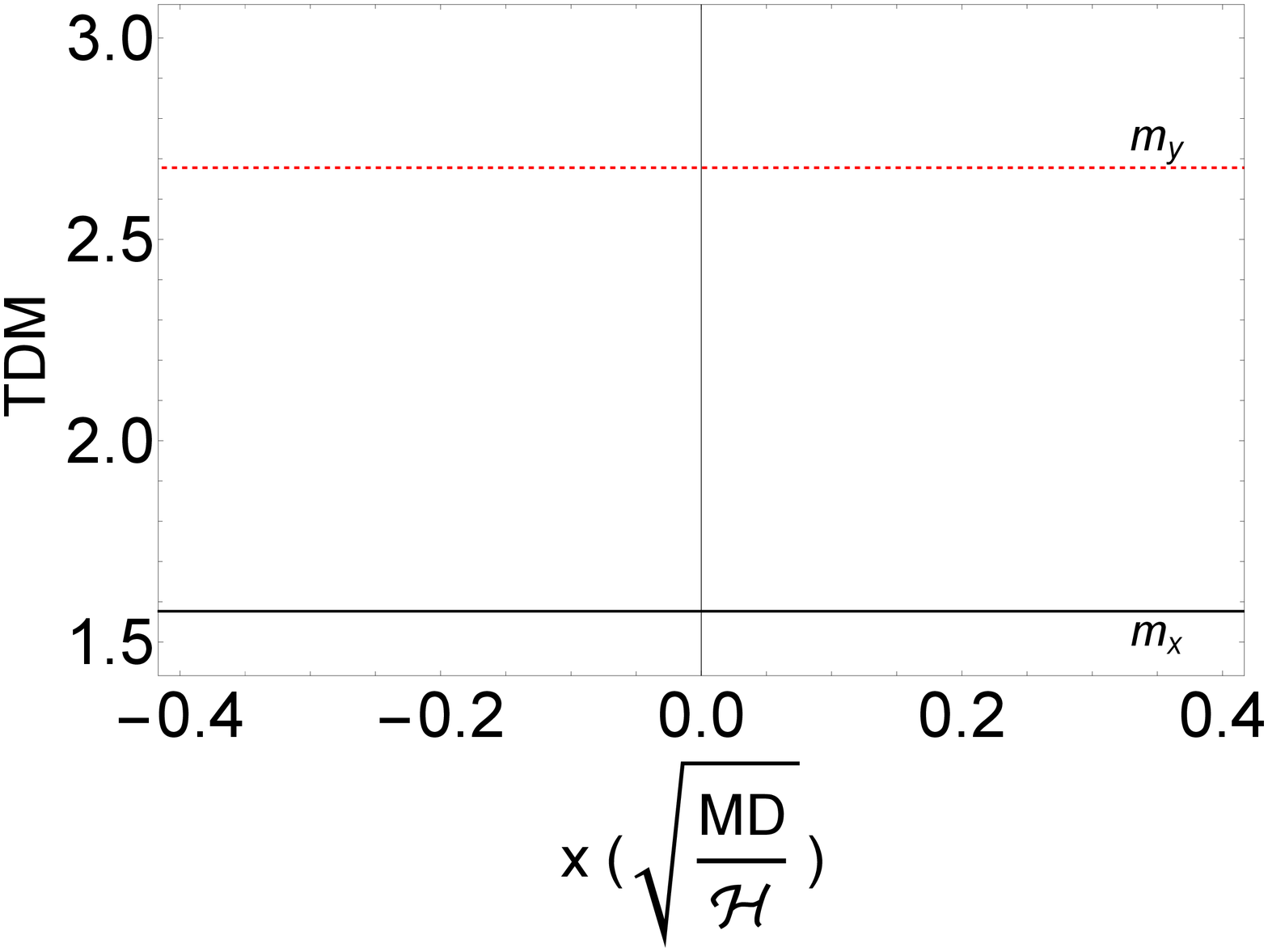}}
\subfigure[]
	{\label{fig:TDM:b}
	\includegraphics[width=5cm]{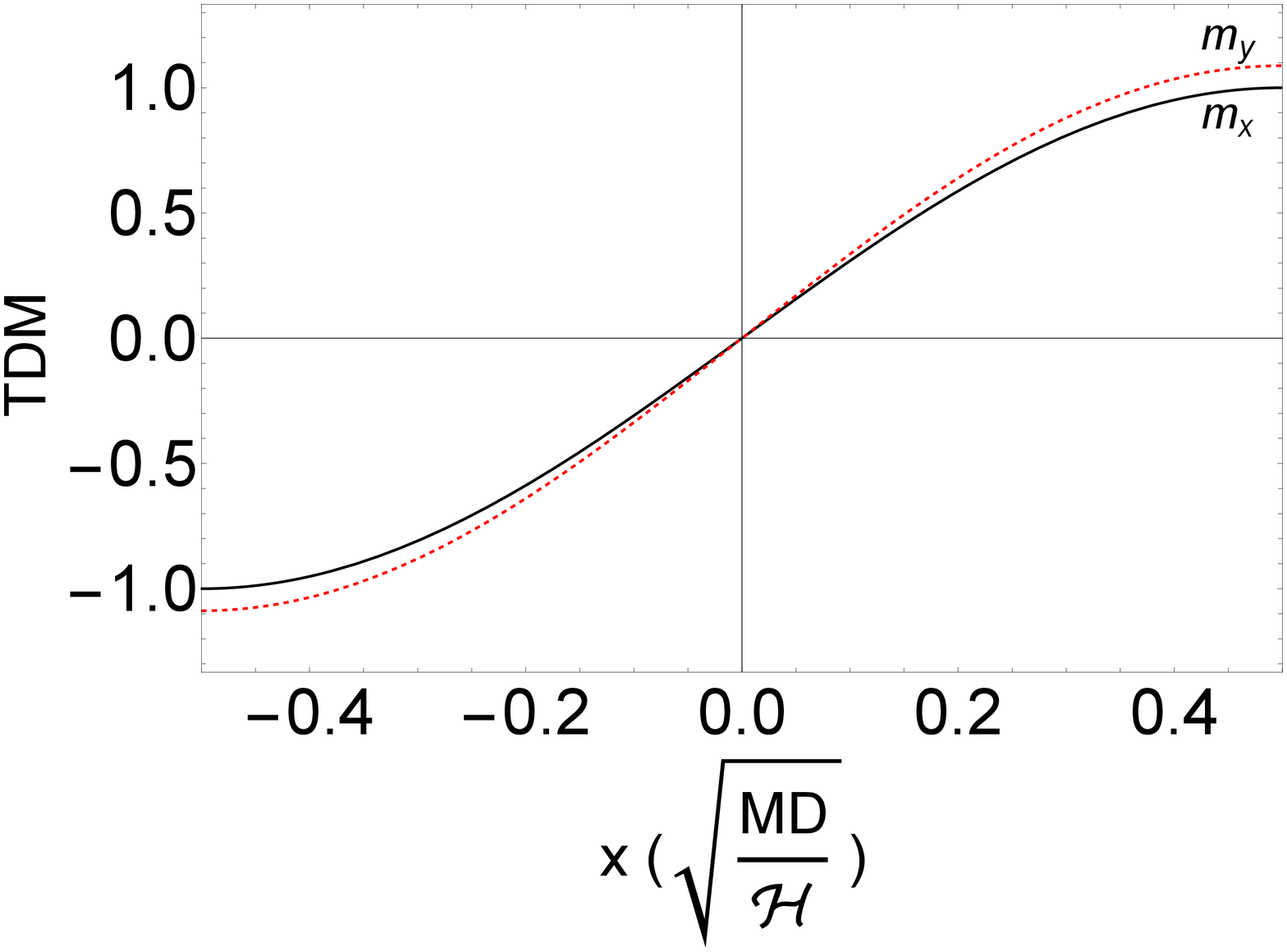}}
	\caption{For the case $\chi=2$ and $\theta=0$ (a) TDM for the lowest mode at $k_{\Vert}=0.1$ and $a_{1x}=1$.
        (b) TDM for the first excited mode at $k_{\Vert}=0.1$ and $b_{1x}=1$. These figures agree with the figures from \cite{Li 2018}.}
	\label{fig:TDM}
\end{figure*}

The graphs of position of minima and the value of frequency in minimum for the lowest mode vs $d$ for thin films are shown in Fig.\ref{fig:frequency}. In the same figures \ref{fig:frequency:a} and \ref{fig:frequency:b}, we compared our results with calculations of the same values  by Kreisel \textit {et al.} \cite {Kreisel 2009}. Finally, the graphs of $k_{xn}$ for the lowest mode vs $d$ at fixed $\kpar$ and $\theta=0$ are shown in Fig. \ref{fig:frequency:c}.  An example of TDM for lowest mode and first excited mode in thin films is shown in Fig. \ref{fig:TDM}.

All ground state spectra cross at the point $\kpar=0,\omega\approx\sqrt{1+\chi}$ ($\sqrt{3}\approx 1.73$ for $\chi=2$), 
exactly the same result as for the thick film. This is manifestation of a general property of films with arbitrary thickness: at $\mathbf{k}_{\Vert}=0$, the transverse wave vector of the lowest transverse mode is also equal to zero. The frequency of the lowest mode equals to $\omega_0=\sqrt{1+\chi}$ (ferromagnetic resonance frequency).

We consider first the limiting case of ultrathin films $d\rightarrow 0$ when $\theta=0$.  It will be shown that only wave vectors of the lowest transverse mode with $\nu=-, n=0$ remains finite in this limit. All excited transverse state with other $\nu$ or $n$ have wave vectors that go to infinity as $1/d$. 
We just take into account the simplest scenario of waves propagating along magnetization and magnetic field in order to simplify calculations. The transverse mode then has a definite parity. 

In such a case, the non-zero amplitudes are $\mathbf{a}_i$ for even modes and $\mathbf{b}_i$ for odd modes. For finite wave vectors  $\mathbf{k}_i$ in the taken limit, $\sin k_{ix}d/2\approx k_{ix}d/2$ and $\cos k_{ix}d/2\approx 1$ are appropriate values. This fact simplifies the SBC (\ref{eq:SBC}) and self-consistency equations (\ref{eq:self-consist}). The second simplification results from the fact that the relationship between the $x$ and $y$ components of the vectors $\mathbf{a}_i$ and $\mathbf{b}_i$ is reduced to $a_{iy}=\frac{\omega}{1+k_i^2}a_{ix}$ and $b_{iy}=\frac{\omega}{1+k_i^2}b_{ix}$, respectively.  Here  we denote three kernels of cubic equation for $k^2$ (45) as $k_1^2, k_2^2, k_3^2$ and corresponding vector amplitudes at $\sin(k_{ix}x)$ and $\cos(k_{ix}x)$ as $\mathbf{a}_i, \mathbf{b}_i$. Let us remind that
$k_1^2>0$, whereas $k_2^2, k_3^2<0$.  After all these simplifications, the quantization of an even mode is described by the system of three equations with three independent amplitudes $a_{ix}$ :

\begin{equation}\label{d=0-eq}
\left \lbrace \begin{array}{ll}
\sum_{i=1}^{3}k_{ix}^2a_{ix}&=0 \\ 
\sum_{i=1}^{3}\frac{k_{ix}^2}{1+k_i^2}a_{ix}&=0\\ 
\sum_{i=1}^{3}\frac{a_{ix}}{k_{i}^2}&=0
\end{array}
\right.
\end{equation}
Zeros of determinant of this system determine quantized values of $k_{xn}^2$. 
In order to transform this determinant into an explicit function of $k_{xn}$ one should employ the relations $k_1^2 = k_{xn}^2+k_z^2$, 
\begin{equation}\label{k23-k1}
k_{2,3}^2=-1-\frac{\chi}{2}-\frac{k_1^2}{2}\pm\sqrt{\left(1+\frac{\chi}{2}+\frac{k_1^2}{2}\right)^2-\frac{\chi k_z^2}{k_1^2}}
\end{equation}
and  $k_{ix}^2=k_i^2-k_z^2$. The only positive root of this equation at small $k_z\ll 1$ is

\begin{equation}\label{low-kz-asympt}
k_{xn}\approx\left(\frac{\chi}{2+\chi}\right)^{1/4} \sqrt{k_z}
\end{equation} 

\begin{figure}
\centering
{\includegraphics[width=7cm]{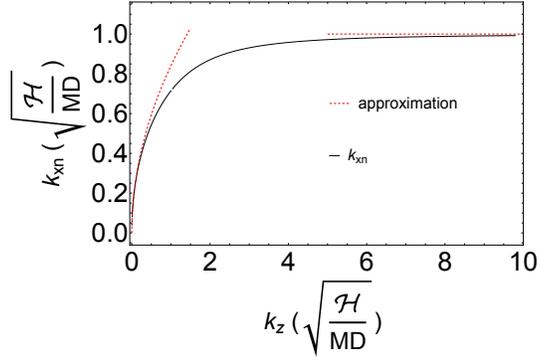}}
\caption{Plot of $k_{xn}$ at $d\rightarrow0$ and approximation to it when $\chi=2$ and $\theta=0$. This figure agrees with the figure from \cite{Li 2018}.}
	\label{fig:kx}
\end{figure}
\begin{figure}
\centering
{\includegraphics[width=7.2cm]{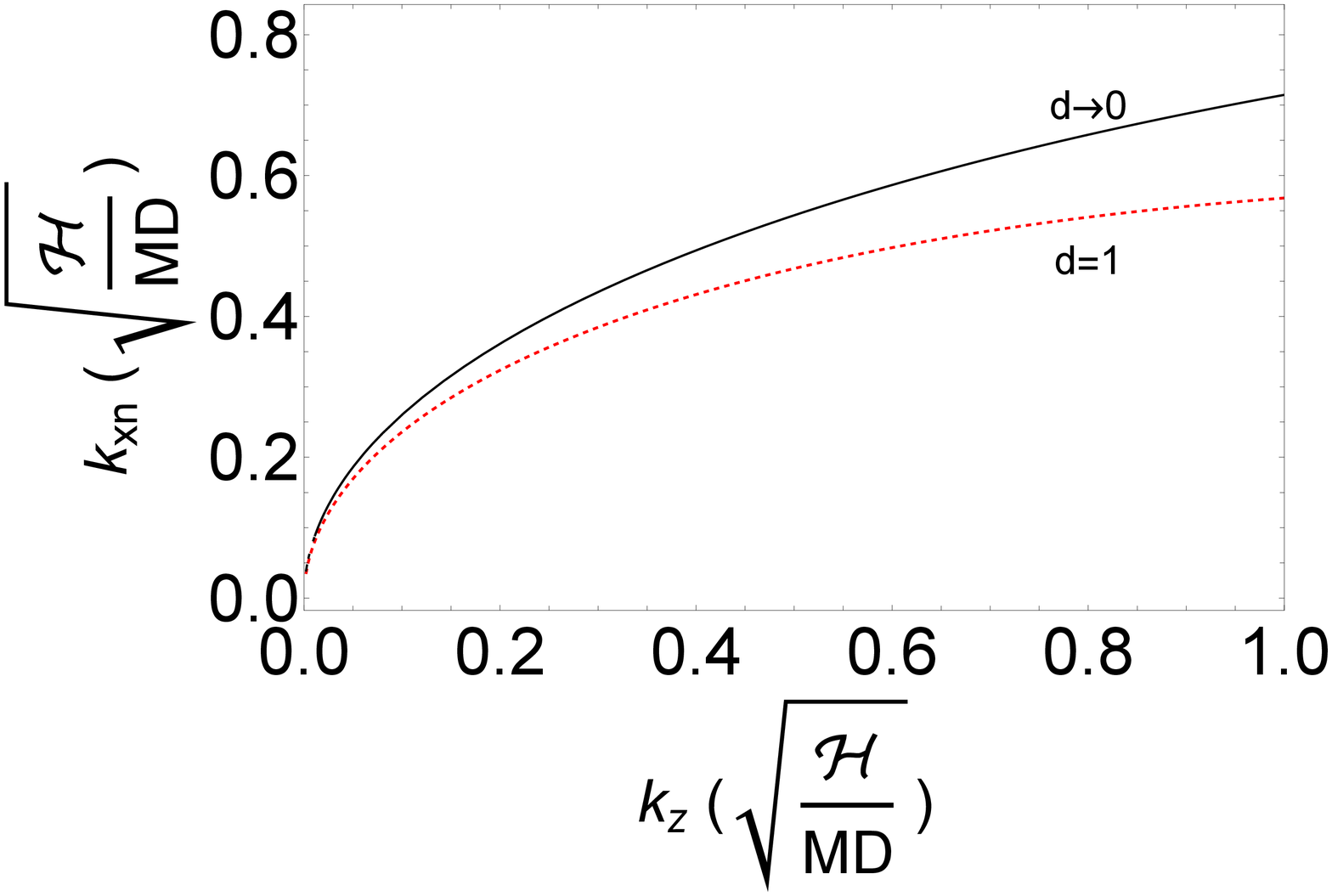}}
\caption{$k_{xn}$ vs. $k_z$ for the lowest mode at $d\rightarrow0$ and $d=1$ at $\chi=2$ and $\theta=0$.This figure agrees with the figure from \cite{Li 2018}.}
	\label{fig:d1}
\end{figure}

At large $k_z$, $k_{xn}$ asymptotically approaches a constant value $k_{xn}\approx \sqrt{\chi/2}$. Both these asymptotic values agree very well with numerical calculations of the dependence of $k_{xn}$ on $k_z$ at $d\rightarrow 0$ (see Fig.\ref{fig:kx}). 
The fact that $k_{xn}=0$ at $k_z=0$ is confirmed by the asymptotic behavior of $k_{xn}$ at small $k_z$. As a result, both in the limit of small $d$ and the limit of large $d$, the value of frequency at $\kpar =0$ is $\sqrt{1+\chi}$. On Fig. \ref{fig:d1}, the plots of $k_{xn}$ vs. $k_z$ at $d=1$ and $d=0$ are compared.

We can now demonstrate the general proposition that, regardless of thickness, the frequency of the lowest mode at $\mathbf{k}_{\Vert}=0$ equals $\sqrt{1+\chi}$ . Set $k_y=0$ and consider $k_z\ll 1/d^2$. We will show that the same equation (\ref{low-kz-asympt}) determines the first quantized value $k_{xn}$, but the arguments must be modified. In order to prove the result (\ref{low-kz-asympt}), let us assume that the initial quantized value of $k_{xn}$ obeys the strong inequalities $k_z\ll k_{xn}\ll 1$.
Then eq. (\ref{k23-k1}) implies that $k_{2x}^2\approx -\chi k_z^2/\left[\left(2+\chi\right)k_{xn}^2\right]$ 
has small magnitude, whereas $k_{3x}^2\approx -2-\chi$ has the magnitude of the order of unity. Let us first consider the SBC (\ref{eq:SBC}) that in considered situation take form 
\begin{eqnarray}
k_{xn}^2 a_{1x} + k_{2x}^2 a_{2x} -\sqrt{2+\chi}\frac{2\sinh\sqrt{2+\chi}d/2}{d} a_{3x}=0 \nonumber\\
k_{xn}^2 a_{1x} + k_{2x}^2 a_{2x} +\frac{2\sqrt{2+\chi}\sinh\sqrt{2+\chi}d/2}{(1+\chi)d} a_{3x}= 0
\end{eqnarray}
These equations imply $a_{3x}=0$. Then they become identical and define the ratio $a_{2x}/a_{1x}=-k_{xn}^2/k_{2x}^2$. Next consider the self-consistency equations that in the same limit have a form:
\[ \frac{a_{1x}}{k_{1}^2}+\frac{a_{2x}}{k_{2}^2}=0
\] 
Using the previously found ratio $a_{1x}/a_{2x}$, we again obtain eq. (\ref{low-kz-asympt}) for this more general situation. It shows that in the limit $k_z\rightarrow 0$, 
the limit of ratio $k_z^2/k_{xn}^2$ is also zero and limiting value of $\omega$ is $\sqrt{1+\chi}$ independently on thickness. Note that in the limit $\mathbf{k}_{\Vert}=0$
the magnetization in the lowest spin-wave mode does not depend on transverse coordinate.

Although thin films are more sensitive to the exact form of the SBC than thick films, changing forms of these requirements have no effect on the symmetry or general features of solutions. 
An important problem is how the wave vector $k_{z\mathrm{min}}$ corresponding to the minimum of energy changes with thickness.
 For thick films it behaves as $1/\sqrt{d}$ \cite{Sonin 2017} and grows when film becomes thinner. However, in the case of ultrathin films, it decreases linearly with thickness.

It means that the wave vector $k_{z\mathrm{min}}$ as function of $d$ has a maximum.  According to numerical calculations shown in Fig. \ref{fig:frequency:a} for $\chi = 2.5$  the maximum is located at $d\approx 6$, and the maximum value of $k_{z\mathrm{min}}$ is around 0.3. For $d=5\mu m$ and $\chi=2$, $k_{z\mathrm{min}}$ is around 0.02. Thus, by decreasing thickness from 5$\mu m$ to $15-30$ nm, the wave vector $k_{z\mathrm{min}}$ may be modified by a factor of roughly 15. The size of any soliton-like formation constructed of magnons that may be utilized for information transfer without dissipation or with very little dissipation has an upper limit determined by the minimal wavelength of a magnon, according to \cite{Sun 2017}. 

\section{Interaction of magnons.}

Previously we considered only quadratic in amplitudes part of the
Hamiltonian. Here we take into account higher order contributions,
i.e, we consider the magnon interaction. The expansion will be
limited by the terms of the third and the fourth order. The expansion
must be applied only to the exchange (\ref{eq:H-ex-psi}) and dipolar
(\ref{eq:H-dip-psi}) Hamiltonians since the Zeeman Hamiltonian is purely
quadratic.

\subsection{Third order terms.}

Let us first write out the 3rd order terms of the Hamiltonian, which
come solely from the dipolar part:
\begin{equation}
H_{d3}=-\frac{\mu_{B}\sqrt{2\mu_{B}M}}{2}\iint\left(\left|\psi\right|^{2}+\frac{1}{4}\left|\psi'\right|^{2}\right)\partial_{z}\left(\psi'\partial'_{-}+\psi^{\prime*}\partial'_{+}\right)\frac{dVdV'}{\left|\mathbf{r-r'}\right|}.\label{eq:H-cubic-psi}
\end{equation}
In terms of the Fourier transforms defined by eq. (\ref{eq:psi-Fourier})
and employing the identity
\begin{equation}
\frac{1}{\left|\mathbf{r-r'}\right|}=\frac{4\pi}{A}\sum_{\mathbf{q}}e^{i\mathbf{q}(\mathbf{r}-\mathbf{r}')}G_{q}\left(x-x'\right),\label{eq:Green-1-identity}
\end{equation}
where the 1d Green function is defined by eq. (\ref{eq:Green-1d}),
we find:
\begin{equation}
\begin{array}{c}
H_{d3}=-\frac{2\pi\mu_{B}\sqrt{2\mu_{B}M}}{\sqrt{A}}\iintop_{-\infty}^{\infty}dxdx^{\prime}\\
\sum_{\mathbf{q}_{1},\mathbf{q}_{2},\mathbf{q}_{3},\mathbf{q}}\left(\chi_{\mathbf{q}_{1}}\chi_{\mathbf{q}_{2}}^{*}\delta_{\mathbf{q}_{1}-\mathbf{q}_{2}+\mathbf{q}}\delta_{\mathbf{q}_{3}-\mathbf{q}}\right.\\
\left.+\frac{1}{4}\chi'_{\mathbf{q}_{1}}\chi_{\mathbf{q}_{2}}^{\prime*}\delta_{\mathbf{q}}\delta_{\mathbf{q}_{1}-\mathbf{q}_{2}+\mathbf{q}_{3}-\mathbf{q}}\right)iq_{z}\times\\
\left[\chi'_{\mathbf{q}_{3}}\left(d_{x'}-q_{y}\right)+\chi_{-\mathbf{q}_{3}}^{\prime*}\left(d_{x'}+q_{y}\right)\right]G_{q}(x-x')
\end{array}\label{eq:Hd3-q}
\end{equation}
The second term in the sum contains the factor $\delta_{\mathbf{q}}$
that makes $q_{y}=q_{z}=0.$ Thus, the square bracket in this equation
is equal to $\left(\chi'_{\mathbf{q}_{3}}+\chi_{-\mathbf{q}_{3}}^{\prime*}\right)d_{x^{\prime}}$.
Acting to $G_{q}(x-x')$, the operator $d_{x^{\prime}}$ transforms
it into $q\mathrm{sign\left(\mathit{x-x^{\prime}}\right)\mathit{G_{q}(x-x')=\frac{\mathrm{sign}\left(\mathit{x-x^{\prime}}\right)}{2}}}$.
Thus, the second term in the sum is zero. The Kronecker $\delta-$symbols
in the first term imply that $\mathbf{q=\mathbf{q}_{3}=\mathbf{q}_{2}-\mathbf{q}_{1}}$.
Thus, the dipolar Hamiltonian of the third order is simplified to
\begin{equation}
\begin{array}{c}
H_{d3}=-\frac{2\pi\mu_{B}\sqrt{2\mu_{B}M}}{\sqrt{A}}\iintop_{-\infty}^{\infty}dxdx^{\prime}\\
\sum_{\mathbf{q}_{1},\mathbf{q}_{2}}\chi_{\mathbf{q}_{1}}\chi_{\mathbf{q}_{2}}^{*}i\left(q_{2z}-q_{1z}\right)\left[\chi_{\mathbf{q}_{2}-\mathbf{q}_{1}}^{\prime}\left(d_{x^{\prime}}-q_{2y}+q_{1y}\right)\right.\\
\left.+\chi_{\mathbf{q}_{1}-\mathbf{q}_{2}}^{\prime*}\left(d_{x^{\prime}}+q_{2y}-q_{1y}\right)\right]G_{\left|\mathbf{q}_{1}-\mathbf{q}_{2}\right|}\left(x-x^{\prime}\right)
\end{array}\label{eq:H-d3-q-fin}
\end{equation}

\subsubsection{Third order non-linearity in terms of quantized magnon amplitudes.}

In this section we perform the Bogoliubov transformation (\ref{eq:inv-bog-uv})
from transverse modes $\chi_{\mathbf{q}}\left(x\right)$ to the quantized
amplitudes of magnons $\eta_{\mathbf{q},n},\eta_{\mathbf{q},n}^{*}$.
After some algebra we arrive at a cubic form for these amplitudes
limited by the requirement of the momentum conservation (translational
invariance):
\begin{equation}
\begin{array}{c}
H_{d3}=-\frac{2\pi\mu_{B}\sqrt{2\mu_{B}M}}{\sqrt{A}}\sum_{\mathbf{q}_{1}n_{1};\mathbf{q}_{2}n_{2};\mathbf{q}_{3}n_{3}}\delta_{\mathbf{q}_{1}-\mathbf{q}_{2}+\mathbf{q}_{3}}\\
\left(I_{d3}^{\left(+++\right)}\eta_{\mathbf{q}_{1}n_{1}}\eta_{\mathbf{-q}_{2}n_{2}}\eta_{\mathbf{q}_{3}n_{3}}+I_{d3}^{\left(++-\right)}\eta_{\mathbf{q}_{1}n_{1}}\eta_{\mathbf{-q}_{2}n_{2}}\eta_{\mathbf{-q}_{3}n_{3}}^{*}\right.\\
\left.I_{d3}^{\left(+-+\right)}\eta_{\mathbf{q}_{1}n_{1}}\eta_{\mathbf{q}_{2}n_{2}}^{*}\eta_{\mathbf{q}_{3}n_{3}}+I_{d3}^{\left(-++\right)}\eta_{\mathbf{-q}_{1}n_{1}}^{*}\eta_{\mathbf{-q}_{2}n_{2}}\eta_{\mathbf{q}_{3}n_{3}}+c.c\right)
\end{array},\label{eq:Hd3-eta}
\end{equation}
where the eight coefficients $I_{d3}^{\left(\rho\sigma\tau\right)}$
with $\rho,\sigma,\tau$ taking values $+,-$ are matrix elements
of the three transverse modes: the first is $u_{\mathbf{q}_{1}n_{1}}^{*}\left(x\right)$
for $\rho=+$ and $u_{-\mathbf{q}_{1}n_{1}}$ for $\rho=-$; the second
is $v_{\mathbf{-q}_{2}n_{2}}^{*}\left(x\right)$ for $\sigma=+$ and
$v_{\mathbf{q}_{2}n_{2}}$ for $\sigma=-$; the third is given by
\[
iq_{3z}\left[u_{\mathbf{q}_{3}n_{3}}^{*}\left(x^{\prime}\right)\left(d_{x^{\prime}}-q_{3y}\right)-v_{\mathbf{q}_{3}n_{3}}^{*}\left(x^{\prime}\right)\left(d_{x^{\prime}}+q_{3y}\right)\right]G_{q_{3}}\left(x-x^{\prime}\right)
\]
for $\tau=+$ and
\[
iq_{3z}\left[u_{\mathbf{-q}_{3}n_{3}}\left(x^{\prime}\right)\left(d_{x^{\prime}}+q_{3y}\right)-v_{\mathbf{-q}_{3}n_{3}}\left(x^{\prime}\right)\left(d_{x^{\prime}}-q_{3y}\right)\right]G_{q_{3}}\left(x-x^{\prime}\right)
\]
for $\tau=-$.
\newline
The matrix element is the double integral over $x$ and $x^{\prime}$ from the products of any set of these three modes.
\newline
For the reader convenience we place below explicit expressions for
the integrals $I_{d3}^{\left(\rho\sigma\tau\right)}$ with all three
indices $+$ and with two $+$ and one $-$:
\begin{equation}
\begin{array}{c}
I_{d3}^{\left(+++\right)}=-iq_{3z}\iint dxdx^{\prime}u_{\mathbf{q}_{1}n_{1}}^{*}v_{-\mathbf{q}_{2}n_{2}}^{*}\times\\
\left[u_{\mathbf{q}_{3}n_{3}}^{\prime*}\left(d_{x'}-q_{3y}\right)-v_{\mathbf{q}_{3}n_{3}}^{\prime*}\left(d_{x'}+q_{3y}\right)\right]G_{q_{3}}(x-x')\\
I_{d3}^{\left(++-\right)}=-iq_{3z}\iint dxdx^{\prime}u_{\mathbf{q}_{1}n_{1}}^{*}v_{-\mathbf{q}_{2}n_{2}}^{*}\times\\
\left[u_{-\mathbf{q}_{3}n_{3}}^{\prime}\left(d_{x'}+q_{3y}\right)-v_{-\mathbf{q}_{3}n_{3}}^{\prime}\left(d_{x'}-q_{3y}\right)\right]G_{q_{3}}(x-x')\\
I_{d3}^{\left(+-+\right)}=iq_{3z}\iint dxdx^{\prime}u_{\mathbf{q}_{1}n_{1}}^{*}v_{\mathbf{q}_{2}n_{2}}\times\\
\left[u_{\mathbf{q}_{3}n_{3}}^{\prime*}\left(d_{x'}-q_{3y}\right)-v_{\mathbf{q}_{3}n_{3}}^{\prime*}\left(d_{x'}+q_{3y}\right)\right]G_{q_{3}}(x-x')\\
I_{d3}^{\left(-++\right)}=iq_{3z}\iint dxdx^{\prime}u_{\mathbf{-q}_{1}n_{1}}v_{-\mathbf{q}_{2}n_{2}}^{*}\times\\
\left[u_{\mathbf{q}_{3}n_{3}}^{\prime*}\left(d_{x'}-q_{3y}\right)-v_{\mathbf{q}_{3}n_{3}}^{\prime*}\left(d_{x'}+q_{3y}\right)\right]G_{q_{3}}(x-x')
\end{array}.\label{eq: Id3}
\end{equation}
In order to obtain the Hamiltonian $H_{d3}$ (\ref{eq:Hd3-eta}) and
coefficients $I_{d3}^{\left(\sigma\rho\tau\right)}$ we have used
the fact that some terms (e.g. the term with $\eta_{-\mathbf{q}_{1}n_{1}}^{*}\eta_{\mathbf{q}_{2}n_{2}}^{*}\eta_{-\mathbf{q}_{3}n_{3}}^{*}$)
can be expressed as complex conjugates of others (e.g. the term with
$\eta_{\mathbf{q}_{1}n_{1}}\eta_{-\mathbf{q}_{2}n_{2}}\eta_{\mathbf{q}_{3}n_{3}}$)
by permutation of the summation indices $\mathbf{q}_{1}\leftrightarrow\mathbf{q}_{2}$
that implies $\mathbf{q}_{3}\rightarrow-\mathbf{q}_{3}$ . Later we
will use this kind of relations when calculating 4th-order terms. Note also that the three terms involving one complex
conjugated function in eq. (\ref{eq:Hd3-eta}) can also be received
each from other by renaming the summation indices. Thus, these three
sums are identical. On the other hand two last of them are complex
conjugates each to other. Therefore, all these sums are real.

\subsubsection{Cherenkov radiation of a low energy magnon by the high energy
magnons.}

In the theory of BECM the life-time of the condensate magnons is dominantly
determined by their merging with a high energy magnon and by the inverse
process of the Cherenkov radiation of the condensate magnon by a high
energy magnons. Here we consider a more general problem when the high
energy magnon emits or absorbs a low energy magnon. The high-energy
magnon is assumed to have the exchange dominated dispersion $\omega_{\mathbf{q},k_{x}}=\gamma\ell^{2}k^{2},$
whereas the low-energy magnon dispersion is given by eq. (\ref{eq:spectrum}).
In the Bogoliubov coefficients $u_{\mathbf{q}n}$ the coefficients $a,b$ dominate for $\nu=+$, $c,d$ dominate for $\nu =-$, whereas $v_{\mathbf{q}n}=0$ .
For low-energy magnons generally the coefficients $a,b,c,d$ are of
the same order of magnitude. They are defined by eqs. (\ref{eq:u-ansatz},\ref{eq:v-ansatz}).
For thick films in the integrals (\ref{eq: Id3}) defining the matrix
elements of the Cherenkov or inverse Cherenkov process, the terms
corresponding to evanescent waves can be neglected.

\subsection{Fourth order terms.}

Here we consider the 4th order terms of the Hamiltonian. In terms
of general magnon wave function $\psi\left(\mathbf{r}\right)$ they are:
\begin{equation}
\begin{array}{cc}
H_{4}= & H_{ex4}+H_{d4}\\
H_{ex4}= & \frac{\mu_{B}^{2}\ell^{2}}{2}\int\left[-\left|\psi\right|^{2}|\nabla\psi|^{2}+\frac{1}{2}\left(\nabla\left(\left|\psi\right|^{2}\right)\right)^{2}\right]dV,\\
H_{d4}= & \frac{\mu_{B}^{2}}{2}\iint\left[\left|\psi\right|^{2}\left|\psi'\right|^{2}\partial_{z}\partial'_{z}-\frac{1}{4}\left|\psi\right|^{2}\left(\psi\partial_{-}+\psi^{*}\partial_{+}\right)\left(\psi'\partial'_{-}+\psi^{\prime*}\partial'_{+}\right)\right]\frac{dVdV'}{\left|\mathbf{r-r'}\right|}
\end{array}\label{eq:H4-psi}
\end{equation}

\subsubsection{Fourth order Hamiltonian in terms of magnon amplitudes $\chi_{\mathbf{q}}\left(\mathbf{r}\right)$.}

Employing Fourier transformation to the wave vector representation
(\ref{eq:psi-Fourier}), we find the following expressions for $H_{ex4}$
and $H_{d4}$:
\begin{equation}
\begin{array}{cc}
&H_{ex4}=\frac{\mu_{B}^{2}\ell^{2}}{2A^{2}}\int\sum_{\mathbf{q}_{1},\mathbf{q}_{2},\mathbf{q}_{3},\mathbf{q}_{4}}\left[-\chi_{\mathbf{q}_{1}}\chi_{\mathbf{q}_{2}}^{*}\left(d_{x}\chi_{\mathbf{q}_{3}}d_{x}\chi_{\mathbf{q}_{4}}^{*}+\mathbf{q}_{3}\mathbf{q}_{4}\chi_{\mathbf{q}_{3}}\chi_{\mathbf{q}_{4}}^{*}\right)\right.\\
 & +\frac{1}{2}d_x(\chi_{\mathbf{q}_1}\chi^*_{\mathbf{q}_2})d_x(\chi_{\mathbf{q}_3}\chi^*_{\mathbf{q}_4})\left.+\frac{1}{2}(\mathbf{q}_{1}-\mathbf{q}_{2})(\mathbf{q}_{3}-\mathbf{q}_{4})\chi_{\mathbf{q}_{1}}\chi_{\mathbf{q}_{2}}^{*}\chi_{\mathbf{q}_{3}}\chi_{\mathbf{q}_{4}}^{*}\right]e^{i(\mathbf{q}_{1}-\mathbf{q}_{2}+\mathbf{q}_{3}-\mathbf{q}_{4})\mathbf{r}}dV\\
 & =\frac{\mu_{B}^{2}\ell^{2}}{4A}\int\sum_{\mathbf{q}_1,\mathbf{q}_2,\mathbf{q}_3,\mathbf{q}_4}\left[d_x\chi_{\mathbf{q}_1}\chi^*_{\mathbf{q}_2}d_x\chi_{\mathbf{q}_3}\chi^*_{\mathbf{q}_4}
+ \chi_{\mathbf{q}_1}d_x\chi^*_{\mathbf{q}_2}\chi_{\mathbf{q}_3}d_x\chi^*_{\mathbf{q}_4}\right.\\
&\left.-(\mathbf{q}_1^2+\mathbf{q}_2^2)\chi_{\mathbf{q}_1}\chi^*_{\mathbf{q}_2}\chi_{\mathbf{q}_3}\chi^{*}_{\mathbf{q}_4}\right]\delta_{\mathbf{q}_1-\mathbf{q}_2+\mathbf{q}_3-\mathbf{q}_4}dx\\
& =\frac{\mu_{B}^{2}\ell^{2}}{4A}\int\sum_{\mathbf{q}_1,\mathbf{q}_2,\mathbf{q}_3,\mathbf{q}_4}\left[d_x\chi_{\mathbf{q}_1}\chi^*_{\mathbf{q}_2}d_x\chi_{\mathbf{q}_3}\chi^*_{\mathbf{q}_4}
+ \chi_{\mathbf{q}_1}d_x\chi^*_{\mathbf{q}_2}\chi_{\mathbf{q}_3}d_x\chi^*_{\mathbf{q}_4}\right.\\
&\left.-\frac{1}{2}(\mathbf{q}_1^2+\mathbf{q}_2^2+\mathbf{q}_3^2+\mathbf{q}_4^2)\chi_{\mathbf{q}_1}\chi^*_{\mathbf{q}_2}\chi_{\mathbf{q}_3}\chi^{*}_{\mathbf{q}_4}\right]\delta_{\mathbf{q}_1-\mathbf{q}_2+\mathbf{q}_3-\mathbf{q}_4}dx
\end{array}\label{eq:H-ex4-chi}
\end{equation}
\begin{equation}
\begin{array}{cc}
H_{d4}= & \frac{2\pi\mu_{B}^{2}}{A^{3}}\iint\sum_{\mathbf{q}_{1},\mathbf{q}_{2},\mathbf{q}_{3},\mathbf{q}_{4},\mathbf{q}}\left\{ q_{z}^{2}\chi_{\mathbf{q}_{1}}\chi_{\mathbf{q}_{2}}^{*}\chi'_{\mathbf{q}_{3}}\chi_{\mathbf{q}_{4}}^{\prime*}e^{i[(\mathbf{q}_{1}-\mathbf{q}_{2})\mathbf{r}+(\mathbf{q}_{3}-\mathbf{q}_{4})\mathbf{r}'+\mathbf{q}(\mathbf{r}-\mathbf{r}')]}-\right.\\
 & \left.\frac{1}{4}\chi_{\mathbf{q}_{1}}\chi_{\mathbf{q}_{2}}^{*}\left[\chi_{\mathbf{q}_{3}}\left(d_{x}+q_{y}\right)+\chi_{-\mathbf{q}_{3}}^{*}\left(d_{x}-q_{y}\right)\right]\left[\chi'_{\mathbf{q}_{4}}\left(d_{x'}-q_{y}\right)+\chi_{-\mathbf{q}_{4}}^{\prime*}\left(d_{x'}+q_{y}\right)\right]\right.\\
 & \left.\times e^{i[(\mathbf{q}_{1}-\mathbf{q}_{2})\mathbf{r}+\mathbf{q}_{3}\mathbf{r}+\mathbf{q}_{4}\mathbf{r}'+\mathbf{q}(\mathbf{r}-\mathbf{r}')]}\right\} G_{q}(x-x')dVdV'
\end{array}\label{eq:H-d4-chi1}
\end{equation}
After integration over $y,z$ and $y^{\prime},z^{\prime}$ the
4th-order dipolar Hamiltonian transforms into the sum over momenta and
integral over transverse coordinates:
\begin{equation}
\begin{array}{cc}
H_{d4}= & \frac{2\pi\mu_{B}^{2}}{A}\iint\sum_{\mathbf{q}_{1},\mathbf{q}_{2},\mathbf{q}_{3},\mathbf{q}_{4},\mathbf{q}}\left\{ q_{z}^{2}\chi_{\mathbf{q}_{1}}\chi_{\mathbf{q}_{2}}^{*}\chi'_{\mathbf{q}_{3}}\chi_{\mathbf{q}_{4}}^{\prime*}\delta_{\mathbf{q}_{1}-\mathbf{q}_{2}+\mathbf{q}}\delta_{\mathbf{q}_{3}-\mathbf{q}_{4}-\mathbf{q}}\right.\\
 & \left.-\frac{1}{4}\chi_{\mathbf{q}_{1}}\chi_{\mathbf{q}_{2}}^{*}\left[\chi_{\mathbf{q}_{3}}\left(d_{x}+q_{y}\right)+\chi_{-\mathbf{q}_{3}}^{*}\left(d_{x}-q_{y}\right)\right]\left[\chi'_{\mathbf{q}_{4}}\left(d_{x'}-q_{y}\right)+\chi_{-\mathbf{q}_{4}}^{\prime*}\left(d_{x'}+q_{y}\right)\right]\right.\\
 & \left.\times\delta_{\mathbf{q}_{1}-\mathbf{q}_{2}+\mathbf{q}_{3}+\mathbf{q}}\delta_{\mathbf{q}_{4}-\mathbf{q}}\right\} G_{q}(x-x')dxdx'.
\end{array}\label{eq:H-d4-chi2}
\end{equation}
In these calculation we used the symmetry with respect to permutations
of running momenta participating in the sum and the relation between
Fourier component of $1/\left|\mathbf{r-r}^{\prime}\right|$ and one-dimensional
Green function $G_{q}\left(x-x^{\prime}\right)$ (see eq. (\ref{eq:Green-1d})).

\subsubsection{Fourth order Hamiltonian in terms of the magnon amplitudes
$\eta_{\mathbf{q}\nu n}$.\label{subsec:amplitudes}}

Employing the Bogoliubov transformation (\ref{eq:unitarity-uv-dual}),
we represent the 4-th order Hamiltonian in terms of the homogeneous
fourth order polynomials of the form (the subscripts $\mathbf{q}_i,n_i$ in the coefficients $I_4$ are omitted for brevity):
\begin{equation}
H_{4}=\sum_{\mathbf{q}_{i}n_{k}\rho_{l}\left(i,k,l=1...4\right)}I_{4}^{\left(\rho_{1}\rho_{2}\rho_{3}\rho_{4}\right)}\left[\prod_{j=1}^{4}\eta_{\mathbf{q}_{j}n_{j}}^{\left(\rho_{j}\right)}\right]\delta_{\mathbf{q}_{1}+\mathbf{q}_{2}+\mathbf{q}_{3}+\mathbf{q}_{4}},\label{eq:H4-eta-gen}
\end{equation}
where $\eta_{\mathbf{q}n}^{\left(+\right)}=\eta_{\mathbf{q}n};\eta_{\mathbf{q}n}^{\left(-\right)}=\eta_{\mathbf{q}n}^{*}$.
It is obvious that the matrix $I_{4}$ can be made invariant under
permutation of four its composite indices $\gamma_{j}=\left(\rho_{j}\mathbf{q}_{j}n_{j}\right);j=1,2,3,4$
since the product in eq. (\ref{eq:H4-eta-gen}) is invariant under
such permutation. Therefore, it is more reasonable to denote the matrix
elements of the matrix $I_{4}$ as $\left(I_{4}\right)_{\gamma_{1}\gamma_{2}\gamma_{3}\gamma_{4}}$.
The table of coefficients $\left(I_{4}\right)_{\gamma_{1}\gamma_{2}\gamma_{3}\gamma_{4}}$
is given in the Appendix {[}Hamiltonian of the 4-th order{]}.

\subsubsection{Interaction of condensate magnons in thick films.\label{subsec:Interaction}}

Here we show the results of calculations of the interaction between
condensate of magnons that have momenta either $\mathbf{Q}=Q\hat{z}$
or $-\mathbf{Q}$. When the condensate exists, the chemical potential
$\mu$ is equal to the minimal magnon energy $\Delta$. Therefore
the wave functions of the condensates $\psi_{\pm\mathbf{Q}}$ do not
depend on time (we remind that the time dependence of the wave function
is given by $\exp\left[-\frac{i\left(\Delta-\mu\right)t}{\hbar}\right]$).
Further for brevity we denote the wave functions of the two condensates
as $\psi_{\pm}$ and present them in terms of the densities of condensates
$n_{\pm}$ and their time-independent phases $\phi_{\pm}$ as
\begin{equation}
\psi_{\pm}=\sqrt{n_{\pm}}e^{i\phi_{\pm}}f\left(x\right),\label{eq:condensates-n-phi}
\end{equation}
where $f\left(x\right)=\sqrt{2}\cos\frac{\pi x}{d}$ is the transverse
wave function corresponding to the ground state of a magnon. The total
wave function is
\begin{equation}
\begin{array}{c}
\psi\left(\mathbf{r}\right)=\psi_{+}e^{i\mathbf{Qr}}+\psi_{-}e^{-i\mathbf{Qr}}=\\
\left[\sqrt{n_{+}}e^{i\left(Qz+\phi_{+}\right)}+\sqrt{n_{-}}e^{i\left(-Qz+\phi_{-}\right)}\right]f\left(x\right)
\end{array}.\label{eq:psi-tot}
\end{equation}
Introducing notation $n=n_{+}+n_{-}$ for the total density of condensate
and $\Phi\left(z\right)=2Qz+\phi_{+}-\phi_{-}$ for the phase difference
of the two condensates, we find the square of modulus of the wave
function:
\begin{equation}
\left|\psi\left(\mathbf{r}\right)\right|^{2}=\left[n+2\sqrt{n_{+}n_{-}}\cos\Phi\left(z\right)\right]f^{2}\left(x\right).\label{eq:psi-mod-square}
\end{equation}
The square of gradient of the wave function is
\begin{equation}
\begin{array}{c}
\left|\nabla\psi\right|^{2}=Q^{2}\left[n-2\sqrt{n_{+}n_{-}}\cos\Phi\left(z\right)\right]f^{2}\left(x\right)\\
+\left[n+2\sqrt{n_{+}n_{-}}\cos\Phi\left(z\right)\right]\left(\frac{df}{dx}\right)^{2}
\end{array}.\label{eq:grad-square}
\end{equation}
The fourth order exchange Hamiltonian contains two terms $-\left|\psi\right|^{2}\left|\nabla\psi\right|^{2}$
and $\frac{1}{2}\left(\nabla\left|\psi\right|^{2}\right)^{2}$. Assuming
that densities of condensates $n_{\pm}$ and their phases $\phi_{\pm}$
vary in plane on the distances much larger than period of density
oscillation $L=2\pi/Q$, the density of interaction energy of condensates
is equal to the exact value of interaction energy averaged over period
of oscillation $L$ integrated over the transverse coordinate $x$.
For thick films the terms in $\nabla\psi$ and $\nabla\left|\psi\right|^{2}$containing
derivatives $\frac{df}{dx}$ can be neglected in comparison with the terms
containing derivatives over $z$ or equivalently the value $Q$ since
$Qd\gg1$. Performing simple operations of averaging and integration
for exchange interaction we find:
\begin{equation}
\frac{\overline{H_{ex4}}}{V}=-\frac{3\mu_{B}^{2}\ell^{2}}{16}Q^{2}\left(n^{2}-6n_{+}n_{-}\right)\label{eq:H-ex4-av}
\end{equation}
Analyzing in similar way the interaction energy generated by dipolar
Hamiltonian of the 4-th order, we should find the average of the integrand
in the third equation (\ref{eq:H4-psi}). To make it, we will use
the identity:
\begin{equation}
\frac{1}{\left|\mathbf{r-r}^{\prime}\right|}=\frac{1}{\pi}\iintop_{-\infty}^{\infty}dq_{y}dq_{z}e^{i\mathbf{q}_{\Vert}\left(\mathbf{r}_{\Vert}-\mathbf{r}_{\Vert}^{\prime}\right)}G_{q_{\Vert}}\left(x-x^{\prime}\right),\label{eq:1/r-Gq}
\end{equation}
where the subscript $\Vert$ at a vector means that it is parallel
to the surfaces of the film, i.e., they have only $y$ and $z-$components;
we remind that the 1-dimensional Green function of the Helmholtz equation
$G_{q}\left(x\right)$ is defined by eq. (\ref{eq:Green-1d}). The
proof of the identity (\ref{eq:1/r-Gq}) is given in the Appendix
{[}1/r-G-identity{]}. Thus, the dipolar Hamiltonian of the 4-th order
can be rewritten as follows:
\begin{equation}
\begin{array}{c}
H_{d4}=\frac{\mu_{B}^{2}}{2\pi}\iiint dVdV^{\prime}d^{2}q_{\Vert}e^{i\mathbf{q}_{\Vert}\left(\mathbf{r}_{\Vert}-\mathbf{r}_{\Vert}^{\prime}\right)}\\
\left[\left|\psi\right|^{2}\left|\psi'\right|^{2}\partial_{z}\partial'_{z}-\frac{1}{8}\left(\left|\psi\right|^{2}+\left|\psi^{\prime}\right|^{2}\right)\times\right.\\
\left.\left(\psi\partial_{-}+\psi^{*}\partial_{+}\right)\left(\psi'\partial'_{-}+\psi^{\prime*}\partial'_{+}\right)\right]G_{q_{\Vert}}\left(x-x^{\prime}\right)
\end{array}\label{eq:Hd4-gen}
\end{equation}
Note that we symmetrized the integrand over the variables $\mathbf{r}$
and $\mathbf{r}^{\prime}$. Except of the exponential function $e^{i\mathbf{q}_{\Vert}\left(\mathbf{r}_{\Vert}-\mathbf{r}_{\Vert}^{\prime}\right)}$
the integrand does not depend of $y$ and $y^{\prime}$. Therefore
the integration over $y^{\prime}$ gives $2\pi\delta\left(q_{y}\right)$.
The partial derivatives $\partial_{\pm}=\partial_{x}\pm iq_{y}$ become
equal each to other and equal to $\partial_{x}$. The magnitude of
derivatives $\partial_{z},\partial_{z^{\prime}}$ is equal to $Q$,
whereas the magnitude of the derivatives $\partial_{x},\partial_{x^{\prime}}$
is equal to $2\pi/d$. For thick films $Q\gg1/d$, therefore, the first
term in the square brackets of this equation dominates. In this approximation
we find
\begin{equation}
\begin{array}{c}
H_{d4}=\mu_{B}^{2}\intop dV\intop_{-d/2}^{d/2}dx^{\prime}\intop_{-\infty}^{\infty}dz^{\prime}\intop_{-\infty}^{\infty}dq_{z}\\
\left[n+2\sqrt{n_{+}n_{-}}\cos\Phi\left(z\right)\right]\left[n+2\sqrt{n_{+}n_{-}}\cos\Phi\left(z^{\prime}\right)\right]\\
\left[f\left(x\right)f\left(x^{\prime}\right)\right]^{2}q_{z}^{2}e^{iq_{z}\left(z-z^{\prime}\right)}G_{\left|q_{z}\right|}\left(x-x^{\prime}\right).
\end{array}\label{eq:Hd4-inter.}
\end{equation}
Since the integrand does not depend on $y$, the integration over
this variable gives the linear size of sample $L_{y}$. Let us make
change of variables $Z=\frac{z+z^{\prime}}{2},\zeta=z-z^{\prime}$.
The Jacobian of this transformation is 1. The only term in the product
of two square brackets in eq. (\ref{eq:Hd4-inter.}) that together
with exponential factor $e^{iq_{z}\left(z-z^{\prime}\right)}$ gives
non-zero average is $4n_{+}n_{-}\cos\Phi\left(z\right)\cos\Phi\left(z^{\prime}\right)=2n_{+}n_{-}\left[\cos\left(\Phi\left(z\right)+\Phi\left(z^{\prime}\right)\right)+\cos\left(\Phi\left(z\right)-\Phi\left(z^{\prime}\right)\right)\right]$.
From these two terms only the second gives nonzero average over $z$:
\begin{equation}
\intop_{-\infty}^{\infty}d\zeta e^{iq_{z}\zeta}2\cos\left(2Q\zeta\right)=2\pi\left[\delta\left(q_{z}-2Q\right)+\delta\left(q_{z}+2Q\right)\right]\label{eq:zeta-av}
\end{equation}
This result allows us to perform also integration over $q_{z}$. Besides
of that the integrand does not depend on $Z$ and integration over
this variable gives the linear size $L_{z}$. These integrations strongly
simplify the expression for $H_{d4}$:
\begin{equation}
\overline{H_{d4}}=4\pi\mu_{B}^{2}L_{y}L_{z}Qn_{+}n_{-}\iintop_{-d/2}^{d/2}f^{2}\left(x\right)f^{2}\left(x^{\prime}\right)e^{-2Q\left|x-x^{\prime}\right|}dxdx^{\prime}\label{eq:Hd4-int-xx'}
\end{equation}
The calculation of the double integral in eq. (\ref{eq:Hd4-int-xx'})
is elementary and gives:
\begin{equation}
\begin{array}{c}
\iintop_{-d/2}^{d/2}f^{2}\left(x\right)f^{2}\left(x^{\prime}\right)e^{-2Q\left|x-x^{\prime}\right|}dxdx^{\prime}\\
=-\frac{-3 d^5 Q^5-5 \pi ^2 d^3 Q^3+\pi ^4 \left(-2 d Q-e^{-2 d Q}+1\right)}{2 Q^2
   \left(d^2 Q^2+\pi ^2\right)^2},
\end{array}\label{eq:int-f-square-xx'}
\end{equation}
In the limit of thick film $Qd\gg1$ the
leading term is equal to $3d/2Q$. This dependence of the integral
in (\ref{eq:int-f-square-xx'}) on parameters as $d/Q$ could be predicted
without detailed calculation since the exponent $e^{-Q\left|x-x^{\prime}\right|}$
cut in the square of integration a band of the width $\sim1/Q$ along
the diagonal, whereas the average value of $f^{2}$ is 1. However,
strong fluctuations of $f^{2}$ from 0 to 1 with period 1/8 of the
diagonal requires explicit calculation to get exact numerical coefficient
at the leading term:
\begin{equation}
\frac{\overline{H_{d4}}}{V}=6\pi\mu_{B}^{2}n_{+}n_{-}.\label{eq:inter-min-energy}
\end{equation}
Thus, we have found the density of interaction energy between condensates
of different minima (the inter-minima interaction). It can be written
as 
\begin{equation}
U_{4int}=Bn_{+}n_{-}\label{eq:U4-int}
\end{equation}
with $B=6\pi\mu_{B}^{2}>0$. It is repulsion. Note that the terms
of the same form in the exchange interaction energy (\ref{eq:H-ex4-av})
has coefficient $B$ which differs from dipolar value by a factor
$\sim Q^{2}\ell^{2}\sim\ell/d\ll1$ that can be neglected.

Another term that enters $\overline{H_{ex4}}/V$ but is absent in
$\overline{H_{d4}}/V$ is interaction of the condensate magnons within
one minimum
\begin{equation}
U_{4inn}=\frac{A}{2}\left(n_{+}^{2}+n_{-}^{2}\right)\label{eq:U4-inn}
\end{equation}
with $A=-\frac{3}{8}\mu_{B}^{2}Q^{2}\ell^{2}.$ Thus, the interaction
within one minimum is attraction. The magnitude $\left|A\right|$
is much smaller than $B$: $\left|A\right|/B=\frac{\pi^{1/2}}{2^{7/2}}\frac{\ell}{d}$.
For YIG film 5\textgreek{m}m thick at room temperature $\left|A\right|/B=0.012.$

\subsubsection{Quasi-equilibrium state.}

In the experiment by Demokritov\textit{ et al. }\cite{Demokritov 2006} the
low energy magnons in the YIG film were generated by a microstrip
resonator. A photon of frequency $\omega_{res}$ emitted by the resonator
decays into two magnons with practically opposite momenta and frequency
$\omega_{p}=\omega_{res}/2$ (in classical electrodynamics this process
is called parametric resonance or parametric pumping). The resonator
frequency is chosen to be less than $4\Delta/\hbar$, where $\Delta\approx2\mu_{B}\mathcal{H}$
is the minimal energy of magnons (gap in the spectrum). Then the decays
of pumped magnons are forbidden, whereas their collisions with other
low energy magnons remain possible. These collisions establish the
equilibrium. The relaxation time $\tau_{r}$ is just the time between
collisions. An important role is played by the processes of the Cherenkov
radiation of a low-energy magnon by a thermal magnon and inverse process
of the absorption of the low-energy magnon by a thermal magnon. These
processes determine the lifetime of low-energy magnons $\tau_{l}$.
In YIG at room temperature $\tau_{r}\ll\tau_{l}$. It means that during
the relaxation the number of magnons is conserved and they go to equilibrium
with the finite chemical potential $\mu$. The role of pumping is
to restore the stationary number of magnons in exchange of absorbed
ones. We will call such a stationary state quasi-equilibrium.

Let us consider the balance of magnons following Bun'kov and Volovik.\cite{Bunkov 2008}
The occupation number of a low-energy magnon with energy $\varepsilon$
in the quasi-equilibrium state is $n\left(\varepsilon\right)=\frac{T}{\varepsilon-\mu}$.
The occupation number of the magnon with the same energy in equilibrium
without pumping is $n_{0}\left(\varepsilon\right)=\frac{T}{\varepsilon}$.
The total density $n_{pm}\left(T,\mu\right)$ of pumped magnons is
\begin{equation}
n_{pm}\left(T,\mu\right)=\intop_{0}^{\infty}\left[n\left(\varepsilon\right)-n_{0}\left(\varepsilon\right)\right]\bar{g}\left(\varepsilon\right)d\varepsilon,\label{eq:n-pumped-gen}
\end{equation}
where $\bar{g}\left(\varepsilon\right)$ is the magnon density of
state per unit volume. It can be rewritten as
\begin{equation}
n_{pm}\left(T,\mu\right)=\intop_{0}^{\infty}\frac{T\mu}{\varepsilon\left(\varepsilon-\mu\right)}\bar{g}\left(\varepsilon\right)d\varepsilon.\label{eq:n-pumped-T-mu}
\end{equation}
The density of magnons pumped per unit time is determined by the pumped
power $W$ per unit volume as $\frac{2W}{\hbar\omega_{res}}$. In a
stationary state it must be equal to the density of pumped magnons
that disappear per unit time $\frac{n_{pm}}{\tau_{l}}$. Thus, the
established density of pumped magnons is
\begin{equation}
n_{pm}=\frac{2W}{\hbar\omega_{res}}\tau_{l}.\label{eq:n-pm-W-pm}
\end{equation}
Replacing $n_{pm}$ by the integral in the r.-h. side of eq. (\ref{eq:n-pumped-T-mu}),
we obtain equation relating the chemical potential $\mu$ to the pumped
power $W$. This equation implies that $\mu$ grows monotonically with
$W$ growing. At a critical value of the pumped power
\begin{equation}
W^{\left(c\right)}=\frac{\hbar\omega_{res}}{2\tau_{l}}\intop_{\Delta}^{\infty}\frac{T\Delta}{\varepsilon\left(\varepsilon-\Delta\right)}\bar{g}\left(\varepsilon\right)d\varepsilon,\label{eq:W-crit}
\end{equation}
chemical potential reaches its maximum possible value $\mu_{\max}=\Delta$
and the density of pumped magnons reaches its critical value
\begin{equation}
n_{pm}^{(c)}=\intop_{\Delta}^{\infty}\frac{T\Delta}{\varepsilon\left(\varepsilon-\Delta\right)}\bar{g}\left(\varepsilon\right)d\varepsilon.\label{eq:density-crit}
\end{equation}
 Chemical potential cannot grow more since at $\mu>\Delta$, the occupation
number of magnons with energy between $\Delta$ and $\mu$ would be
negative that is nonsense. Therefore, at $W>W^{\left(c\right)}$ the
chemical potential remains unchanged $\mu=\Delta$. The excessive
magnons go to the state with minimal energy $\Delta$ and form the
BEC. The condensate density is
\begin{equation}
n_{c}=\frac{2\left(W-W^{\left(c\right)}\right)}{\hbar\omega_{res}}\tau_{l}.\label{eq:cond-density}
\end{equation}
 All these calculations assumed that the integrals are converging.
There are two possible sources of divergence: large energies $\varepsilon\rightarrow\infty$
and $\varepsilon$ close to $\Delta$ for $W\geq W^{\left(c\right)}$.
For large $\varepsilon$ the exchange interaction dominates, the
magnon energy is quadratic function of momentum and $\bar{g}\left(\varepsilon\right)\propto\sqrt{\varepsilon}, $whereas
the denominator of integrand in eq. (\ref{eq:n-pumped-T-mu}) asymptotically
approaches $\varepsilon^{2}$. Thus, the integral converges at $\varepsilon\rightarrow\infty$.
This result physically means that the pumped magnons after relaxation
remain in the range of low energy $\sim\Delta$. Paradoxically their
energy escapes into the range $\iota\sim T$. Indeed, the pumped energy
is
\begin{equation}
E_{pm}=\intop_{0}^{\infty}\frac{T\mu}{\varepsilon\left(\varepsilon-\mu\right)}\varepsilon\bar{g}\left(\varepsilon\right)d\varepsilon.\label{eq:pumped-energy}
\end{equation}
This integral diverges at $\varepsilon\rightarrow\infty$. It happens
because we applied low-energy Rayleigh-Jeans approximation $n\left(\varepsilon\right)=\frac{T}{\varepsilon-\mu},n_{0}\left(\varepsilon\right)=\frac{T}{\varepsilon}$
for the occupation numbers of magnons, which at high energy must
be replaced by the Planck-Bose-Einstein distribution $n\left(\varepsilon\right)=\left(\exp\frac{\varepsilon-\mu}{T}-1\right)^{-1},n_{0}\left(\varepsilon\right)=\left(\exp\frac{\varepsilon}{T}-1\right)^{-1}$.
Thus, the integral (\ref{eq:pumped-energy}) is cut-off at $\varepsilon\sim T$.
Neglecting $\mu$ in denominator of integrand, we find the rough estimate
of the pumped energy per unit volume $\mu T^{3/2}/\left[\left(\mu_{B}M\right)^{3/2}\ell^{3}\right]$
that corresponds to the change of the magnons temperature by $\delta T \approx \Delta/k_B $. For YIG film in
external magnetic field $\mathcal{H}=600\mathrm{Oe}$ and at room temperature, the resulting
increase of temperature is about $0.04K$.

The convergence at the points of minimum energy $\epsilon = \Delta$ follows from the fact that, in the continuous limit, they are isolated points in 3-dimensional space. Therefore, the density of states near each minimum goes to zero as $\sqrt{\epsilon - \Delta}$.

\subsubsection{Spontaneous violation of the reflection symmetry in the quasi-equilibrium
state.}

In the state of quasi-equilibrium its energy (more accurately its
Helmholtz free energy) must be minimum. At fixed temperature and volume,
the free energy has minimum when the occupation numbers obey the Bose-Einstein
law and excessive magnons occupy the state with minimal energy $\Delta$.
In ferromagnetic films there are two such states. Therefore, the ground
state of the ideal magnon gas is highly degenerate: the condensate
energy $E_{id}=Vn_{c}\Delta$ depends only on the total number of
magnons in condensate $N_{c}=Vn_{c}=N_{+}+N_{-}$ and does not depend
on how these magnons are distributed between two minima. This $N_{c}+1-$fold
degeneration is lifted by magnons interaction.\cite{Sun 2017}

As it was derived in the subsection,\ref{subsec:Interaction} the
4-th order interaction density of energy is
\begin{equation}
U_{4}=\frac{A}{2}\left(n_{+}^{2}+n_{-}^{2}\right)+Bn_{+}n_{-},\label{eq:inter-total}
\end{equation}
with $A<0$ and $B>0$ for thick films. The interaction energy $U_{4}$
has minimum equal to $U_{4}=-\frac{A}{2}n^{2}$ either at $n_{+}=n,n_{-}=0$
or at $n_{+}=0,n_{-}=n$. In both cases the symmetry with respect
to reflection in the plane $z=0$ combined with the time reversal
is violated. Unfortunately such a most asymmetric state contradicts
to the experiment and to a more sophisticated theory.

\begin{figure}
\centering
{\includegraphics[width=7cm]{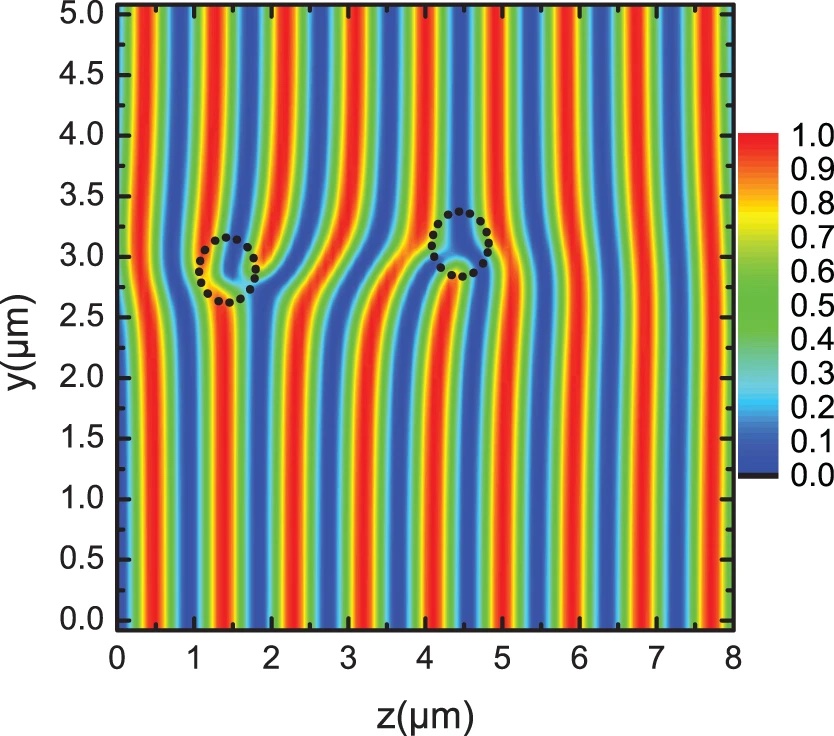}}
\caption{Measurement of the BLS intensity. Dashed circles indicate the positions of two defects causing an appearance of two vortices of positive circulation in different components of the condensate. The vortices show themselves as forks in the interference pattern.Reprinted by permission from Macmillan Publishers Ltd: Scientific Reports \cite{Novik-Boltyk 2012}, Copyright 2012.}
	\label{interference}
\end{figure}

Let us start with the experiment. In 2012 in the work by P. Novik-Boltyk
\textit{et al. }\cite{Novik-Boltyk 2012} the Münster experimental team led by
S. Demokritov discovered a stripe interference structure of the magnetization
$M_{z}$ in the YIG sample (see the interference picture in Fig.
\ref{interference}.) It can be interpreted as the measurement of 
\[
\left|\psi\right|^{2}=\left|\sqrt{n_{+}}e^{iQz+\phi_{+}}+\sqrt{n_{-}}e^{-iQz+\phi_{-}}\right|=n+2\sqrt{n_{+}n_{-}}\cos\left(2Qz+\phi_{+}-\phi_{-}\right).
\]

This equation clearly shows that the interference picture can be observed
only if both $n_{+}$ and $n_{-}$ are not zero. In order to explain
this result, F. Li, W. Saslow and V. Pokrovsky\cite{Li 2013} proposed
to consider the additional term in the 4-th order interaction Hamiltonian
of purely dipolar origin of the form 
\begin{equation}
\frac{C}{2}\left[\left(\psi_{+}^{*}\psi_{+}^{2}\psi_{-}+c.c.\right)+\left(+\leftrightarrow-\right)\right]\label{eq:H4-anomalous}
\end{equation}
where the abbreviation $c.c.$ stays for complex conjugate, $C$ is
a real constants whose magnitude in terms of parameters is of the
same order as $\left|A\right|$, however the numerical constant in
$C$ is by a factor $1/2\pi^{3}\approx0.016$ smaller. This term is
contained in the earlier neglected terms of the 4-th order dipolar
interaction containing derivatives over $x$. The real processes associated
with this term would be decay of one condensate magnon in three and
inverse process of merging three condensate magnon in one. All such
processes are forbidden by the energy conservation. However, they
determine additional (anomalous) 4-order interaction energy:
\begin{equation}
\frac{H_{4an}}{V}=Cn\sqrt{n_{+}n_{-}}\cos\left(\phi_{+}+\phi_{-}\right)\label{eq:H4-an-n-phi}
\end{equation}
Note that this energy depends on a different combination of phases
$\phi_{+}+\phi_{-}$ than the Goldstone phase $\phi_{+}-\phi_{-}$
whose variation does not change energy. The minimum energy is reached
at $\phi_{+}+\phi_{-}=\pi$ or $0$ depending on the sign of the coefficient
$C$. On the line $C=0$ the transition from $0-$ to $\pi-$phase
or vice versa proceeds. In both these phases the minimum anomalous
interaction energy is negative:
\begin{equation}
\min\left(\frac{H_{4an}}{V}\right)=-\left|C\right|n\sqrt{n_{+}n_{-}}.\label{eq:H4-an-min}
\end{equation}
Thus, the total $4-$th order interaction energy acquires the form:
\begin{equation}
U_{4}\equiv\frac{H_{4}}{V}=\frac{A}{2}\left(n_{+}^{2}+n_{-}^{2}\right)+Bn_{+}n_{-}-\left|C\right|n\sqrt{n_{+}n_{-}}\label{eq:4-int-total}
\end{equation}
Its minimization at a fixed $n$ gives:
\begin{equation}
\frac{n}{\sqrt{n_{+}n_{-}}}=\frac{2\left(B-A\right)}{\left|C\right|}.\label{eq:U4-minimization}
\end{equation}
Let us denote $R=\frac{B-A}{\left|C\right|}+\sqrt{\left(\frac{B-A}{\left|C\right|}\right)^{2}-1}$
and $\Theta=\frac{\left|C\right|}{2\left(B-A\right)R}.$ The value
$R$ is very big, whereas the value $\Theta\approx\frac{1}{4R^{2}}$
is very small. The two solutions of this equation are either
\begin{equation}
\begin{array}{c}
n_{+}=\left(1-\Theta\right)n;\\
n_{-}=\Theta n,
\end{array}\label{eq:4-solutions}
\end{equation}
or $n_{+}$ and $n_{-}$ interchange. In each solution
one of two condensate densities is much larger than another, but the
smaller one turns into zero only if $C=0$. The total interaction energy
in this phase is
\begin{equation}
\begin{array}{c}
U_{4}=n^{2}\left[\frac{A}{2}+\frac{\left|C\right|}{2R}\left(1-\Theta\right)-\left|C\right|\sqrt{\Theta\left(1-\Theta\right)}\right]\\
\approx n^{2}\left(\frac{A}{2}-\frac{C^{2}}{4B}\right)<0
\end{array}\label{eq:U4-final}
\end{equation}

\subsubsection{Instability of homogeneous asymmetric phase.}

We have found that the homogeneous phase with the violated reflection
symmetry has negative interaction energy proportional to $n^{2}$.
It means that the interaction energy decreases when the volume occupied
by the condensate decreases. In the weakly non-ideal attractive Bose-gas
of $N$ particles with the coupling constant $g<0$ and mass $m$
of particle this tendency leads to the mechanical instability of the
gas and its collapse at a critical value of number of particles $N_{c}$
. At this value the isothermal compressibility $\kappa{}_{T}=-\frac{1}{V}\left(\frac{\partial V}{\partial P}\right)_{T}$
is zero and at $N>N_{c}$ becomes negative. Due to quantum uncertainty,
the kinetic energy per particle can be written as $K/N=\frac{\hbar^{2}}{2mV^{2/3}}$,
whereas the interaction energy is $U=$$\frac{gN^{2}}{2V}$. Thus,
the total energy is
\begin{equation}
E\left(N,V\right)=N\frac{\hbar^{2}}{2mV^{2/3}}+\frac{gN^{2}}{2V}.\label{eq:3d-Bose-collapse}
\end{equation}
The pressure is
\begin{equation}
P=-\frac{\partial E}{\partial V}=\frac{\hbar^{2}N}{3mV^{5/3}}+\frac{gN^{2}}{2V^{2}}\label{eq:3d-pressure}
\end{equation}
and the compressibility is 
\begin{equation}
\kappa_{T}=\frac{5\hbar^{2}N}{9mV^{5/3}}+\frac{gN^{2}}{V^{2}}\label{eq:3d-compressibility}
\end{equation}
Equation $\kappa_{T}=0$ determines the critical number of particles
$N_{c}=-\frac{5\hbar^{2}V^{1/3}}{9mg}$. At $N>N_{c}$, the compressibility
is negative and the gas becomes mechanically unstable. It starts to
contract. Since this process proceeds simultaneously in the total
volume occupied by the gas, the process will stop when the volume
wil be divided into $N/N_{c}$ cells each containing $N_{c}$ particles
and isolated each from other. The volume of such a cell is $v=VN_{c}/N$, therefore
the critical number in a cell is different than the critical number
in the entire volume. It should be found from equation $N_{c}=\frac{5\hbar^{2}}{9m\left|g\right|}\left(\frac{VN_{c}}{N}\right)^{1/3}$.
It is convenient to express the coupling constant $g$ in terms of
the Born scattering length $a_{s}$ as $g=\frac{\hbar^{2}}{m}a_{s}$.
Then $N_{c}=\frac{5^{3/2}}{27}\frac{1}{n^{1/6}\left|a_{s}\right|^{1/2}}$,
where $n=N/V$ is the average density of particles. For a weakly interacting
Bose gas $n^{1/3}\left|a_{s}\right|\ll1$. Therefore $N_{c}\gg1$.
The collapse was observed in cooled gases of alkali atoms $^{7}$Li
\cite{Bradley 1997} and $^{85}$Rb \cite{Roberts 2001}. At finite temperature
the pressure from excitations must be included. It changes the critical
values for starting the collapse, but the collapse persists. Gases
of cooled attracting alkali atoms after the collapse flew out the
magnetic or laser trap. Our calculations relate to the Bose gas of
quasiparticles that cannot avoid the system in which they exist like
excitons in semiconductors or spin waves in magnets.

Theoretical predictions for starting parameters of collapse in weakly
attracting Bose gas at finite temperature were made by Mueller and
Baym.\cite{Mueller 2000} Dynamic approach to the same problem was developed
by Pitaevskii. \cite{Pitaevskii 1996}

For the magnon condensation in a ferromagnetic film, the
problem is effectively two-dimensional. It is because the minimum
of energy corresponds to the transverse standing wave, 
period of which fits between surfaces of the film. The effective masses
are strongly anisotropic (see subsection \ref{subsec:Wave-vectors-and}).
The curve of constant kinetic energy is the ellipsis $\frac{\hbar^{2}k_{y}^{2}}{2m_{y}}+\frac{\hbar^{2}k_{z}^{2}}{2m_{z}}=K$.
Therefore we expect that the collapsed magnon condensate will be limited
by an ellipsis with semi-axes $R_{y},R_{z}$ whose ratio is $R_{y}/R_{z}=\sqrt{m_{z}/m_{y}}.$
Then the kinetic energy of collapsed condensate can be estimated as
\begin{equation}
K=N\left(\frac{\hbar^{2}}{2m_{y}R_{y}^{2}}+\frac{\hbar^{2}}{2m_{z}R_{z}^{2}}\right)=\frac{N\hbar^{2}}{\sqrt{m_{y}m_{z}}R_{y}R_{z}},\label{eq:collapse-2d-kinetic}
\end{equation}
whereas the condensate potential energy is 
\begin{equation}
U=\frac{gN^{2}}{2\pi R_{y}R_{z}d}.\label{eq:collapse 2d-potential}
\end{equation}
The pressure at zero temperature is 
\begin{equation}
P=\frac{N}{V^{2}}\left(\frac{\pi\hbar^{2}d}{\sqrt{m_{y}m_{z}}}+\frac{g}{2}N\right),\label{eq:pressure-2d}
\end{equation}
where $V=\pi R_{y}R_{z}d$ is the volume of the condensate cloud.
The compressibility of the magnon gas in the film differs from the
pressure only by numerical factor 2. Thus, the pressure and compressibility
simultaneously become zero when $N$ reaches a critical value 
\begin{equation}
N_{c}=-\frac{2\pi\hbar^{2}d}{\sqrt{m_{y}m_{z}}g}=\frac{2\pi d}{\left|a_{s}\right|}.\label{eq:N-crit-2d}
\end{equation}
The film will be divided into $N/N_{c}$ almost isolated cells each
containing $N_{c}$ magnons. Let the cell be a rectangle with the
sides $R_{y},R_{z}$. If $A$ is the area of the sample, then the
area of a cell is $R_{y}R_{z}=\frac{AN_{c}}{N}$ . From this equation
and requirement $R_{y}/R_{z}=\sqrt{m_{z}/m_{y}}$ we find $R_{y}=\left(\frac{m_{z}}{m_{y}}\right)^{1/4}\sqrt{\frac{AN_{c}}{N}};R_{z}=\left(\frac{m_{y}}{m_{z}}\right)^{1/4}\sqrt{\frac{AN_{c}}{N}}$.
According to eq. (\ref{eq:N-crit-2d}), this result can be rewritten
as
\begin{equation}
R_{y}=\left(\frac{m_{z}}{m_{y}}\right)\sqrt{\frac{2\pi}{n\left|a_{s}\right|}};R_{z}=\left(\frac{m_{y}}{m_{z}}\right)^{1/4}\sqrt{\frac{2\pi}{n\left|a_{s}\right|}},\label{eq:cell-sizies-2d}
\end{equation}
where $n=N/(Ad)$ is the average density of magnons. The collapse
destroys the homogeneous coherent condensate transforming it into
a set of isolated islands.

For YIG with $n=10^{18}cm^{-3}$ we find $N_{c}\approx1.14\times10^{5}$,
$R_{y}\approx1.16\times10^{-6}cm;R_{z}\approx0.58\times10^{-7}cm$.
There is no experimental evidence of the cell structure in YIG films.
\subsection{New experiments and our new theoretical ideas about slow inter-minima relaxation and laser effects.}

Two recent articles by the Münster University
experimental team led by S.O. Demokritov \cite{Borisenko 2020-1, Borisenko 2020-2}
revealed several important facts about the Bose-Einstein condensation of magnons (BECM) under permanent pumping first discovered  in 2006 \cite{Demokritov 2006}.  
Existing theories of this phenomenon predict an attractive interaction between magnons \cite{Tupitsyn 2008, Rezende 2009, Li 2013} and a strong spontaneous violation of the reflection symmetry \cite{Li 2013}.  However these theories implicitly assumed that all relaxation processes were fast compared to the lifetime of magnons, whereas one of them, the relaxation between two energy minima, is slow.

We predict the properties of the stationary state of the magnon gas with condensate, that is  far from equilibrium with respect to variables responsible for inter-minima coherence.
The momentum-flip relaxation time is no less than 1 hour, which exceeds even the time of the experiment without considering the lifetime. It means that the equilibrium between condensates in different minima is never reached. As a result, the condensates' stationary state is far from equilibrium. In this regard, it is analogous to the laser stationary state, and, like the laser, the magnon condensate state can produce coherent magnon radiation \cite{Demokritov, Hillebrands}.

The very slow inter-minima relaxation implies that the appearance of the stationary condensate in a ferromagnetic film is a dynamic phase transition. Since the inter-minima equilibrium is not established, the pumping, which is symmetric with respect to the two minima, creates equal numbers of magnons in the two condensates $n_+=n_-=n_c/2$.  Therefore, the inter-minima repulsion energy $Bn_+n_-=Bn_c^2/4$ strongly exceeds the magnitude of in-minimum attraction $\frac{\vert A\vert}{2}(n_{+}^2+n_{-}^2)= \vert A\vert n_c^2 /4$. This consideration explains why experimenters observe repulsion of magnons in the stationary state with the condensate. 
With a somewhat more sophisticated point of view, the mirror symmetry of the pumping does not necessarily lead to the same symmetry of the condensate. In principle, dynamic violation of mirror symmetry is possible. But in this case, there is no reason why it should be strong. This issue requires further theoretical investigation. 

It is difficult to avoid a slight asymmetry of the device in real-world experiments, which favors a slightly asymmetric stationary state. Such a device asymmetry could explain
the asymmetry observed in the experiment II by Borisenko et al. If
the asymmetry is relatively small, then in eq. (\ref{eq:U4-minimization})
the term $Bn_{+}n_{-}$ is dominant and positive, but it completely conceals the possibility of dynamic spontaneous violation of the reflection symmetry.

Because the inter-minima equilibrium is not established, the consistent theory of the stationary state with condensate necessitates solving the Boltzmann kinetic equation for magnons and the Gross-Pitaevskii equations for the two condensates. It can be accomplished using either a variational technique based on the idea of maximal entropy production or by solving a problem with proper initial conditions that asymptotically approaches a stationary state.

This kinetic approach may help to bridge another gap between the existing theories \cite{Bunkov 2008, Sun 2017} and experiment \cite{Demokritov, Hillebrands}.  The theory proofs that the pumped magnons are accumulated in the low-energy region assuming the temperature of accumulated magnons to be the same as initial temperature of the system (room temperature). 
The temperature of low energy magnons is approximately three times higher, according to experimental data. If the temperature is a slow-varying function of energy (momentum) that saturates to the system's room temperature at some intermediate energy between $\mu_B H$  and the room temperature, the controversy may be resolved.

\section*{Acknowledgments.}
  We are thankful to T. Nattermann, W.M. Saslow, Fuxiang Li, Chen Sun together with whom were obtained many results mentioned in this article. Our gratefulness is due to S.O. Demokritov and participants of his experimental team V. E. Demidov, I. Borisenko, B. Divinskii, P. Novik-Boltyk for many useful discussions of the experimental results and cooperation. 
We thank J. Ketterson and J. Lim for explanation of their experiment and discussion of its results. We are indebted to B. Hillebrands, A. Serga and D. Bozhko for discussion of their experiments.
Many theoretical problems were discussed with A.N. Slavin, V.S. L'vov and G. E. Volovik, who also informed us on a vast literature on the subjecr. Our thanks to them. We remember thankfully the discussion with deceased  L.P. Pitaevskii on the instability of attractive Bose condensate.

\section*{Appendix 1. Motion of minima.}
The dependence of frequency on wave vector is determined by equation
(\ref{eq:energy-square-qz-qx}) of the main text. For the reader's convenience we reproduce it:
\begin{equation}
\omega^{2}=\mu_B^2\mathcal{H}^2\left(1+k^{2}\right)\left(1+k^{2}+\chi-\chi\frac{k_{z}^{2}}{k^{2}}\right)\label{eq:frequency-ap}
\end{equation}
Here $k^{2}=k_{\parallel}^{2}+k_{x}^{2}$, where $k_{x}$ is a positive
quantized transverse component of wave vector. Generally to find minimum
of frequency for a given mode with fixed quantum numbers and direction
of propagation, it is necessary to take in account the dependence
of quantized $k_{x}$ on $k_{\Vert}$. This dependence can be neglected
in thick films with $d\gg1$. Indeed according to the main
text, quantized values of $k_{x}$ are equal to $k_{x,\nu ,n}=\frac{2\pi n}{d}+\mu_{\nu, n}$.
Here $\mu_{\nu, n}=\frac{2}{d}\arctan f_{\nu, n}\left(k_{\Vert}\right)$,
where $f_{\nu, n}\left(k_{\Vert}\right)$ is a smooth function. According
to this definition, $\mu_{\nu, n}$ varies in the limits $\left(-\frac{\pi}{d},\frac{\pi}{d}\right)$
when $k_{\Vert}$ changes at least by $1/\sqrt{d}$. Therefore, the
derivative $\frac{dk_{x}}{dk_{\Vert}}\lesssim\frac{1}{\sqrt{d}}\ll1$
and the values $k_{\Vert}$ and $k_{x}$ can be considered as independent.
In this approximation the value of parallel wave vector $k_{\Vert0}$
at which frequency has minimum can be found from equation:
\begin{equation}
\frac{\partial\omega^{2}}{\partial\left(k_{\Vert}^{2}\right)}=2k^{2}+2+\chi\sin^{2}\theta-\frac{\chi k_{x}^{2}\cos^{2}\theta}{k^{4}}=0\label{eq:min-position}
\end{equation}
At small $k_{x}$ i.e. at $n\ll d/2\pi$, the value $k^{2}$ satisfying
eq. (\ref{eq:min-position}) is also small and equal to
\begin{equation}
k_{0}^{2}\approx k_{\Vert0}^{2}\approx\sqrt{\frac{\chi}{2+\chi\sin^{2}\theta}}k_{x}\cos\theta 
\label{eq:k-square-appr}
\end{equation}
It is however much larger than $k_{x}^{2}$. The value of frequency
in minimum is $\omega_{min}\approx\sqrt{1+\chi\sin^{2}\theta}$. The
equation for $k_{0}^{2}$ valid in the range of larger $k_{x}$ comparable
with 1 can be found by the following scaling transformation: 
\begin{equation}
k_{0}^{2}=\frac{2+\chi\sin^{2}\theta}{2}w\left(\xi\right);\:\xi=\frac{4\chi k_{x}^{2}\cos^{2}\theta}{\left(2+\chi\sin^{2}\theta\right)^{3}},\label{eq:k-square-xi}
\end{equation}
where function $w\left(\xi\right)$ obeys cubic equation:
\begin{equation}
w^{3}+w^{2}=\xi\label{eq:w-xi}
\end{equation}
At small $\xi$, this equation gives the result (\ref{eq:k-square-appr}).
This equation shows that at small $k_{x}$, the wave vector corresponding
to minimal frequency $k_{\Vert0}$ grows with $k_{x}$. To study the
motion of minimum in a broader interval of $k_{x}$ it is useful to
look at the derivative $\frac{dk_{\Vert0}^{2}}{d(k_{x}^{2})}$. According
to eq. (\ref{eq:min-position}), it can be expressed as follows:
\begin{equation}
\frac{dk_{\Vert0}^{2}}{d(k_{x}^{2})}=-\frac{\frac{\partial^{2}\omega^{2}}{\partial\left(k_{\Vert}^{2}\right)\partial\left(k_{x}^{2}\right)}}{\frac{\partial^{2}\omega^{2}}{\left(\partial\left(k_{\Vert}^{2}\right)\right)^{2}}}\label{eq:derivative}
\end{equation}
From this equation it follows that maximal value of $k_{\Vert0}$
can be found from equation:
\begin{equation}
\frac{\partial^{2}\omega^{2}}{\partial\left(k_{\Vert}^{2}\right)\partial\left(k_{x}^{2}\right)}=2-\chi\frac{\cos^{2}\theta}{k^{4}}+2\chi\frac{k_{x}^{2}\cos^{2}\theta}{k^{6}}=0\label{eq:maximum-cond}
\end{equation}
It is cubic equation for $k^{2}$. It must be solved together with
equation of frequency minimum (\ref{eq:min-position}). Eliminating
$k_{x}^{2}$ from these two equations, we arrive at a closed equation
for $k^{2}$:
\begin{equation}
6k^{6}+2\left(2+\chi\sin^{2}\theta\right)k^{4}-\chi\cos^{2}\theta k^{2}=0\label{eq:k-square-eq}
\end{equation}
Dividing this equation by $k^{2}\neq0$, we obtain a quadratic equation
for $k^{2}$, whose solution reads:
\begin{equation}
k_{m}^{2}=\frac{\sqrt{\left(2+\chi\sin^{2}\theta\right)^{2}+6\chi\cos^{2}\theta}-\left(2+\chi\sin^{2}\theta\right)}{6}\label{eq:k-square-max}
\end{equation}
The value of $k_{x}^{2}$ corresponding to maximal value of $k_{\Vert0}$
can be found by eliminating $k^{6}$ from eqs. (\ref{eq:min-position},\ref{eq:k-square-eq}).
It reads:
\begin{equation}
\left(k_{x}^{2}\right)_{m}=\frac{1}{3\chi\cos^{2}\theta}\left[\left(2+\chi\sin^{2}\theta\right)k_{m}^{4}+\chi\cos^{2}\theta k_{m}^{2}\right]\label{eq:k-x-max}
\end{equation}
The maximal value of $k_{\Vert0}^{2}$ is equal to
\[
\left(k_{\Vert0}^{2}\right)_{\max}=k_{m}^{2}-\left(k_{x}^{2}\right)_{m}=\frac{2}{3}k_{m}^{2}-\frac{\left(2+\chi\sin^{2}\theta\right)k_{m}^{4}}{3\chi\cos^{2}\theta}
\]
At further increase of $k_{x}$, the position of minimum $k_{\Vert0}$
decreases and finally becomes zero. At this point, $k^{2}=k_{x}^{2}$
and eq. (\ref{eq:min-position}) turns into quadratic equation for
$k_{x}^{2}$. Its solution reads: 
\[
\left(k_{x}^{2}\right)_{f}=\frac{\sqrt{\left(2+\chi\sin^{2}\theta\right)^{2}+8\chi\cos^{2}\theta}-\left(2+\chi\sin^{2}\theta\right)}{4}
\]
At this value of $k_{x}$, minimum merges with a local maximum at
$ $$k_{\Vert}=0$. At larger values of $k_{x}$, the only minimum
of frequency is at $k_{\Vert}=0$.

\section*{Appendix 2. Hamiltonian of the 4-th order.} \label{App: I-4}.
According to the subsection 4.2.2, the 4-th order Hamiltonian is:
\begin{equation}
H_{4}=\sum_{i,k,l=1}^{4}\sum_{\mathbf{q}_{i}n_{k}\rho_{l}}I_{4\mathbf{q}_{1}n_{1},\mathbf{q}_{2}n_{2},\mathbf{q}_{3}n_{3},\mathbf{q}_{4}n_{4}}^{\left(\rho_{1}\rho_{2}\rho_{3}\rho_{4}\right)}\left[\prod_{j=1}^{4}\eta_{\mathbf{q}_{j}n_{j}}^{\left(\rho_{j}\right)}\right]\delta_{\mathbf{q}_{1}+\mathbf{q}_{2}+\mathbf{q}_{3}+\mathbf{q}_{4}},\label{eq:H-4-in}
\end{equation}
where $\eta_{\mathbf{q}n}^{\left(+\right)}=\eta_{\mathbf{q}n},\eta_{\mathbf{q}n}^{\left(-\right)}=\eta_{\mathbf{-q}n}^{*}$
and upper indices $\rho_{l}\left(l=1,2,3,4\right)$ take values $+,-$
independently each from others. 
In terms of complex indices $\gamma_{i}=\left(\rho_{i}\mathbf{q}_{i}n_{i}\right)$
the Hamiltonian $H_{4}$ can be rewritten as
\begin{equation}
H_{4}=\sum_{\gamma_{i}}I_{\gamma_{1}\gamma_{2}\gamma_{3}\gamma_{4}}\eta_{\gamma_{1}}\eta_{\gamma_{2}}\eta_{\gamma_{3}}\eta_{\gamma_{4}}\delta_{\mathbf{q}_{1}+\mathbf{q}_{2}+\mathbf{q}_{3}+\mathbf{q}_{4}}\label{eq:H-4-gamma}
\end{equation}
Since the product of four $\eta_{j}$ is symmetric at any permutation
$P$ of four $j$, it is possible to replace the initial coefficients
$I_{\gamma_{1}\gamma_{2}\gamma_{3}\gamma_{4}}$ by the symmetrized
coefficients
\begin{equation}
I_{\gamma_{1}\gamma_{2}\gamma_{3}\gamma_{4}}^{s}=\frac{1}{24}\sum_{P}I_{\gamma_{P1}\gamma_{P2}\gamma_{P3}\gamma_{P4}},\label{eq:I-symmetrized}
\end{equation}
where $Pj$ means the number appearing on $j-$th place at permutation
$P.$ For example, for the permutation $1,2,3,4\rightarrow4,3,2,1$
one finds $P1=4$, $P2=3$, $P3=2$, $P4=1$.

Let us now analyze what are constraints for symmetrized coefficients
following from the fact that the energy is real. To make notations
more compact further we omit the subscript 4 and round brackets in
upper part of initial coefficients. Then eq. (\ref{eq:H-4-gamma})
turns into
\begin{equation}
H_{4}=\sum_{\mathbf{q}_{k}n_{l}\rho_{m}}I_{\mathbf{q}_{1}n_{1}\mathbf{q}_{2}n_{2}\mathbf{q}_{3}n_{3},\mathbf{q}_{4}n_{4}}^{s\rho_{1}\rho_{2}\rho_{3}\rho_{4}}\prod_{j=1}^{4}\eta_{\mathbf{q}_{j}n_{j}}^{\rho_{j}}.\label{eq:H-4-sym-det}
\end{equation}
Since $\eta_{\mathbf{q}_{j}n_{j}}^{-}=\left(\eta_{\mathbf{-q}_{j}n_{j}}^{+}\right)^{*},$
the energy is real if the following relations are satisfied:
\begin{equation}
I_{\mathbf{q}_{1}n_{1}\mathbf{q}_{2}n_{2}\mathbf{q}_{3}n_{3},\mathbf{q}_{4}n_{4}}^{s-\rho_{1}-\rho_{2}-\rho_{3}-\rho_{4}}=\left(I_{\mathbf{-q}_{1}n_{1}\mathbf{-q}_{2}n_{2}\mathbf{-q}_{3}n_{3},\mathbf{-q}_{4}n_{4}}^{s\rho_{1}\rho_{2}\rho_{3}\rho_{4}}\right)^{*}\label{eq:reality-general}
\end{equation}

However, the initial non-symmetrized coefficients $I_{\mathbf{q}_{1}n_{1}\mathbf{q}_{2}n_{2}\mathbf{q}_{3}n_{3},\mathbf{q}_{4}n_{4}}^{\rho_{1}\rho_{2}\rho_{3}\rho_{4}}$
calculated according to the rules formulated in the subsection 4.2.2.
do not obey these relationships. Nevertheless, not all of them are
independent. In this Appendix we derive the integral presentation
for independent coefficients and find relations that allow to find
the rest of them.

At fixed values $\mathbf{q}_{i},n_{i};i=1,2,3,4$, there are $2^{4}=16$
different combinations of $\rho_{j}=\pm$ that defines coefficients
$I_{\mathbf{q}_{1}n_{1}\mathbf{q}_{2}n_{2}\mathbf{q}_{3}n_{3},\mathbf{q}_{4}n_{4}}^{\rho_{1}\rho_{2}\rho_{3}\rho_{4}}$.
Each of them contains contributions from exchange $I_{e}$ and dipolar
$I_{d}$ interactions, in total 32 coefficients. In each of them $\rho_{j}$
take the same value $+$ or $-$ more than once. It allows to make
the partial symmetrization over repeating indices. For a further compactification
of notations we denote the pair $j\equiv\mathbf{q}_{j}n_{j}$ and
$\overline{j}=-\mathbf{q}_{j}n_{j}$; $j=1,2,3,4$. Then the resulting
relationships for exchange coefficients are:
\begin{equation}
I_{e1234}^{----}=\left(I_{e\overline{4}\overline{3}\overline{2}\overline{1}}^{++++}\right)^{*}\label{eq:e-0}
\end{equation}

\begin{equation}
I_{e1234}^{---+}=\left(I_{e\overline{2}\overline{1}\overline{4}\overline{3}}^{++-+}\right)^{*}=\left(I_{e\overline{4}\overline{3}\overline{2}\overline{1}}^{-+++}\right)^{*}\label{eq:e-1}
\end{equation}
\begin{equation}
I_{e1234}^{+---}=\left(I_{e\overline{2}\overline{1}\overline{4}\overline{3}}^{+-++}\right)^{*}=\left(I_{e\overline{4}\overline{3}\overline{2}\overline{1}}^{+++-}\right)^{*}\label{eq:e-2}
\end{equation}
\begin{equation}
I_{e1234}^{--+-}=\left(I_{e\overline{4}\overline{3}\overline{2}\overline{1}}^{+-++}\right)^{*}=\left(I_{e\overline{2}\overline{1}\overline{4}\overline{3}}^{+++-}\right)^{*}\label{eq:e-3}
\end{equation}
\begin{equation}
I_{e1234}^{-+--}=\left(I_{e\overline{4}\overline{3}\overline{2}\overline{1}}^{++-+}\right)^{*}=\left(I_{e\overline{2}\overline{1}\overline{4}\overline{3}}^{-+++}\right)^{*}\label{eq:e-4}
\end{equation}
\begin{equation}
I_{e1234}^{--++}=I_{e3412}^{++--}=I_{e3214}^{+--+}=I_{e1432}^{-++-}\label{eq:e-2-2}
\end{equation}
Altogether there are 10 equations for 16 exchange coefficients. Thus,
only 6 of them are independent. This 6 coefficients can be chosen
as:
\begin{equation}
\begin{array}{c}
I_{e1234}^{++++}=\frac{\mu_{B}^{2}\ell^{2}}{4A}\intop_{-d/2}^{d/2}\left[\left(d_{x}u_{1}^{*}\right)v_{2}^{*}\left(d_{x}u_{3}^{*}\right)v_{4}^{*}+u_{1}^{*}\left(d_{x}v_{2}^{*}\right)u_{3}^{*}\left(d_{x}v_{4}^{*}\right)\right.\\
\left.-\frac{1}{2}\left(\sum_{j=1}^{4}\mathbf{q}_{j}^{2}\right)u_{1}^{*}v_{2}^{*}u_{3}^{*}v_{4}^{*}\right]dx;
\end{array}\label{eq:++++}
\end{equation}
\begin{equation}
\begin{array}{c}
I_{e1234}^{+++-}=-\frac{\mu_{B}^{2}\ell^{2}}{4A}\intop_{-d/2}^{d/2}\left[\left(d_{x}u_{1}^{*}\right)v_{2}^{*}\left(d_{x}u_{3}^{*}\right)u_{\overline{4}}+u_{1}^{*}\left(d_{x}v_{2}^{*}\right)u_{3}^{*}\left(d_{x}u_{\overline{4}}\right)\right.\\
\left.-\frac{1}{2}\left(\sum_{j=1}^{4}\mathbf{q}_{j}^{2}\right)u_{1}^{*}v_{2}^{*}u_{3}^{*}u_{\overline{4}}\right]dx;
\end{array}\label{eq:+++-}
\end{equation}
\begin{equation}
\begin{array}{c}
I_{e1234}^{-+++}=-\frac{\mu_{B}^{2}\ell^{2}}{4A}\intop_{-d/2}^{d/2}\left[\left(d_{x}v_{\overline{1}}\right)v_{2}^{*}\left(d_{x}u_{3}^{*}\right)v_{4}^{*}+v_{\overline{1}}\left(d_{x}v_{2}^{*}\right)u_{3}^{*}\left(d_{x}v_{4}^{*}\right)\right.\\
\left.-\frac{1}{2}\left(\sum_{j=1}^{4}\mathbf{q}_{j}^{2}\right)v_{\overline{1}}v_{2}^{*}u_{3}^{*}v_{4}^{*}\right]dx;
\end{array}\label{eq:-+++}
\end{equation}
\begin{equation}
\begin{array}{c}
I_{e1234}^{++--}=\frac{\mu_{B}^{2}\ell^{2}}{4A}\intop_{-d/2}^{d/2}\left[\left(d_{x}u_{1}^{*}\right)v_{2}^{*}\left(d_{x}v_{\overline{3}}\right)u_{\overline{4}}+u_{1}^{*}\left(d_{x}v_{2}^{*}\right)v_{\overline{3}}\left(d_{x}u_{\overline{4}}\right)\right.\\
\left.-\frac{1}{2}\left(\sum_{j=1}^{4}\mathbf{q}_{j}^{2}\right)u_{1}^{*}v_{2}^{*}v_{\overline{3}}u_{\overline{4}}\right]dx;
\end{array}\label{eq:++--}
\end{equation}
\begin{equation}
\begin{array}{c}
I_{e1234}^{+-+-}=\frac{\mu_{B}^{2}\ell^{2}}{4A}\intop_{-d/2}^{d/2}\left[\left(d_{x}u_{1}^{*}\right)u_{\overline{2}}\left(d_{x}u_{3}^{*}\right)u_{\overline{4}}+u_{1}^{*}\left(d_{x}u_{\overline{2}}\right)u_{3}^{*}\left(d_{x}u_{\overline{4}}\right)\right.\\
\left.-\frac{1}{2}\left(\sum_{j=1}^{4}\mathbf{q}_{j}^{2}\right)u_{1}^{*}u_{\overline{2}}u_{3}^{*}u_{\overline{4}}\right]dx;
\end{array}\label{eq:+-+-}
\end{equation}
\begin{equation}
\begin{array}{c}
I_{e1234}^{-+-+}=\frac{\mu_{B}^{2}\ell^{2}}{4A}\intop_{-d/2}^{d/2}\left[\left(d_{x}v_{\overline{1}}\right)v_{2}^{*}\left(d_{x}v_{\overline{3}}\right)v_{4}^{*}+v_{\overline{1}}\left(d_{x}v_{2}^{*}\right)v_{\overline{3}}\left(d_{x}v_{4}^{*}\right)\right.\\
\left.-\frac{1}{2}\left(\sum_{j=1}^{4}\mathbf{q}_{j}^{2}\right)v_{\overline{1}}v_{2}^{*}v_{\overline{3}}v_{4}^{*}\right]dx.
\end{array}\label{eq:-+-+}
\end{equation}
There only 8 relationships for the dipolar part of 4-th order Hamiltonian:
\begin{equation}
I_{d1,2,3,4}^{----}=\left(I_{d\overline{2},\overline{1},\overline{3},\overline{4}}^{++++}\right)^{*}\label{eq:r-ship-d}
\end{equation}
\begin{equation}
I_{d1,2,3,4}^{-+--}=\left(I_{d\overline{2},\overline{1},\overline{3},\overline{4}}^{-+++}\right)^{*}\label{eq:r-ship-d}
\end{equation}
\begin{equation}
I_{d1,2,3,4}^{+---}=\left(I_{d\overline{2},\overline{1},\overline{3},\overline{4}}^{+-++}\right)^{*}\label{eq:r-ship-d}
\end{equation}
\begin{equation}
I_{d1,2,3,4}^{--+-}=\left(I_{d\overline{2},\overline{1},\overline{3},\overline{4}}^{++-+}\right)^{*}\label{eq:r-ship-d}
\end{equation}
\begin{equation}
I_{d1,2,3,4}^{---+}=\left(I_{d\overline{2},\overline{1},\overline{3},\overline{4}}^{+++-}\right)^{*}\label{eq:r-ship-d}
\end{equation}
\begin{equation}
I_{d1,2,3,4}^{--++}=\left(I_{d\overline{2},\overline{1},\overline{3},\overline{4}}^{++--}\right)^{*}\label{eq:r-ship-d}
\end{equation}
\begin{equation}
I_{d1,2,3,4}^{+--+}=\left(I_{d\overline{2},\overline{1},\overline{3},\overline{4}}^{+-+-}\right)^{*}\label{eq:r-ship-d}
\end{equation}
\begin{equation}
I_{d1,2,3,4}^{-++-}=\left(I_{d\overline{2},\overline{1},\overline{3},\overline{4}}^{-+-+}\right)^{*}\label{eq:r-ship-d}
\end{equation}
Thus,
only 8 of them are independent. This 8 coefficients can be chosen
as:
\begin{equation}
\begin{array}{c}
I_{d1,2,3,4}^{++++}=\frac{\pi\mu_{B}^{2}}{2A}\iint dxdx^{\prime}\times\\
\left[\left(q_{1z}+q_{2z}\right)^{2}\left(u_{1}^{*}v_{2}^{*}+v_{1}^{*}u_{2}^{*}\right)\left(u_{3}^{\prime*}v_{4}^{\prime*}+v_{3}^{\prime*}u_{4}^{\prime*}\right)G_{\left|\mathbf{q}_{1}+\mathbf{q}_{2}\right|}\left(x-x^{\prime}\right)\right.\\
\left.-u_{1}^{*}v_{2}^{*}u_{3}^{*}u_{4}^{\prime*}\left(d_{x}+q_{4y}\right)^{2}G_{\left|\mathbf{q}_{4}\right|}\left(x-x^{\prime}\right)\right.\\
\left.+u_{1}^{*}v_{2}^{*}u_{3}^{*}v_{4}^{\prime*}\left(d_{x}^{2}-q_{4y}^{2}\right)G_{\left|\mathbf{q}_{4}\right|}\left(x-x^{\prime}\right)\right.\\
\left.+u_{1}^{*}v_{2}^{*}v_{3}^{*}u_{4}^{\prime*}\left(d_{x}^{2}-q_{4y}^{2}\right)G_{\mathbf{q}_{4}}\left(x-x^{\prime}\right)\right.\\
\left.-u_{1}^{*}v_{2}^{*}v_{3}^{*}v_{4}^{\prime*}\left(d_{x}-q_{4y}\right)^{2}G_{\mathbf{q}_{4}}\left(x-x^{\prime}\right)\right]
\end{array}
\end{equation}

\begin{equation}
\begin{array}{c}
I_{d1,2,3,4}^{-+++}=-\frac{\pi\mu_{B}^{2}}{2A}\iint dxdx^{\prime}\times\\
\left[\left(q_{1z}+q_{2z}\right)^{2}\left(v_{\overline{1}}v_{2}^{*}+v_{1}^{*}v_{\overline{2}}\right)\left(u_{3}^{\prime*}v_{4}^{\prime*}+v_{3}^{\prime*}u_{4}^{\prime*}\right)G_{\left|\mathbf{q}_{1}+\mathbf{q}_{2}\right|}\left(x-x^{\prime}\right)\right.\\
\left.-v_{\overline{1}}v_{2}^{*}u_{3}^{*}u_{4}^{\prime*}\left(d_{x}+q_{4y}\right)^{2}G_{\left|\mathbf{q}_{4}\right|}\left(x-x^{\prime}\right)\right.\\
\left.+v_{\overline{1}}v_{2}^{*}u_{3}^{*}v_{4}^{\prime*}\left(d_{x}^{2}-q_{4y}^{2}\right)G_{\left|\mathbf{q}_{4}\right|}\left(x-x^{\prime}\right)\right.\\
\left.+v_{\overline{1}}v_{2}^{*}v_{3}^{*}u_{4}^{\prime*}\left(d_{x}^{2}-q_{4y}^{2}\right)G_{\mathbf{q}_{4}}\left(x-x^{\prime}\right)\right.\\
\left.-v_{\overline{1}}v_{2}^{*}v_{3}^{*}v_{4}^{\prime*}\left(d_{x}-q_{4y}\right)^{2}G_{\mathbf{q}_{4}}\left(x-x^{\prime}\right)\right]
\end{array}
\end{equation}

\begin{equation}
\begin{array}{c}
I_{d1,2,3,4}^{+-++}=-\frac{\pi\mu_{B}^{2}}{2A}\iint dxdx^{\prime}\times\\
\left[\left(q_{1z}+q_{2z}\right)^{2}\left(u_{1}^{*}u_{\overline{2}}+v_{\overline{1}}u_{2}^{*}\right)\left(u_{3}^{\prime*}v_{4}^{\prime*}+v_{3}^{\prime*}u_{4}^{\prime*}\right)G_{\left|\mathbf{q}_{1}+\mathbf{q}_{2}\right|}\left(x-x^{\prime}\right)\right.\\
\left.-u_{1}^{*}u_{\overline{2}}u_{3}^{*}u_{4}^{\prime*}\left(d_{x}+q_{4y}\right)^{2}G_{\left|\mathbf{q}_{4}\right|}\left(x-x^{\prime}\right)\right.\\
\left.+u_{1}^{*}u_{\overline{2}}u_{3}^{*}v_{4}^{\prime*}\left(d_{x}^{2}-q_{4y}^{2}\right)G_{\left|\mathbf{q}_{4}\right|}\left(x-x^{\prime}\right)\right.\\
\left.+u_{1}^{*}u_{\overline{2}}v_{3}^{*}u_{4}^{\prime*}\left(d_{x}^{2}-q_{4y}^{2}\right)G_{\mathbf{q}_{4}}\left(x-x^{\prime}\right)\right.\\
\left.-u_{1}^{*}u_{\overline{2}}v_{3}^{*}v_{4}^{\prime*}\left(d_{x}-q_{4y}\right)^{2}G_{\mathbf{q}_{4}}\left(x-x^{\prime}\right)\right]
\end{array}
\end{equation}

\begin{equation}
\begin{array}{c}
I_{d1,2,3,4}^{++-+}=-\frac{\pi\mu_{B}^{2}}{2A}\iint dxdx^{\prime}\times\\
\left[\left(q_{1z}+q_{2z}\right)^{2}\left(u_{1}^{*}v_{2}^{*}+v_{1}^{*}u_{2}^{*}\right)\left(v_{\overline{3}}^{\prime}v_{4}^{\prime*}+v_{3}^{\prime*}v_{\overline{4}}^{\prime}\right)G_{\left|\mathbf{q}_{1}+\mathbf{q}_{2}\right|}\left(x-x^{\prime}\right)\right.\\
\left.-u_{1}^{*}v_{2}^{*}v_{\overline{3}}u_{4}^{\prime*}\left(d_{x}+q_{4y}\right)^{2}G_{\left|\mathbf{q}_{4}\right|}\left(x-x^{\prime}\right)\right.\\
\left.+u_{1}^{*}v_{2}^{*}v_{\overline{3}}v_{4}^{\prime*}\left(d_{x}^{2}-q_{4y}^{2}\right)G_{\left|\mathbf{q}_{4}\right|}\left(x-x^{\prime}\right)\right.\\
\left.+u_{1}^{*}v_{2}^{*}u_{\overline{3}}u_{4}^{\prime*}\left(d_{x}^{2}-q_{4y}^{2}\right)G_{\mathbf{q}_{4}}\left(x-x^{\prime}\right)\right.\\
\left.-u_{1}^{*}v_{2}^{*}u_{\overline{3}}v_{4}^{\prime*}\left(d_{x}-q_{4y}\right)^{2}G_{\mathbf{q}_{4}}\left(x-x^{\prime}\right)\right]
\end{array}
\end{equation}

\begin{equation}
\begin{array}{c}
I_{d1,2,3,4}^{+++-}=-\frac{\pi\mu_{B}^{2}}{2A}\iint dxdx^{\prime}\times\\
\left[\left(q_{1z}+q_{2z}\right)^{2}\left(u_{1}^{*}v_{2}^{*}+v_{1}^{*}u_{2}^{*}\right)\left(u_{3}^{\prime*}u_{\overline{4}}^{\prime}+u_{\overline{3}}^{\prime}u_{4}^{\prime*}\right)G_{\left|\mathbf{q}_{1}+\mathbf{q}_{2}\right|}\left(x-x^{\prime}\right)\right.\\
\left.-u_{1}^{*}v_{2}^{*}u_{3}^{*}v_{\overline{4}}^{\prime}\left(d_{x}+q_{4y}\right)^{2}G_{\left|\mathbf{q}_{4}\right|}\left(x-x^{\prime}\right)\right.\\
\left.+u_{1}^{*}v_{2}^{*}u_{3}^{*}u_{\overline{4}}^{\prime}\left(d_{x}^{2}-q_{4y}^{2}\right)G_{\left|\mathbf{q}_{4}\right|}\left(x-x^{\prime}\right)\right.\\
\left.+u_{1}^{*}v_{2}^{*}v_{3}^{*}v_{\overline{4}}^{\prime}\left(d_{x}^{2}-q_{4y}^{2}\right)G_{\mathbf{q}_{4}}\left(x-x^{\prime}\right)\right.\\
\left.-u_{1}^{*}v_{2}^{*}v_{3}^{*}u_{\overline{4}}^{\prime}\left(d_{x}-q_{4y}\right)^{2}G_{\mathbf{q}_{4}}\left(x-x^{\prime}\right)\right]
\end{array}
\end{equation}

\begin{equation}
\begin{array}{c}
I_{d1,2,3,4}^{++--}=\frac{\pi\mu_{B}^{2}}{2A}\iint dxdx^{\prime}\times\\
\left[\left(q_{1z}+q_{2z}\right)^{2}\left(u_{1}^{*}v_{2}^{*}+v_{1}^{*}u_{2}^{*}\right)\left(v_{\overline{3}}^{\prime}u_{\overline{4}}^{\prime}+u_{\overline{3}}^{\prime}v_{\overline{4}}^{\prime}\right)G_{\left|\mathbf{q}_{1}+\mathbf{q}_{2}\right|}\left(x-x^{\prime}\right)\right.\\
\left.-u_{1}^{*}v_{2}^{*}v_{\overline{3}}v_{\overline{4}}^{\prime}\left(d_{x}+q_{4y}\right)^{2}G_{\left|\mathbf{q}_{4}\right|}\left(x-x^{\prime}\right)\right.\\
\left.+u_{1}^{*}v_{2}^{*}v_{\overline{3}}u_{\overline{4}}^{\prime}\left(d_{x}^{2}-q_{4y}^{2}\right)G_{\left|\mathbf{q}_{4}\right|}\left(x-x^{\prime}\right)\right.\\
\left.+u_{1}^{*}v_{2}^{*}u_{\overline{3}}v_{\overline{4}}^{\prime}\left(d_{x}^{2}-q_{4y}^{2}\right)G_{\mathbf{q}_{4}}\left(x-x^{\prime}\right)\right.\\
\left.-u_{1}^{*}v_{2}^{*}u_{\overline{3}}u_{\overline{4}}^{\prime}\left(d_{x}-q_{4y}\right)^{2}G_{\mathbf{q}_{4}}\left(x-x^{\prime}\right)\right]
\end{array}
\end{equation}

\begin{equation}
\begin{array}{c}
I_{d1,2,3,4}^{+-+-}=\frac{\pi\mu_{B}^{2}}{2A}\iint dxdx^{\prime}\times\\
\left[\left(q_{1z}+q_{2z}\right)^{2}\left(u_{1}^{*}u_{\overline{2}}+u_{\overline{1}}u_{2}^{*}\right)\left(u_{3}^{\prime*}u_{\overline{4}}^{\prime}+u_{\overline{3}}^{\prime}u_{4}^{\prime*}\right)G_{\left|\mathbf{q}_{1}+\mathbf{q}_{2}\right|}\left(x-x^{\prime}\right)\right.\\
\left.-u_{1}^{*}u_{\overline{2}}u_{3}^{*}v_{\overline{4}}^{\prime}\left(d_{x}+q_{4y}\right)^{2}G_{\left|\mathbf{q}_{4}\right|}\left(x-x^{\prime}\right)\right.\\
\left.+u_{1}^{*}u_{\overline{2}}u_{3}^{*}u_{\overline{4}}^{\prime}\left(d_{x}^{2}-q_{4y}^{2}\right)G_{\left|\mathbf{q}_{4}\right|}\left(x-x^{\prime}\right)\right.\\
\left.+u_{1}^{*}u_{\overline{2}}v_{3}^{*}v_{\overline{4}}^{\prime}\left(d_{x}^{2}-q_{4y}^{2}\right)G_{\mathbf{q}_{4}}\left(x-x^{\prime}\right)\right.\\
\left.-u_{1}^{*}u_{\overline{2}}v_{3}^{*}u_{\overline{4}}^{\prime}\left(d_{x}-q_{4y}\right)^{2}G_{\mathbf{q}_{4}}\left(x-x^{\prime}\right)\right]
\end{array}
\end{equation}

\begin{equation}
\begin{array}{c}
I_{d1,2,3,4}^{-+-+}=\frac{\pi\mu_{B}^{2}}{2A}\iint dxdx^{\prime}\times\\
\left[\left(q_{1z}+q_{2z}\right)^{2}\left(v_{\overline{1}}v_{2}^{*}+v_{1}^{*}v_{\overline{2}}\right)\left(v_{\overline{3}}^{\prime}v_{4}^{\prime*}+v_{3}^{\prime*}v_{\overline{4}}^{\prime}\right)G_{\left|\mathbf{q}_{1}+\mathbf{q}_{2}\right|}\left(x-x^{\prime}\right)\right.\\
\left.-v_{\overline{1}}v_{2}^{*}v_{\overline{3}}u_{4}^{\prime*}\left(d_{x}+q_{4y}\right)^{2}G_{\left|\mathbf{q}_{4}\right|}\left(x-x^{\prime}\right)\right.\\
\left.+v_{\overline{1}}v_{2}^{*}v_{\overline{3}}v_{4}^{\prime*}\left(d_{x}^{2}-q_{4y}^{2}\right)G_{\left|\mathbf{q}_{4}\right|}\left(x-x^{\prime}\right)\right.\\
\left.+v_{\overline{1}}v_{2}^{*}u_{\overline{3}}u_{4}^{\prime*}\left(d_{x}^{2}-q_{4y}^{2}\right)G_{\mathbf{q}_{4}}\left(x-x^{\prime}\right)\right.\\
\left.-v_{\overline{1}}v_{2}^{*}u_{\overline{3}}v_{4}^{\prime*}\left(d_{x}-q_{4y}\right)^{2}G_{\mathbf{q}_{4}}\left(x-x^{\prime}\right)\right]
\end{array}
\end{equation}
All the integrals participating in $I_d$ can be calculated explicitly since the integrand is the product of sines, cosines and exponential function of $\vert x - x^{\prime}\vert$. However, the large number of different combinations of sines and cosines and the necessity to use different exponents depending on the sign of $x -x^{\prime}$ makes real calculation sufficiently tiresome to charge a computer with this task. For the coefficients $I_e$ the calculations are much simpler since they include only sines and cosines and integrals over one variable $x$. However,  6 independent coefficients $I_e$ contain about 30 different integrals, so that charging computer with this task is again justified.

\section*{Appendix 3. 1/r-G-identity .}
From the Fourier transfromation of $\frac{1}{\left|\mathbf{r-r}^{\prime}\right|}$
we have
\begin{eqnarray*}
\frac{1}{\left|\mathbf{r-r}^{\prime}\right|}&=&\frac{1}{(2\pi)^3}\iiintop_{-\infty}^{\infty}d\mathbf{q}e^{i\mathbf{q}\mathbf{r}}\frac{4\pi}{q^2}\\
&=&\frac{1}{(2\pi)^3}\iiintop_{-\infty}^{\infty}d\mathbf{q}e^{i\mathbf{q}_{\Vert}\left(\mathbf{r}_{\Vert}-\mathbf{r}_{\Vert}^{\prime}\right)+iq_x\left(x-x^{\prime}\right)}\frac{4\pi}{q_{\Vert}^2+q_x^2}\\
&=& \frac{1}{(2\pi)^3}\iintop_{-\infty}^{\infty}dq_ydq_ze^{i\mathbf{q}_{\Vert}\left(\mathbf{r}_{\Vert}-\mathbf{r}_{\Vert}^{\prime}\right)}\intop_{-\infty}^{\infty}dq_x e^{iq_x\left(x-x^{\prime}\right)}\frac{4\pi}{q_{\Vert}^2+q_x^2}
\end{eqnarray*}
Since $\intop_{-\infty}^{\infty}dq_x e^{iq_x\left(x-x^{\prime}\right)}\frac{4\pi}{q_{\Vert}^2+q_x^2}=\frac{4\pi^2}{q_{\Vert}}e^{-q_{\Vert}|x-x^{\prime}}|=8\pi^2G_{q_{\Vert}}$
Then we get the 1/r-G-identity
\begin{equation}
\frac{1}{\left|\mathbf{r-r}^{\prime}\right|}=\frac{1}{\pi}\iintop_{-\infty}^{\infty}dq_{y}dq_{z}e^{i\mathbf{q}_{\Vert}\left(\mathbf{r}_{\Vert}-\mathbf{r}_{\Vert}^{\prime}\right)}G_{q_{\Vert}}\left(x-x^{\prime}\right)
\end{equation}


\begin{thebibliography}{99}

\bibitem{Landau 1935} L.D. Landau and E.M. Lifshitz, Phys.
Zs. Sowiet. \textbf{8}, 153, 1935.

\bibitem{Landau 1984} L.D. Landau and E.M. Lifshitz, Electrodynamics
of Continuous Media, Elsevier, 2nd Edition, 1984, Ch. 5.

\bibitem{Schlöman 1959} E. Schlöman, Phys. Rev. \textbf{116},
828 (1959).

\bibitem{Patton 2010}Pavol Krivosik and Carl E. Patton, Phys.
Rev. B \textbf{82}, 184428 (2010).

\bibitem{Damon 1961} R.W. Damon and J.R. Eshbach, J. Phys.
Chem. Solids \textbf{19}, 308 (1961).

\bibitem{Gann 1967} V.V. Gann, Sov. Phys. Solid State \textbf{8},
2537.

\bibitem{Wolfram 1970} T. Wolfram and R.R. De Wames, Phys.
Rev. Lett. \textbf{24}, 1489 (1970).

\bibitem{Kalinikos 1980} B.A. Kalinikos, IEEE Proc. H \textbf{127},
4 (1980).

\bibitem{Kalinikos 1986}B.A. Kalinikos and A.N. Slavin, J.
Solid State Phys. \textbf{19}, 7013 (1986).

\bibitem{Arias 2016} R.E. Arias, Phys. Rev. B \textbf{94},
134408 (2016).

\bibitem{Li 2018} Gang.Li, Chen Sun, T. Nattermann and V.L.
Pokrovsky, Phys. Rev. B \textbf{98}, 014436 (2018).

\bibitem{Demokritov 2006} S.O. Demokritov, V.E. Demidov,
O. Dzyapko, G.A. Melkov, A.A. Serga, B. Hillebrands, and A.N. Slavin,
Nature (London) \textbf{443}, 430 (2006)

\bibitem{Goldstein 2011} H. Goldstein, C.P. Poole and J.
Safko, Classical Mechanics, 3d edition, Pearson Education, 2011.

\bibitem{Kolokolov 1983} I.V. Kolokolov, V.S. L'vov and V.B.
Cherepanov, Zh. Eksp. Theor. Fiz. \textbf{84}, 1043 (1983) {[}Sov.
Phys. JETP \textbf{57}, 605 (1983).

\bibitem{Sonin 2017} E.B. Sonin, Phys. Rev. B \textbf{95},
144432 (2017).

\bibitem {Kreisel 2009} A. Kreisel, F.Sauli, L. Bartosch, and P. Kopietz, Eur. Phys. J B \textbf{71}, 59 (2009).

\bibitem{Serga} A. A. Serga, C. W. Sandweg, V. I. Vasyuchka, M. B. Jungfleisch, B. Hillebrands, A. Kreisel, P. Kopietz, and M. P. Kostylev. Phys. Rev. B $\textbf{86}$, 134403 (2012)

\bibitem{Demidov:2008cg} V. E. Demidov, O. Dzyapko, S. O. Demokritov, G. A. Melkov, and A. N. Slavin, Phys. Rev. Lett. $\textbf{100}$, 047205 (2008).

\bibitem{Lim 2018}
J. Lim, W. Bang, J. Trossman, A. Kreisel, M.B. J\"ungfleisch, A. Hoffmann, C.S. Tsal and J.B. Ketterson, Study of micron scale dispersion of spin waves in Yttrium Iron Garnet film. Absrract of presentation at March APS Meeting 2018, Los Angeles. 



\bibitem{Sun 2017} Chen Sun, Thomas Nattermann and Valery
L Pokrovsky, J. Phys. D: Appl. Phys. \textbf{50}, 143002 (2017).

\bibitem{Bunkov 2008} Y.M. Bunkov and G.E. Volovik, J. Low
Temp. Phys. \textbf{150}, 135 (2008).

\bibitem{Novik-Boltyk 2012} P. Novik-Boltyk, O. Dzyapko,
V.E. Demidov, N.G. Berloff and S.O. Demokritov, Sci. Rep. \textbf{2},
482 (2012).

\bibitem{Li 2013}
    Fuxiang Li, Wayne M. Saslow, and Valery L. Pokrovsky
    Sci Rep $\textbf{3}$, 1372 (2013)

\bibitem{Bradley 1997} C.C. Bradley, C.A. Sackett, and R.G.
Hulet, Phys. Rev. Lett. \textbf{78}, 985, (1997); C.C. Bradley, C.A.
Sackett, and R.G. Hulet, Phys. Rev. A \textbf{55}, 3951, (1997).

\bibitem{Roberts 2001} J. L. Roberts, N. R. Claussen, S.
L. Cornish, E. A. Donley, E. A. Cornell, and C. E. Wieman Phys. Rev.
Lett. \textbf{86}, 4211 (2001).

\bibitem{Mueller 2000} Eric J. Mueller and Gordon Baym, Phys.
Rev. A \textbf{62}, 053605 (2000).

\bibitem{Pitaevskii 1996}L.P. Pitaevskii, Physics Letters
A \textbf{221}, 14 (1996).

\bibitem{Borisenko 2020-1} Borisenko, I., Divinskiy, B., Demidov, V.\textit{et al.} Direct evidence of spatial stability of Bose-Einstein condensate of magnons. Nat Commun $\textbf{11}$, 1691 (2020). 
\bibitem{Borisenko 2020-2} Borisenko, I.V., Demidov, V.E., Pokrovsky, V.L. \textit{et al.} Spatial separation of degenerate components of magnon Bose-Einstein condensate by using a local acceleration potential. Sci Rep $\textbf{10}$, 14881 (2020).


\bibitem{Tupitsyn 2008} I. S. Tupitsyn, P. C. E. Stamp, and A. L. Burin. Stability of Bose-Einstein Condensates of Hot Magnons in Yttrium Iron Garnet Films. Phys. Rev. Lett. $\textbf{100}$, 257202 (2008).

\bibitem{Rezende 2009} S.M. Rezende. Theory of coherence in Bose-Einstein condensation phenomena in a microwave-driven interacting magnon gas. Phys. Rev. B $\textbf {79}$, 174411 (2009).

\bibitem{Demokritov} Divinskiy, B., Merbouche, H., Demidov, V.E. et al. Evidence for spin current driven Bose-Einstein condensation of magnons. Nat Commun \textbf{12}, 6541 (2021). 

\bibitem{Hillebrands}  Noack, Timo B. and Vasyuchka, Vitaliy I. and Pomyalov, Anna and L'vov, Victor S. and Serga, Alexander A. and Hillebrands, Burkard, Evolution of room-temperature magnon gas: Toward a coherent Bose-Einstein condensate, Phys. Rev. B \textbf{104}, L100410 (2021).




\end{thebibliography}
\end{document}